\documentclass[twocolumn,10pt,a4paper]{article}%
\usepackage[T1]{fontenc}
\usepackage{amsmath}
\usepackage{amsthm}
\usepackage{amssymb}
\usepackage{upgreek}
\usepackage{graphicx}
\usepackage[caption=false]{subfig}
\usepackage{color}
\usepackage{resizegather}
\usepackage{enumerate}
\usepackage{hhline}
\usepackage[shortlabels]{enumitem}
\usepackage{bm}
\usepackage{array}
\usepackage[ruled,vlined]{algorithm2e}
\usepackage{soul}
\usepackage{algorithmic}
\usepackage{hyperref}
\usepackage{tikz} 
\usetikzlibrary{arrows.meta} 
\usepackage{pifont}
\usepackage{multirow}

\usepackage[left=1.5cm, right=1.5cm, top=2cm, bottom=2cm, columnsep=0.6cm]{geometry}
\usepackage[round, authoryear]{natbib}
\usepackage{authblk}
\usepackage{titlesec}

\let\oldappendix\appendix
\renewcommand{\appendix}{
  \oldappendix
  \titleformat{\section}[block]
    {\normalfont\Large\bfseries}
    {Appendix\ \thesection: }
    {0pt}
    {\Large}
}

\theoremstyle{definition}
\newtheorem{defn}{Definition}{\bfseries}{\rmfamily}

\newtheorem{remk}{Remark}{\bfseries}{\rmfamily}

\newcommand{\remarkend}{\unskip\nobreak\hspace{1em}\penalty0\hbox{}\hfill$\triangle$}
\newcommand{\abs}[1]{\left|#1\right|}
\newcommand{\hidden}{H$[i]$dden\xspace}

\makeatletter
\def\linkref[#1]#2{%
  \extract@labelname#2\@nil{#1}%
}

\def\extract@labelname\ref#1\@nil#2{%
  \hyperref[#1]{#2~\ref*{#1}}%
}
\makeatother

\theoremstyle{plain}
\newtheorem{thm}{Theorem}

\newcommand{\agg}{^\mathrm{Agg}}
\newcommand{\tps}{^\mathrm{T}}
\renewcommand{\Re}{\operatorname{Re}}
\renewcommand{\Im}{\operatorname{Im}}

\DeclareMathOperator{\Mod}{\mathrm{mod}}
\renewcommand{\pmod}[1]{\,(\Mod #1)}

\newcommand{\enc}{\mathrm{Enc}}
\newcommand{\dec}{\mathrm{Dec}}

\begin{document}

\begin{titlepage}
\thispagestyle{empty}

\noindent {\huge \textbf{Citation for published version}} \\[2em]

\large

\noindent \textbf{Please cite this article as:}\\[0.75em]
D. Megías, ``Complex domain approach for reversible data hiding and homomorphic encryption: General framework and application to dispersed data,'' \textit{Journal of Information Security and Applications}, vol. 102, Art. No. 104596, 2026.\\[2em]
  
\noindent \textbf{DOI:}\\[0.75em] 
\url{https://doi.org/10.1016/j.jisa.2026.104596}\\[2em]
    
\noindent\textbf{Document version:}\\[0.75em] 
This is the \textbf{accepted manuscript} version.\\[2em]
    
\noindent \textbf{Copyright and Reuse:}\\[0.75em] 
{\small \copyright~2026. This manuscript version is made available under the CC-BY-NC-ND 4.0 license (\url{https://creativecommons.org/licenses/by-nc-nd/4.0/)}}.
\end{titlepage}
\clearpage
\pagenumbering{arabic} 

\title{Complex domain approach for reversible data hiding and homomorphic encryption: General framework and application to dispersed data}
\author{David Meg\'ias}
\affil{UOC-TECH Research Centre, Faculty of Computer Science, Telecommunication and Multimedia, Universitat Oberta de Catalunya (UOC), CYBERCAT-Center for Cybersecurity Research of Catalonia,Rambla del Poblenou, 154, Barcelona, 08018, Spain\\ \vspace{4mm}
e-mail:\texttt{dmegias@uoc.edu}}

\twocolumn[
  \begin{@twocolumnfalse}
    \maketitle
    \begin{abstract} \noindent Ensuring the trustworthiness of data from distributed and resource-constrained environments, such as Wireless Sensor Networks or IoT devices, is critical. Existing Reversible Data Hiding (RDH) methods for scalar data suffer from low embedding capacity and poor intrinsic entanglement between host data and watermark. This paper introduces Hiding in the Imaginary Domain with Data Encryption (\hidden), a novel framework based on complex number arithmetic for simultaneous information embedding and encryption. The \hidden framework offers perfect reversibility, highly scalable watermark capacity decoupled from the host data's range, and intrinsic data-watermark entanglement. The paper further introduces two protocols: \hidden-EG, for joint reversible data hiding and encryption, and \hidden-AggP, for privacy-preserving aggregation of watermarked data, based on partially homomorphic encryption. Rigorous comparative evaluation against recent state-of-the-art baselines demonstrates that the framework significantly expands embedding capacity and achieves an exponential improvement in False Data Injection (FDI) resilience ---ranging from $10^5$ to $10^{19}$ under standard configurations--- effectively neutralizing targeted structural vulnerabilities such as homomorphic injection attacks. Furthermore, empirical performance analysis ---encompassing hardware-agnostic cryptographic metrics and direct IoT edge hardware emulation (ESP32)--- confirms the practical deployment feasibility of the framework within standard sensor duty cycles. Ultimately, these protocols provide efficient and resilient solutions for data integrity, provenance (or group-level integrity in the aggregated case), and confidentiality, serving as a foundation for new schemes based on the algebraic properties of the complex domain.

    \vspace{4mm}
        
    \noindent\textbf{Keywords:} Reversible data hiding, Complex domain, Gaussian integers,  Homomorphic encryption, Data aggregation, Dispersed data, Paillier, ElGamal, Data integrity

    \vspace{4mm}
    \end{abstract}

  \end{@twocolumnfalse}
]

\section{Introduction}\label{Xsec1-1}
\label{sec:introduction}

The ubiquitous generation and pervasive use of digital data are remarkable characteristics of the modern era. From vast networks of Internet of Things (IoT) devices and industrial sensors to intricate smart city infrastructures and critical healthcare systems, data fuels decision-making, automates processes, and enables unprecedented levels of connectivity. This reliance on data, often collected from distributed and potentially vulnerable environments, is critically linked to its trustworthiness \citep{yan2014survey}. Ensuring the confidentiality, integrity, authenticity, and privacy of information has become a priority, as compromised data can lead to erroneous conclusions, system failures, and severe breaches of privacy or security \citep{liang2016review}.

Within this complex digital landscape, the concept of \textbf{Reversible Data Hiding (RDH)} \citep{shi:2016,kumar2022reversible} emerges as a specialized and increasingly popular approach to data security. RDH involves embedding secret information (a ``watermark'') directly into a host data item in such a way that, after extraction of the watermark, the original host data can be perfectly and exactly restored to its original state. This unique property, combining data embedding with guaranteed reversibility, makes RDH appropriate in sensitive applications such as medical imaging, legal documentation, and military intelligence, where any permanent alteration of the original data is unacceptable. Unlike traditional data encryption, which primarily ensures confidentiality, RDH focuses on providing intrinsic integrity, authenticity, and provenance, within the data\break itself.

In the context of individual, scalar, or time-series numerical data, such as those originating from Wireless Sensor Networks (WSNs) \citep{shi:2013}, IoT devices \citep{rahman:2021}, or critical financial and medical monitoring systems \citep{bahlearao:2019}, RDH offers unique advantages. Foremost among these is its ability to ensure robust data integrity and authenticity at the granular level of single readings or sequential observations. By embedding a watermark that can be extracted and verified, receivers (e.g., a Data Collector in a WSN) can confirm that the data originated from a legitimate source and has not been tampered with in transit. This inherent capability to facilitate data provenance and traceability is particularly valuable, allowing for forensic analysis of data origin and processing paths in distributed and often resource-constrained environments.

Beyond these general benefits, RDH offers distinct advantages compared to other commonly employed security primitives in certain contexts. Unlike cryptographic hashing or digital signatures, which typically append a separate authentication tag or signature to the data, RDH embeds the integrity verification mechanism directly \textit{within} the data itself. This can simplify data handling and ensure the integrity proof travels jointly with the content. While conventional encryption ensures confidentiality, it does not directly provide data integrity; and Physical Unclonable Functions (PUFs) \citep{gao2020physical} focus on device authentication rather than data integrity. RDH uniquely combines content-dependent integrity (through watermark verification) with the guarantee of perfect original data restoration, a feature not offered by one-way hash functions or irreversible authentication tags. This intrinsic embedding can be particularly beneficial in constrained environments or specialized data formats where appending additional metadata is undesirable or challenging.

From an implementation perspective, basic RDH techniques can often be quite lightweight, requiring minimal computational resources. This aspect is particularly useful for deployment on resource-constrained devices like low-power sensors. However, it is worth pointing out that this ``lightweight'' advantage primarily applies to simple, non-encrypted RDH schemes. When RDH is combined with advanced cryptographic primitives, such as the homomorphic encryption schemes explored in this paper, the computational overhead significantly increases, with cryptographic operations becoming the dominant factor. Nevertheless, the enhanced security and privacy offered by such combination often justify this increased complexity, addressing critical requirements for secure and trustworthy data\break ecosystems.

Despite these advantages, a significant limitation of existing RDH techniques for individual or dispersed data, particularly in high-security contexts, is their low embedding capacity. Typically, these methods allow for the embedding of only 1 bit per data sample or even less (as is often the case with difference expansion methods). This low capacity severely limits the granularity of tampering detection and the strength of integrity guarantees for independent, single data items. For instance, if only 1 bit is embedded in a single measurement, the probability of an attacker randomly guessing the correct bit is 1/2, offering very weak protection. While the security scales with the number of samples, making the probability of deceiving detection negligible for a sufficient number of measurements $M$ (e.g., $2^{-M}$), this aggregate protection is not robust in situations where the frequency of measurements or the total number of independent data points is small. This highlights a gap in current RDH capabilities for truly strong, individual data item integrity in such sparse or low-frequency scenarios.

In a parallel development within cryptography, \textbf{Homomorphic Encryption (HE)} \citep{acar2018survey} has emerged as a transformative technology for data privacy \citep{iezzi2020practical}. HE schemes allow computations to be performed directly on encrypted data without ever needing to decrypt it \citep{zhang2020lvpda}. The output of such a computation is an encrypted result that, when decrypted, is identical to the result of the computation performed on the unencrypted data. This property enables the processing of sensitive information by untrusted third parties (such as cloud servers) while ensuring that the data's confidentiality remains fully protected throughout its\break lifecycle.

While HE is broadly applied to solve privacy challenges in cloud computing, its principles can be integrated with other security primitives to achieve more complex security goals. For instance, combining data integrity mechanisms like RDH with HE's ability to operate on encrypted data allows for the creation of protocols that simultaneously guarantee confidentiality, authenticity, and data integrity. This powerful synergy is especially relevant in systems that require both privacy during data processing and proof of data origin and integrity, setting the stage for more robust and trustworthy data ecosystems.

\subsection{Plan and contributions of the paper}\label{Xsec2-1.1}
\label{sec:contributions}

This paper introduces a novel framework for RDH and encryption, taking advantage of the algebraic properties of the complex domain (Gaussian integers). Termed \textbf{Hiding in the Imaginary Domain with Data Encryption (\hidden)}, the proposed approach fundamentally addresses key limitations of existing RDH techniques when applied to individual, scalar, or dispersed data. Unlike methods that merely append or isolate hidden bits in low-significance positions (e.g., modulo-based schemes), \hidden uses complex number arithmetic to entangle the original data and the watermark, providing a richer embedding space and assuring a larger tampering detection capacity.

The main contributions of the paper are summarized below:

\begin{itemize}
 \item A novel core complex-domain reversible information hiding scheme for individual data, offering intrinsic data-watermark entanglement, perfect reversibility and, highly scalable watermark capacity decoupled from the host data's range.

 \item A distinct application of the core \hidden embedding scheme to aggregated data, which prepares individual data items for privacy-preserving and efficient summation of \textit{watermarked data} while retaining extractability.

 \item A novel application of the extension of ElGamal encryption to the Gaussian integers within the \textbf{\hidden-EG} protocol, providing joint reversible information hiding and encryption specifically for individual data points.

 \item A novel component-wise application of Paillier encryption to complex-valued data, establishing a privacy-preserving additive homomorphism for secure aggregation.

 \item The \textbf{\hidden-AggP} protocol, enabling joint reversible information hiding, encryption, and privacy-preserving data aggregation from multiple sources, built upon the component-wise Paillier application.

 \item A comprehensive comparative evaluation against state-of-the-art scalar RDH-HE schemes, demonstrating a security improvement of up to nineteen orders of magnitude in FDI resilience (under reasonable settings) alongside practical hardware-level feasibility on IoT edge devices (ESP32).
\end{itemize}

Through these contributions, the paper demonstrates a new paradigm for reversibly embedding authenticating information directly into sensitive data, providing strong cryptographic protection while maintaining the relevant property of perfect host data reversibility. Please note that, while the \hidden-EG protocol provides rigorous per-node provenance for individual readings, the \hidden-AggP protocol is designed to prioritize group-level integrity and scalable homomorphic aggregation, effectively securing the aggregate dataset against structural tampering.

The rest of the paper is organized as follows. \linkref[Section]{\ref{sec:background}} provides the necessary background on RDH and HE. \linkref[Section]{\ref{sec:reversible}} details the proposed complex-domain reversible watermarking scheme for both individual and aggregated data. \linkref[Section]{\ref{sec:homomorphic}} presents the extension of ElGamal and Paillier encryption to the complex domain. \linkref[Section]{\ref{sec:protocols}} details the \hidden-EG and \hidden-AggP protocols for joint watermarking and encryption and aggregation, respectively. A comprehensive formal security and privacy analysis of the proposed protocols is provided in \linkref[Section]{\ref{sec:security}}. \linkref[Section]{\ref{sec:performance}} presents a rigorous performance and comparative analysis against state-of-the-art baselines, evaluating embedding capacity, cryptographic complexity, structural security, and practical IoT edge feasibility. Finally, \linkref[Section]{\ref{sec:conclusion}} concludes the paper and outlines future research directions.

\section{Background}\label{Xsec3-2}
\label{sec:background}

This section provides the foundational concepts and related work that are essential for understanding the proposed framework. The core of this work lies at the intersection of two distinct but complementary fields of research: RDH and HE. While RDH offers powerful mechanisms for embedding information within data for integrity and provenance, HE provides the unique capability to perform computations on encrypted data, thereby ensuring confidentiality. The following sections detail the relevant background on these two key areas.

\subsection{Reversible data hiding}\label{Xsec4-2.1}
\label{sec:reversible_data_hiding}

Reversible Data Hiding \citep{shi:2016,kumar2022reversible} is a specialized field within information hiding that focuses on embedding secret information (a watermark) into host data in such a way that the original host data can be perfectly and exactly recovered after the watermark has been extracted. This property is essential in applications that do not allow the permanent alteration of the original content.

Historically, RDH techniques have been extensively developed and applied primarily to multimedia content, including digital images, audio, and video \citep{shi:2016}. Methods like Least Significant Bit (LSB) modification (when combined with original LSB storage) \citep{celik:2002}, Difference Expansion (DE) \citep{tian:2003}, Histogram Shifting (HS) \citep{ni:2006}, and Prediction-Error Expansion (PEE) \citep{ou:2013} use the inherent statistical redundancy and the human perceptual system's tolerance for slight alterations in these rich data formats. These approaches can embed information in an imperceptible manner while ensuring perfect reversibility, meaning the host data is faithfully restored to its original state after watermark\break extraction.

In contrast, the application of RDH to individual, scalar, or time-series numerical data (e.g., sensor readings, financial figures, or simple numerical sequences) presents distinct and more challenging problems. Unlike multimedia, such data typically exhibits very low inherent redundancy, is highly sensitive to even minor numerical changes (as there is no ``perceptual'' domain to hide in), and has a very limited host capacity. This scarcity of ``hiding space'' and the demand for exact numerical preservation make traditional RDH techniques, often designed for media, largely unsuitable.

For this class of individual or dispersed data, existing reversible embedding methods tend to fall into a few categories. One common approach is \textbf{Modulo or Remainder-based embedding}, where the secret information $w$ (an integer in range $0, \ldots, q-1$) is embedded into a host integer $d$ using an operation like $d' = q \cdot d + w$ \citep{kabir2025rewap}. A common special case involves embedding a single bit ($w \in \{0,1\}$) by setting $q=2$, resulting in $d' = 2d + w$ \citep{kabir2023watercrypt}. Upon extraction, $w$ is simply $d' \pmod q$ and $d$ is $\lfloor d' / q\rfloor$. When $q$ is a power of 2 ($q=2^B$), these extraction operations can be efficiently performed using bitwise logic (e.g., $w = d' \& (q-1)$ and $d = d' \gg B$).

While perfectly reversible and simple, a significant characteristic of this method is that the host data's magnitude is directly scaled or inflated by a factor of $q$, making the numerical modification immediately obvious and leading to a poor entanglement between the original data and the watermark.

While modulo-based embedding places the watermark in a very distinct (e.g., least significant) position within the watermarked data, one might consider more intricate bit-level operations to ``entangle'' the bits of the original data and the watermark more thoroughly within the resulting embedded value. Such approaches would likely involve complex sequences of bitwise operations, permutations, or scrambling algorithms. However, these methods, even if designed to be perfectly reversible, tend to lack mathematical elegance and direct interpretability. The resulting watermarked data would exhibit a less intuitive relationship to its original host and watermark, making such bit-mixing techniques generally less desirable compared to approaches that achieve blending through inherent algebraic properties.

Another relevant approach for sequential data is \textbf{Difference Expansion (DE)}, or its more general form, \textbf{Prediction-Error Expansion} \citep{shi:2013}. These methods exploit relationships between consecutive data samples by computing prediction errors (e.g., $e_k = d_k - d_{k-1}$) and then embedding information by expanding these errors.

These existing methods for individual/scalar data often suffer from relatively low embedding capacity \cite{kabir2024privacy}. This highlights the need for novel approaches that can integrate data hiding more subtly or within a more complex framework, especially when combined with other security operations like encryption.

It is important to position the proposed framework within the specific constraints of non-redundant, time-dispersed scalar data. Unlike multimedia content, which contains spatial or temporal redundancy that can be exploited for reversible embedding, independent scalar samples provide no ``room'' for a watermark. In existing literature, achieving reversibility in such data typically requires data expansion (e.g., $d' = d \cdot q + w$), which increases the bit-depth of the scalar (from $A$ to $A+B$ bits, as detailed in \linkref[Remark]{\ref{rem:expansion}}).

In light of these constraints, we categorize the proposed framework as a form of \textit{Domain-Extending RDH}. Unlike traditional \textit{Domain-Preserving RDH}, which seeks to embed data within the original real-valued representation, Domain-Extending RDH maps the host data into a higher-dimensional algebraic structure, such as the complex plane $\mathbb{Z}[i]$. While this transition from 1D to 2D is conceptually analogous to bit-depth expansion ($d' = d + iw$), it is a deliberate design choice. By extending the domain, we create the necessary ``algebraic room'' to accommodate the watermark while ensuring the structure remains invariant under the homomorphic transformations required for secure data aggregation.

\subsection{Homomorphic encryption}\label{Xsec5-2.2}

Homomorphic encryption \citep{acar2018survey} is a unique and powerful class of cryptographic schemes that allows a user to perform computations on encrypted data without first having to decrypt it. The result of these operations is an encrypted output which, when decrypted, is the same as if the operations had been performed on the original, unencrypted data. This remarkable capability is useful for maintaining data privacy and confidentiality when computations are outsourced to untrusted third parties, such as cloud services or remote data aggregators.

HE schemes are typically classified based on the types and number of operations they support. Partially Homomorphic Encryption (PHE) schemes, such as \citep{cominetti2020fast}, support an unlimited number of a single type of operation, either addition or multiplication. Somewhat Homomorphic Encryption (SHE) schemes, which include the pioneering work by Boneh-Goh-Nissim (BGN) in \citep{boneh2005evaluating}, can perform a limited number of both additions and multiplications. Finally, Fully Homomorphic Encryption (FHE) schemes, such as Ring Learning with Errors (RLWE) \citep{brakerski2011fully}, support an unlimited number of both operations, allowing for arbitrary computations on encrypted data. In this paper, two classical PHE schemes, \citet{elgamal1985public} and \citet{paillier1999public}, are applied for their distinct homomorphic properties that enable the proposed protocols for both individual data encryption and privacy-preserving data aggregation.

\subsubsection{ElGamal encryption}\label{Xsec6-2.2.1}

ElGamal encryption \citep{elgamal1985public} is an asymmetric key encryption algorithm based on the Diffie-Hellman key exchange scheme. Its security relies on the difficulty of the discrete logarithm problem (DLP) in the multiplicative group of integers modulo a prime number.

The cryptosystem works as follows. Let $p$ be a prime number and $g$ be a generator of the cyclic multiplicative group $\mathbb{Z}_p^*$, often also denoted as $\left({\mathbb Z/p\mathbb Z}\right){}^{\times}$. The ElGamal encryption scheme consists of three algorithms:

\begin{itemize}
 \item Key generation: The key generation process involves selecting a random integer $a$ such that $1 \leq a \leq p-1$ as the private key. Then, $K = g^a \pmod{p}$ is computed. The public key is formed as the triple $\mathbf{K}=(p, g, K)$.
 \item Encryption: To encrypt a message $m \in \mathbb{Z}_p^*$, a random integer $b$, such that $1 \leq b \leq p-1$, is selected as the ephemeral key. The ciphertext is then computed as $y_1 = g^b \pmod{p}$ and $y_2 = m \cdot K^b \pmod{p}$, resulting in the pair $(y_1, y_2)$. The encryption function can be denoted as follows:
 \[(y_1,y_2)=\enc_{\mathbf K}(m, b) = \left(g^b \pmod{p}, m \cdot K^b \pmod{p}\right).\]
 \item Decryption: The decryption of the ciphertext $(y_1, y_2)$ using the private key $a$ involves computing the shared secret $s = y_1^a \pmod{p}$ and then decrypting the message as $m = y_2 \cdot s^{-1} \pmod{p}$. This decryption function can be formally denoted as follows: \[\dec_{a}(y_1, y_2) = y_2 \cdot (y_1^a){}^{-1} \pmod{p}.\]
\end{itemize}

The security of ElGamal encryption relies on the difficulty of the DLP in $\mathbb{Z}_p^*$. If an attacker can efficiently solve the DLP, they can compute the private key $a$ from the public key $K$ and then decrypt any ciphertext. However, for sufficiently large prime numbers $p$, the DLP is believed to be computationally infeasible.

ElGamal encryption exhibits a homomorphic property under component-wise multiplication of ciphertexts. This means that given two ciphertexts $\enc_{\mathbf K}(m_1, b_1) = (y_1, y_2)$ and $\enc_{\mathbf K}(m_2, b_2) = (y'_1, y'_2)$, multiplying them component-wise as $(y_1 \cdot y'_1, y_2 \cdot y'_2)$ results in a ciphertext that decrypts to the product of the original plaintexts:
\[\dec_a\left(\enc_{\mathbf K}(m_1, b_1) \cdot \enc_{\mathbf K}(m_2, b_2)\right) = m_1 \cdot m_2 \pmod{p}.\]
This property has applications in various cryptographic protocols and it is exploited to combine reversible watermarking and encryption in this paper.

\subsubsection{Paillier encryption}\label{Xsec7-2.2.2}
\label{sec:paillier}

\citet{paillier1999public} is a probabilistic asymmetric encryption algorithm. Unlike ElGamal, which is homomorphic with respect to multiplication, Paillier encryption is additively homomorphic, meaning that sums of plaintexts can be computed from their ciphertexts.

The cryptosystem works as follows. It operates in the group $\mathbb{Z}_{n^2}^*$, where $n$ is a product of two large prime numbers. The Paillier encryption scheme consists of three algorithms:

\begin{itemize}
 \item Key generation:
 \begin{enumerate}
 \item Select two large random prime numbers $p$ and $q$ such that $p \neq q$ and $\mathrm{gcd}(pq, (p-1)(q-1)) = 1$.
 \item Compute $n = pq$ and $L_P = \mathrm{lcm}(p-1, q-1)$, where ``$\mathrm{lcm}(\cdot)$'' stands for the least common multiple.
 \item Select a random integer $g \in \mathbb{Z}_{n^2}^*$ such that $g$ has order a multiple of $n$ modulo $n^2$. A common choice is $g = n+1$.
 \item Compute $M_P = (L(g^{L_P} \pmod{n^2})){}^{-1} \pmod n$, where the function $L(x)$ is defined as
 \[
 L(x) = \frac{x-1}{n}.
 \]
 \end{enumerate}
 The public key is formed as the pair $\mathsf{K_{pub}}=(n, g)$, and the private key is $\mathsf{K_{priv}}=(L_P, M_P)$.

 \item Encryption: To encrypt a message $m$, where $0 \leq m < n$, a random integer $r$ is selected such that $0 < r < n$ and $\mathrm{gcd}(r, n) = 1$. The ciphertext $c$ is then computed as:
 \[c = g^m \cdot r^n \pmod{n^2}.\]
 The encryption function can be denoted as \[\enc_{\mathsf{K_{pub}}}(m, r) = g^m \cdot r^n \pmod{n^2}.\]

 \item Decryption: The decryption of a ciphertext $c$ using the private key $(L_P, M_P)$ involves computing:
 \[m = L(c^{L_P} \pmod{n^2}) \cdot M_P \pmod n.\]
 This decryption function can be formally denoted as $\dec_{\mathsf{K_{priv}}}(c) = L(c^{L_P} \pmod{n^2}) \cdot M_P \pmod n$.
\end{itemize}

The security of Paillier encryption relies on the computational difficulty of the composite residuosity problem. This problem is considered hard for sufficiently large $n$, making it computationally infeasible for an attacker to recover the plaintext $m$ from a ciphertext $c$ without knowing the private key $(L_P, M_P)$.

A core feature of Paillier encryption is its additive homomorphic property. This means that operations can be performed on ciphertexts such that the decryption of the result corresponds to the addition of the original plaintexts.
\begin{enumerate}
 \item Addition of encrypted numbers: Given two ciphertexts $c_1 = \enc_{\mathsf{K_{pub}}}(m_1, r_1)$ and $c_2 = \enc_{\mathsf{K_{pub}}}(m_2, r_2)$, their product modulo $n^2$ decrypts to the sum of their plaintexts:
 \[\dec_{\mathsf{K_{priv}}}(c_1 \cdot c_2 \pmod{n^2}) = (m_1 + m_2) \pmod n.\]
 \item Multiplication by a plaintext constant: Given a ciphertext $c_1 = \enc_{\mathsf{K_{pub}}}(m_1, r_1)$ and a plaintext constant $k$, the ciphertext raised to the power of $k$ modulo $n^2$ decrypts to the product of the plaintext and the constant:
 \[\dec_{\mathsf{K_{priv}}}(c_1^k \pmod{n^2}) = k \cdot m_1 \pmod n.\]
\end{enumerate}
This additive homomorphism is particularly useful for applications requiring privacy-preserving aggregation, where sums of data points need to be computed without revealing individual values. In the context of this paper, it will be exploited for secure data aggregation in a new protocol.

\section{Reversible watermarking approach on the complex domain}\label{Xsec8-3}
\label{sec:reversible}

This section presents a reversible watermarking scheme that can be used for isolated data measurements (e.g., readings of a smart meter or sensor data).

\subsection{Basic definitions}\label{Xsec9-3.1}\label{sec:basic}

First, we define the different elements of the reversible watermarking scheme.

\begin{defn}[Fine-grained original data]\label{def:data}
$d_{j,k}$ is the scalar data produced by the $j$th device (e.g., a sensor) in the \mbox{$k$th} time sample (time $t_k$).

Without loss of generality, we assume that the original data is an integer: $d_{j,k}\in \mathbb{Z}$, since it is represented on a digital device and can always be mapped to an integer (even if it is a floating-point number).
\label{Xenun32-1}
\end{defn}

\begin{defn}[Watermark]\label{def:wm}
$w_k$ is the watermark to be embedded at time $k$ in the fine-grained data $d_{j,k}$. Similarly, we assume $w_k\in \mathbb{Z}$.
\label{Xenun33-2}
\end{defn}

\begin{remk}The only assumption we are making with respect to the watermark $w_{k}$ in the core embedding scheme is that it is an integer. While this basic algebraic mapping imposes no theoretical bounds on the integer size, practical implementations are strictly constrained. Beyond basic data structure limits (e.g., standard 32-bit or 64-bit integers), integrating embedding with the homomorphic encryption protocols described in \linkref[Section]{\ref{Xsec22-5}} rigorously bounds the maximum capacity. Specifically, the watermark size is constrained by the chosen cryptographic modulus ($p$ or $n$) to prevent wrap-around and guarantee perfect reversibility, as formalized in \linkref[Remarks]{\ref{rem:ElGamal}} and \ref{rem:Paillier}. In addition, the maximum practical capacity may be limited by the acceptable communication overhead and the Maximum Transmission Unit (MTU) of the network, as evaluated in \linkref[Section]{\ref{sec:performance}}.
\remarkend \label{Xenun7-1}
\end{remk}

\begin{defn}(Reversible embedding and blind extraction functions)
\label{def:embext}
The embedding process is carried out using an embedding function $E_w$ in the following way:
\[
d'_{j,k}=E_w(d_{j,k},w_k,K_w),
\]
where $K_w$ is a watermarking key required for both the embedding and extraction processes. The reversible extraction or recovery function $R_w$ allows recovery of \textbf{both} the original data and the watermark, as follows:
\[
\left[d_{j,k}, w_k\right]\tps=R_w(d'_{j,k},K_w),
\]
for the watermarked data $d'_{j,k}$, where the superscript ``$\mathrm{T}$'' denotes the transposition operator.

Note that the extraction is assumed to be blind, since the original data is not used in the function.
\label{Xenun34-3}
\end{defn}

\begin{remk}\label{rem:expansion}
Unlike reversible data hiding for multimedia content or highly redundant data ---where source redundancy allows for techniques such as difference expansion, histogram shifting, or prediction error expansion to hide information without necessarily increasing carrier size--- embedding a watermark $w_k$ in a single scalar $d_{j,k}$ such that both can be perfectly recovered requires data expansion. For instance, as seen in \cite{kabir2025rewap}, embedding a $B$-bit watermark using an operation like $d'_{j,k} = 2^B \cdot d_{j,k} + w_k$ increases the bit-length of the original data by exactly $B$ bits. From an information-theoretic perspective, if the host data consists of $A$ bits and the watermark comprises $B$ bits, and both are independent and occupy their full possible ranges, the combined state space consists of $2^{A+B}$ distinct values. In such cases, the result must be encoded with at least $A+B$ bits for perfect recovery. While ``slack'' in a data's restricted range (e.g., values $d_{j,k}\in [101,200]$ encoded in a 1-byte container) can theoretically host bits without expansion, scalar reversible embedding of independent, full-range variables fundamentally requires increasing the bit-length of the transmitted values.
\remarkend\end{remk}

\begin{defn}(Aggregated data and watermarked aggregated data)\label{def:aggregated}
The aggregated data $S_k$, for the devices $j=1,2,\dots,N$ in the $k$th time sample is defined as follows:
\[
S_k = \sum_{j=1}^{N}d_{j,k}.
\]
Similarly, the aggregated watermarked data $S'_k$ can be defined as detailed below:
\[
S'_k = \sum_{j=1}^{N}d'_{j,k} = \sum_{j=1}^{N}E_w(d_{j,k},w_k,K_w).
\]
Note that the watermark embedded in the different devices is the same ($w_k$), although it can vary at every $t_k$.
\label{Xenun35-4}
\end{defn}

\begin{defn}(Reversible blind aggregated data and watermark extraction)
An aggregated data and watermark extraction function $R\agg_w$ allows extracting both $S_k$ and $w_k$ from $S'_k$ as follows:
\[
\left[S_k, w_k\right]\tps=R\agg_w(S'_k,K_w,[N]),
\]
where ``$[N]$'' indicates that the extraction function may require, as an additional parameter, the number of devices involved in the aggregation.
\label{Xenun36-5}
\end{defn}

\subsection{Proposed reversible watermarking scheme}\label{Xsec10-3.2}
This section details the embedding, extraction, and aggregated extraction functions for the proposed watermarking scheme.

First, let the watermarking key $K_w$ be a Gaussian integer $\lambda_k=a_k+i b_k$, where $i=\sqrt{-1}$ is the positive imaginary unit, $a_k,b_k\in\mathbb{Z}$, and $a_k, b_k\neq0$. The last condition guarantees that $\lambda_k$ is not either just real or imaginary, and also that $\lambda_k\neq0$. Thus, there exists a complex number $\lambda_k^{-1}$ such that $\lambda_k\lambda_k^{-1}=1$, with:
\[
\lambda_k^{-1} = \frac{1}{\lambda_k}=\frac{\overline{\lambda_k}}{\lambda_k\overline{\lambda_k}}=\frac{a_k-ib_k}{a_k^2+b_k^2},
\]
where the complex conjugate of a Gaussian integer $x+iy$ is denoted as $\overline{x+iy} = x-iy$.

Hence, $\lambda_k$ is simply a complex number with integer real $\Re(\cdot)$ an imaginary $\Im(\cdot)$ parts (\textit{i.e.,} a Gaussian integer, by definition): $\lambda_k\in\mathbb{Z}[i]$. In general, $\lambda_k^{-1}$ is not a Gaussian integer, but a Gaussian rational: $\lambda_k^{-1}\in\mathbb{Q}[i]$, which means that both real and imaginary parts of $\lambda_k^{-1}$ are rational numbers, or $\Re(\lambda_k^{-1}),\Im(\lambda_k^{-1})\in\mathbb{Q}$:
\[
\Re(\lambda_k^{-1})=\frac{a_k}{a_k^2+b_k^2},\:
\Im(\lambda_k^{-1})=\frac{-b_k}{a_k^2+b_k^2}.
\]

The secret watermarking key $\lambda_k$, termed as challenge factor, could be fixed or change at each time interval, hence the subscript ``$k$'' used in its definition. The changes on $\lambda_k$ can be scheduled according to the desired security level.

The embedding function, based on complex number arithmetic operations, is defined as follows. Let $\delta_{j,k}$ be the Gaussian integer $\delta_{j,k} = d_{j,k} + i w_k$.

\begin{remk} Note that $\delta_{j,k}$ is simply a complex number (Gaussian integer), whose real part is the original data and the imaginary part is the embedded watermark:
 $\Re(\delta_{j,k})=d_{j,k}$ and $\Im(\delta_{j,k})=w_k$.

 As detailed in \linkref[Definitions]{\ref{def:data}} and \ref{def:wm}, both the data and the watermark are assumed to be integers and, hence, the complex number $\delta_{j,k}$ is indeed a Gaussian integer. \remarkend
\label{Xenun9-3}
\end{remk}

Now, the watermarked data\footnote{We use the notation $\delta'_{j,k}$ instead of $d'_{j,k}$ to remark that the value is complex (Gaussian integer) and not a regular integer.} $\delta'_{j,k}$ can be obtained applying the following embedding function:
\begin{equation}\label{eq:embedding}
\begin{split}
E_w(d_{j,k},w_k,\lambda_k)&=\lambda_k \cdot \delta_{j,k}\\& = (a_k+i b_k)(d_{j,k} + i w_k) \\ & = (a_k d_{j,k}-b_k w_k)+i (a_k w_k+b_k d_{j,k}).
\end{split}
\end{equation}

\begin{remk}$\delta'_{j,k}$ is, again, a Gaussian integer, since it is the product of two Gaussian integers ($\lambda_k$ and $\delta_{j,k}$). Hence, the real and the imaginary parts of the watermarked data are integers.
This is particularly interesting if we want to carry out further operations on this data, such as public-key encryption. \remarkend
\label{Xenun10-4}
\end{remk}

This embedding approach illustrates the core principle of Domain-Extending RDH: rather than seeking to minimize ``distortion'' within the original 1D scalar domain, we prioritize structural integrity across the transformation. By mapping the regular integer-valued data to the real part and the watermark to the imaginary part of a Gaussian integer, we effectively create a ``two-channel'' container. This ensures that the host data and the watermark remain algebraically distinct but mathematically coupled, allowing them to undergo identical homomorphic rotations and summations without interfering with one another's reversibility.

\begin{remk}From a purely mathematical point of view, there is no need to constrain the data, the watermark and the watermarking key to be integers or Gaussian integers. Consequently, a formulation of the scheme based on real and complex numbers (without constraints) is perfectly possible. However, in practical implementations, using floating point numbers would produce round-off errors in simple computations (such as multiplications or divisions). In addition, floating point numbers are not handled directly in some operations, such as public-key encryption. For these reasons, the proposed method is defined using integers in this paper.
\remarkend
\label{Xenun11-5}
\end{remk}

\begin{remk}While the embedding function $E_w$ transforms scalar inputs $d_{j,k}$ and $w_k$ into a complex Gaussian integer $\delta_{j,k}'$, this transition does not require specialized complex-valued hardware or unconventional data architectures. As detailed in \linkref[Remark]{\ref{rem:vector}}, this approach is algebraically equivalent to a linear transformation of a $2 \times 2$ vector. In practical digital systems, a Gaussian integer can be represented as a simple structure or array consisting of two standard integers for the real and imaginary components. Although this admittedly doubles the number of components per sample, such data expansion is a fundamental requirement for achieving perfect reversibility and high-capacity embedding in non-redundant scalar data.

Note that the sensor only performs basic integer additions and multiplications to embed the watermark, while the complex division operations required for extraction, as shown in Expression (\ref{eq:extraction}), are relegated to the Data Collector. This ensures that the proposed scheme remains computationally efficient and compatible with standard, resource-constrained IoT and sensor data pipelines. \remarkend
\label{Xenun12-6}
\end{remk}

\begin{remk}\label{rem:vector}
 It may be argued that the proposed approach is just a linear transformation of the vector \linebreak $[d_{j,k}, w_k]{^\mathrm{T}}$ to a new vector $v$, obtained multiplying by a particular $2\times 2$ matrix, as follows:
\[\left[\begin{array}{c}
 v_1 \\
 v_2
 \end{array}\right] =
 \left[\begin{array}{cc}
 a_k & -b_k \\
 b_k & a_k
 \end{array}\right] =
 \left[\begin{array}{c}
 d_{j,k} \\
 w_k
 \end{array}\right],
 \]
 where $v_1$ corresponds to $\Re(\delta'_{j,k})$ and $v_2$ to $\Im(\delta'_{j,k})$.

While this is true, the idea behind the proposed scheme is to exploit the algebraic operations that are already available for complex numbers. There are libraries for complex number operations in most programming languages and, hence, they can be directly applied in any practical implementation of the method. This prevents the explicit use of matrices and matrix inverses, which may be cumbersome.
 Obviously, the method can be further generalized in its vector form, as discussed in \linkref[Section]{\ref{sec:generalized}}. \remarkend
\end{remk}

The recovery of the original data and the watermark from the watermarked data is straightforward, since the reversible extraction function $R_w$ can be defined as follows:
\begin{equation}\label{eq:extraction}
 \left[\begin{array}{c}
 d_{j,k} \\
 w_k
 \end{array}
 \right] =
 \left[\begin{array}{c}
 \Re(\lambda_k^{-1}\delta'_{j,k})\\
 \Im(\lambda_k^{-1}\delta'_{j,k})
 \end{array}
 \right].
\end{equation}

Consequently, we only need to multiply $\delta'_{j,k}$ by $\lambda_k^{-1}$ (the complex inverse of $\lambda_k$) and, then, take the real and the imaginary parts to recover both the original data and the watermark.
\begin{remk} Expression (\ref{eq:extraction}) perfectly matches the formulation given for $R_w$ in \linkref[Definition]{\ref{def:embext}}, since it takes, as input parameters, $\delta'_{j,k}$ (analogous to $d'_{j,k}$) and $K_w=\lambda_k$. \remarkend
\label{Xenun14-8}
\end{remk}

\begin{remk}\label{rem:factor}
The proposed method is vulnerable against Gaussian integer factorization. Even with just one eavesdropped value, a malicious observer may try to compute the Gaussian integer factors of the watermarked data and obtain knowledge about the original data, the watermark, or the watermarking key. \textbf{Consequently, for practical applications, it is not recommended to rely solely on the challenge $\lambda_{k}$ as a security measure.}

The proposed scheme is intended to function within a layered security architecture. While we focus on its synergy with homomorphic encryption (\linkref[Section]{\ref{sec:protocols}}), in extremely resource-constrained environments where public-key operations are prohibitive, we recommend protecting the Gaussian pair with a lightweight symmetric cipher, such as Ascon \citep{Turan2025} or AES-128. This ensures that confidentiality is enforced by established standards, while \hidden provides the intrinsic integrity and reversibility that traditional ciphers lack.

The use of symmetric ciphers, however, comes with some additional complexity--symmetric key generation and management--- that is avoided if public-key PHE encryption is used, as shown in \linkref[Section]{\ref{sec:elgamal_protocol_description}} (at the price of computational overhead and data expansion).
\remarkend
\end{remk}

\begin{remk}Although the data and the watermark are clearly separated in $\delta_{j,k}$ (the real part contains the data and the imaginary part is the watermark), this is no longer true after multiplying by $\lambda_k$. In the watermarked data $\delta'_{j,k}=\lambda_k \cdot \delta_{j,k}$, the data $d_{j,k}$ and the watermark $w_k$ contribute to both the real and the imaginary parts. To recover the data and the watermark, knowledge about the secret key $\lambda_k$ is necessary (to compute $\lambda_k^{-1}$).

Hence, an eavesdropper who accessed $\delta'_{j,k}$ would not be able to determine the data or the watermark directly, but may try to infer information from Gaussian integer factorization, as discussed in \linkref[Remark]{\ref{rem:factor}}. In addition, if the eavesdropper had access to the watermarked values corresponding to different time samples ($k$) or different devices ($j$), they could try to obtain even more information about $\lambda_k$, especially if it is not changed frequently. 
\label{Xenun16-10} \remarkend
\end{remk}

\subsection{Data aggregation}\label{Xsec11-3.3}
\label{sec:aggregation}

When a collection of devices $j=1,2,\dots,N$ transmit data, it is often desirable, for privacy reasons, to work on the aggregated data rather than individual readings. Let $d_{j,k}$ be the sensor data (an integer) from device $j$ at time $t_k$, and $S_k$ (\linkref[Definition]{\ref{def:aggregated}}) be the total aggregated data.

For a data aggregation protocol, each device $j$ at time $t_k$ first combines
the watermark $w_k$ with its data to form a complex number $\delta_{j,k} = d_{j,k} + i w_k$. This individual watermarked data is then scaled by the challenge factor $\lambda_k$ (which is provided by the Data Collector for the specific $k$th round) to produce $\delta'_{j,k} = \lambda_k \delta_{j,k}$.

The total aggregated watermarked data, denoted as $\sigma'_k$, is the sum of these individual scaled and watermarked complex numbers across all devices:
\[
\sigma'_k=\sum_{j=1}^N \delta'_{j,k} = \sum_{j=1}^N \lambda_k \delta_{j,k}.
\]
Since $\lambda_k$ is a common factor for all devices in a given round, we can factor it out of the summation:
\[
\sigma'_k = \lambda_k \sum_{j=1}^N\delta_{j,k}=\lambda_k \sum_{j=1}^N(d_{j,k}+iw_k) =\lambda_k \sigma_k,
\]
with~$\sigma_k=\sum_{j=1}^N(d_{j,k}+iw_k)$. By separating the real and imaginary components of the sum, and noting that the watermark $w_k$ is the same for all devices in a given round, the expression for $\sigma'_k$ can be rearranged as:
\begin{equation}\label{eq:agg}
\sigma'_k = \lambda_k \left( \sum_{j=1}^N d_{j,k} + i \sum_{j=1}^N w_k \right) = \lambda_k ( S_k + i N w_k ).
\end{equation}
where $S_k$ is the aggregated real data and $N w_k$ is obtained as the aggregated imaginary part.

Similarly to the non-aggregated case, the result $\sigma'_k$ is a complex number where both the aggregated data $S_k$ and the watermark $w_k$ (scaled by $N$) affect its real and imaginary parts. It is not possible to separate these values without the secret challenge factor $\lambda_k$.

The extraction of both the aggregated data $S_k$ and the watermark $w_k$ from $\sigma'_k$ is straightforward. We can first multiply $\sigma'_k$ by the inverse of $\lambda_k$ in the complex domain ($\lambda_k^{-1}$). Then, the real part of the result directly yields the aggregated original data ($S_k$), and the imaginary part, divided by the number of devices ($N$), gives the watermark ($w_k$). This extraction process can be formally expressed as:
\begin{equation}
\left[
\begin{array}{c}
S_k \\ w_k
\end{array}
\right]= \left[\begin{array}{c}
\Re(\lambda_k^{-1}\sigma'_k) \\
\displaystyle \frac{1}{N} \Im(\lambda_k^{-1}\sigma'_k)
\end{array}\right].
\label{Xeqn4-4}
\end{equation}
Hence, both the aggregated data and the watermark can be easily recovered from the aggregated-watermarked data.

A numerical example with small numbers is provided in \linkref[Section]{\ref{sec:numerical}} to illustrate how the embedding and recovery methods work, both for single data and for data aggregration.

\begin{remk}This particular version of the extraction function $R\agg_w$ does require $N$ as an input: $R\agg_w(\sigma'_k,\lambda_k,N)$. However, if further constraints in the definition of the watermark and the number of devices are introduced, an alternative aggregated extraction function that would not require $N$ as an input may be obtained.

For example, if the number of devices $N$ was limited to a maximum of $100$: $N\leq 100$ (which may be enough in many applications), it would suffice to choose the watermark $w_k$ as any integer with no prime factors lower than 100. A partial factorization of $\Im(\lambda_k^{-1}\sigma'_k)=Nw_k$, for primes below 100, would clearly reveal both $N$ and $w_k$. All prime factors of $\Im(\lambda_k^{-1}\sigma'_k)$ lower than 100, with their respective exponents, would yield $N$. Once $N$ is known (by multiplying all those powers of small prime factors), then $w_k$ could be finally obtained by dividing $\Im(\lambda_k^{-1}\sigma'_k)=Nw_k$ by $N$.

In any case, knowledge about the number of participants is, in general, acceptable in many real-life situations.\remarkend
\label{Xenun17-11}
\end{remk}

\begin{remk}The transition from a single scalar to a complex pair doubles the nominal number of components per sample. However, as established in \linkref[Section]{\ref{sec:basic}}, this expansion is an information-theoretic necessity for achieving perfect reversibility in non-redundant scalar data. Unlike traditional security primitives that append separate authentication tags, such as HMACs or digital signatures, the proposed scheme embeds the integrity proof directly within the data's algebraic structure. This ``all-in-one'' approach ensures that the watermark and host data are mathematically inseparable during transmission, simplifying provenance tracking in both individual and aggregated data scenarios. \remarkend
\label{Xenun18-12}
\end{remk}

\subsection{Generalized vector-based approach for RDH}\label{Xsec12-3.4}\label{sec:generalized}
As discussed in \linkref[Remark]{\ref{rem:vector}}, a similar RDH scheme could be obtained using vectors and matrices instead of complex numbers. In this case, the complex number would be a particular example of the vector form (but with some advantages as already discussed).

In the generalized case, the watermarked data $d'_{j,k}$ can be defined in vector form, as follows:
\[
d'_{j,k}=
 \left[\begin{array}{cc}
 m_{11,k} & m_{12,k} \\
 m_{21,k} & m_{22,k}
 \end{array}\right]
 \left[\begin{array}{c}
 d_{j,k} \\
 w_k
 \end{array}\right]= \left[\begin{array}{c}
 m_{11,k}d_{j,k} + m_{12,k} w_k \\
 m_{21,k}d_{j,k} + m_{22,k} w_k
 \end{array}\right],
\]
where a $2\times2$ matrix makes the appropriate transformation.

For reversibility, it is enough that the matrix be invertible, i.e., that its determinant is different from zero: $m_{11,k}m_{22,k}-m_{12,k}m_{21,k} \neq 0$. If this condition is satisfied, the original data and the watermark can be extracted from $d'_{j.k}$ as follows:
\[
\left[\begin{array}{c}
 d_{j,k} \\
 w_k
 \end{array}\right]=\left[\begin{array}{cc}
 m_{11,k} & m_{12,k} \\
 m_{21,k} & m_{22,k}
 \end{array}\right]^{-1}d'_{j,k}.
\]
Since this is a linear transformation, it would also work for aggregated data, similarly as in the complex numbers case.

The property of ``distributing'' the watermark and the data in the two components of the resulting vector ($d'_{j,k}$) will be satisfied as far as none of the elements of the matrix is zero. Again, if all the values $m_{11,k}, \dots, m_{22,k}$ are integers, the obtained vector will also have integer components ($d'_{j,k}\in\mathbb{Z}^2$), which may be convenient for different applications, especially those that require public-key encryption.

\begin{remk} This vector form of the reversible watermarking system, however, requires four parameters ($m_{11,k}, \dots, m_{22,k}$) instead of two: the real ($a_k$) and imaginary ($b_k$) parts of a Gaussian integer $\lambda_k$. While using four parameters offers no clear advantage over the two-parameter (complex number) case, it does require a larger secret watermarking key at both the embedding and extraction ends. This increased key size, without any compensating benefits, can be considered a disadvantage of the generalized approach, leading to the preference for the complex number-based formulation in this paper. The only advantage of such a scheme would rely on the longer apparent key size, but this can be easily replicated with two values instead of four just by selecting a larger range for each of them. \remarkend
\label{Xenun19-13}
\end{remk}

While the generalized vector-based approach is algebraically valid for simple data embedding, it presents critical structural limitations when integrated with
homomorphic encryption. Standard partially homomorphic cryptosystems rely fundamentally on commutative operations, whereas general matrix multiplication is inherently non-commutative. While certain subrings of $2\times2$ matrices do commute, they fail to satisfy the core requirements of our RDH framework. For instance, diagonal matrices commute but process the data and watermark entirely independently, failing to achieve the intrinsic algebraic entanglement required for structural security. Other commutative 2D structures, such as those isomorphic to dual or split-complex numbers, suffer from zero divisors; this means many potential challenge factors would be mathematically non-invertible, thereby breaking the perfect reversibility of the host data. To simultaneously achieve commutativity (for HE compatibility), algebraic diffusion (for entanglement), and guaranteed invertibility (for exact extraction), the transformation matrix must be strictly constrained to the form:
\[
\begin{bmatrix} a & -b \\ b & a \end{bmatrix}.
\]
Because this specific matrix ring is mathematically isomorphic to the complex numbers, the complex-domain formulation (Gaussian integers) emerges not merely as a convenient reparameterization, but as the unique algebraic structure that allows a 2×2 matrix representation to preserve the commutative properties required for standard homomorphic cryptographic primitives.

Obviously, the method could be further extended by introducing larger vectors and matrices. However, no clear advantages would be obtained with that additional complexity.

\begin{remk}It is important to emphasize that while the complex plane ($\mathbb{Z}[i]$) and the generalized matrix-vector representations discussed in this section are highly efficient vehicles for Domain-Extending RDH, they do not constitute the only theoretically viable algebraic options. Other algebraic structures, such as polynomial rings (e.g., $\mathbb{Z}[x]/\langle f(x)\rangle$) or higher-dimensional geometric algebras, could fundamentally provide the multi-channel capacity required to decouple the watermark from the host data's range. However, the Gaussian integer framework is prioritized in this work due to its pragmatic advantages: it allows for direct, low-overhead implementation using standard complex arithmetic available in most programming environments, and it facilitates seamless integration with existing partially homomorphic cryptosystems without the need for the custom algebraic infrastructure that non-scalar structures (such as polynomials) would require.
\label{Xenun20-14} \remarkend
\end{remk}

\subsection{Numerical example for watermark embedding, extraction and data recovery}\label{Xsec13-3.5}\label{sec:numerical}
This section presents an example of how the method would work with simple data. In this example, three devices produce the following data: $d_{1,1}=5$, $d_{2,1}=8$, and $d_{3,1}=17$ at time $t_1$. All of them must embed the watermark $w_1=4$, and the secret embedding key is $\lambda_1=3+2i$. With this setting, the inverse $\lambda_1^{-1}$ can be obtained as follows:
\[
\lambda_1^{-1}=\frac{3-2i}{3^2+2^2}=\frac{3}{13}-\frac{2}{13}i.
\]
The corresponding watermarked data are the following:
\[
\begin{split}
\delta'_{1,1}&=(3+2i)(5+4i)=7+22i,\\
\delta'_{2,1}&=(3+2i)(8+4i)=16+28i,\\
\delta'_{3,1}&=(3+2i)(17+4i)=43+46i.
\end{split}
\]

\begin{remk}As discussed in \linkref[Remark]{\ref{rem:factor}}, a simple Gaussian integer factorization of eavesdropped data could reveal much information. For example, a factorization calculator for $7+22i$ yields $7+22i=(3+2i)(5+4i)$, which leaks $\lambda_k=3+2i$, in one factor, and the original data and watermark, in the other factor. While an individual eavesdropped data would not be enough to identify $\lambda_k$ and $d_{j,k}+iw_k$ from the computed factors, it is obvious that a small number of leaked data would suffice for an attacker. Additional security measures are required for the transmission of watermarked data. In addition, it is recommended to modify $\lambda_k$ frequently (e.g., at each time sample). \remarkend
\label{Xenun21-15}
\end{remk}

\begin{remk}As in the regular integer case (without complex numbers), the use of large prime factors for $\lambda_k$ would make the factorization problem harder for an attacker and, consequently, would increase security, unless both the data and the watermark are small. \remarkend
\label{Xenun22-16}
\end{remk}

At the receiver end, the following computation would be performed:
\[
\begin{split}
\lambda_1^{-1}\delta'_{1,1}&=\left(\frac{3}{13}-\frac{2}{13}i\right)(7+22i)=5+4i,\\
\lambda_1^{-1}\delta'_{2,1}&=\left(\frac{3}{13}-\frac{2}{13}i\right)(16+28i)=8+4i,\\
\lambda_1^{-1}\delta'_{3,1}&=\left(\frac{3}{13}-\frac{2}{13}i\right)(43+46i)=17+4i.
\end{split}
\]
Thus, taking the real and the imaginary parts after the multiplication recovers both the original data (5, 8 or 17) and the watermark (4), respectively.

Reversible extraction is also possible from aggregated data. The aggregated data for these three devices at time $t_1$ can be computed as:
\[
S_1=\sum_{j=1}^3d_{j,1}=d_{1,1}+d_{1,2}+d_{1,3}=5+8+17 =30.
\]
Similarly, the aggregated watermarked data can be obtained as follows:
\[
\begin{split}
\sigma'_1&=\sum_{j=1}^3\delta'_{j,1}=\delta'_{1,1}+\delta'_{1,2}+\delta'_{1,3}\\&=(7+22i)+(16+28i)+(43+46i)\\&=66+96i.
\end{split}
\]
The receiver end would compute the following:
\[
\lambda_1^{-1}\sigma'_1=\left(\frac{3}{13}-\frac{2}{13}i\right)(66+96i)=30+12i,
\]
whose real part provides the aggregated data: $S_1=\Re(30+12i)=30$, whereas the imaginary part is $\Im(30+12i)=12=3\cdot 4=Nw_1$. If the imaginary part is divided by the number of devices ($N=3$), the embedded watermark ($w_1=4$) is finally extracted.

\section{Homomorphic encryption on the complex domain}\label{Xsec14-4}
\label{sec:homomorphic}

This section presents the extension of ElGamal and Paillier cryptosystem to use complex-valued data.

The extension of ElGamal encryption to Gaussian integers, first presented in \cite{El-Kassar2001El-Gamal}, requires a careful examination of the mathematical basis of the cryptosystem. The first building block of the standard ElGamal cryptosystem is the multiplicative cyclic group $\mathbb Z^*_p$, for a prime number $p$, and a generator $g$ for such a group. In the Gaussian integer case, we must have a prime number $\pi$ in $\mathbb Z[i]$ and a consistent definition of the modular arithmetic $\pmod{\pi}$.

For a better understanding of the adaptation of this cryptosystem to the complex domain, a brief introduction to modular arithmetic over Gaussian integers is provided in \linkref[Appendix]{\ref{app:modular}}. A more detailed and comprehensive reference on Gaussian integers can be found in \citep{conrad2016gaussian}.

\subsection{ElGamal encryption over Gaussian integers}\label{Xsec15-4.1}
\label{sec:GaussianElGamal}

As shown in \cite{El-Kassar2001El-Gamal}, the principles of ElGamal encryption can be extended to other finite cyclic groups, including the multiplicative group of Gaussian integers modulo a Gaussian prime. This adaptation exploits the rich algebraic structure of Gaussian integers while maintaining the security reliance on the DLP.

We define the ``Gaussian version'' of ElGamal encryption as follows. Let $p$ be a Gaussian prime (preferably a regular prime integer of the form $4n-1$, as discussed in \linkref[Appendix]{\ref{app:modular}}). Let $\gamma$ be a generator of the cyclic multiplicative group of Gaussian integers modulo $p$, denoted as ${\mathbb{Z}[i]}^*_p$. The ElGamal encryption scheme over Gaussian integers consists of the same three fundamental algorithms:

\begin{itemize}
 \item Key generation: The key generation process involves selecting a random integer $a$ such that $1 \leq a \leq \mathcal{N}-1$ as the private key, where $\mathcal{N}$ is the order of the group ${\mathbb{Z}[i]}^*_p$. For a prime integer $p$ (of the form $4n-1$), $\mathcal{N} = p^2-1$. Then, the public key $\mathcal{K} = \gamma^a \pmod{p}$ is computed. The public key is formed as the triple $\mathbf{K}=(p, \gamma, \mathcal{K})$.
 \item Encryption: To encrypt a message $\mu \in {\mathbb{Z}[i]}^*_p$, a random integer $b$, such that $1 \leq b \leq \mathcal{N}-1$, is selected as the ephemeral key. The ciphertext is then computed as $\psi_1 = \gamma^b \pmod{p}$ and $\psi_2 = \mu \cdot \mathcal{K}^b \pmod{p}$, resulting in the pair $(\psi_1, \psi_2)$. The encryption function can be denoted as follows:
 \[(\psi_1, \psi_2)=\enc_{\mathbf K}(\mu, b) = \left(\gamma^b \pmod{p}, \mu \cdot \mathcal{K}^b \pmod{p}\right).\]
 \item Decryption: The decryption of the ciphertext $(\psi_1, \psi_2)$ using the private key $a$ involves computing the shared secret $\tau = \psi_1^a \pmod{p}$. Then, the message is decrypted as $\mu = \psi_2 \cdot \tau^{-1} \pmod{p}$. This decryption function can be formally denoted as follows:
 \[\dec_{a}(\psi_1, \psi_2) = \psi_2 \cdot (\psi_1^a){}^{-1} \pmod{p}.\]
\end{itemize}
The selection of a suitable generator $\gamma$ for the group ${\mathbb{Z}[i]}^*_p$ is a practical consideration, similar to the regular integer case. This can be efficiently achieved through probabilistic methods involving testing candidate elements.

The security of this Gaussian integer ElGamal variant fundamentally relies on the difficulty of the DLP in the group ${\mathbb{Z}[i]}^*_p$. If an attacker can efficiently solve the DLP in this group, they could compute the private key $a$ from the public key $\mathcal{K}$ and subsequently decrypt any ciphertext. For sufficiently large Gaussian primes $p$, the DLP is believed to be computationally infeasible.

As a key property, the order of this group, $\mathcal{N} = p^2-1$, is significantly larger than $p-1$ for the same numerical prime $p$ used in the regular integer ElGamal. This relationship, $\mathcal{N} = (p-1)(p+1)$, is particularly advantageous for security. If a prime $p$ is chosen such that $p-1$ possesses a large prime factor (a standard requirement for robust security in integer-based ElGamal against algorithms like Pohlig-Hellman \citep{pohlig:1978}), then this same large prime factor will inherently be a factor of $\mathcal{N}$. Therefore, the Gaussian integer ElGamal variant naturally inherits at least the same level of security (in terms of resistance to Pohlig-Hellman attacks) as its regular integer counterpart for a given $p$. Furthermore, the existence of the additional factor $(p+1)$ in $\mathcal{N}$ can potentially introduce even larger prime factors, further enhancing the security of the scheme. Thus, careful selection of the Gaussian prime $p$ to ensure that the largest prime factor of $\mathcal{N}$ is sufficiently large guides robust parameter choices.

Similar to its regular integer counterpart, this Gaussian integer ElGamal encryption exhibits a homomorphic property under component-wise multiplication of ciphertexts. This means that given two ciphertexts \linebreak $\enc_{\mathbf K}(\mu_1, b_1) = (\psi_1, \psi_2)$ and $\enc_{\mathbf K}(\mu_2, b_2) = (\psi'_1, \psi'_2)$, multiplying them component-wise as $(\psi_1 \cdot \psi'_1, \psi_2 \cdot \psi'_2)$ results in a ciphertext that decrypts to the product of the original plaintexts:
\[\dec_a\left(\enc_{\mathbf K}(\mu_1, b_1) \cdot \enc_{\mathbf K}(\mu_2, b_2)\right) = \mu_1 \cdot \mu_2 \pmod{p}.\]
This property enables its use in various cryptographic protocols and specifically finds application in the proposed method for combining reversible watermarking and encryption in this paper.

\subsubsection{Numerical example}\label{Xsec16-4.1.1}\label{sec:numericalEG}

This example demonstrates the ElGamal encryption process and its homomorphic property using Gaussian integers. The multiplicative group of Gaussian integers modulo $p=23$, denoted as ${\mathbb{Z}[i]}^*_p$, is selected. The order of this group is $\mathcal{N} = 23^2-1 = 528$.

For clarity and consistency with standard programmatic implementations, all Gaussian integers in this example are represented using the standard positive residue class modulo $p$ (\textit{i.e.,} $\{0, 1, \dots, p-1\}$). While a centered representation (e.g., $\{-(p-1)/2, \dots, (p-1)/2\}$, as formally detailed in \linkref[Appendix]{\ref{app:modular}}) could alternatively be used, both mappings are algebraically equivalent in $\mathbb{Z}[i]{}_p$.

For the parameters, the prime modulus $p = 23$ and the generator $\gamma = 1+2i$ are taken, and the private key is chosen as $a = 7$. Now, $\mathcal{K}$ is computed as \[\mathcal{K} = \gamma^a \pmod{p} = (1+2i){}^7 \pmod{23} = 6+2i,\] and the public key is thus formed as the triple $\mathbf{K}=(p, \gamma, \mathcal{K}) = (23, 1+2i, 6+2i)$.

To encrypt a first message $\mu_1 = 5+4i$, an ephemeral key $b_1 = 5$ is chosen. The ciphertext components are computed as:
\[\resizebox{\columnwidth}{!}{
$\begin{aligned}
\psi_1 &= \gamma^{b_1} \pmod{p} = (1+2i){}^5 \pmod{23} = 18+8i, \\
\psi_2 &= \mu_1 \cdot \mathcal{K}^{b_1} \pmod{p} = (5+4i) \cdot (6+2i){}^5 \pmod{23} = 21+11i.
\end{aligned}$}
\]
The resulting ciphertext is $(\psi_1, \psi_2) = (18+8i, 21+11i)$. Decryption of this ciphertext using the private key $a=7$ successfully yields $\mu_1 = (5+4i) \pmod{23}$.

For a second message $\mu_2 = 3+2i$, an ephemeral key $b_2 = 7$ is chosen. The ciphertext components are obtained as:
\[\resizebox{\columnwidth}{!}{$
\begin{aligned}
\psi'_1 &= \gamma^{b_2} \pmod{p} = (1+2i){}^7 \pmod{23} = 6+2i,\\
\psi'_2 &= \mu_2 \cdot \mathcal{K}^{b_2} \pmod{p} = (3+2i) \cdot (6+2i){}^7 \pmod{23} = 21+16i.
\end{aligned}$}
\]
 The resulting ciphertext is $(\psi'_1, \psi'_2) = (6+2i, 21+16i)$. Decryption of this ciphertext using the private key $a=7$ successfully yields $\mu_2 = (3+2i) \pmod{23}$.

The homomorphic property is demonstrated by multiplying the two ciphertexts component-wise:
$(\psi_1, \psi_2) = (18+8i, 21+11i)$ and $(\psi'_1, \psi'_2) = (6+2i, 21+16i)$.
The resulting ciphertext components, obtained by component-wise multiplication, are:
\[\resizebox{\columnwidth}{!}{$
\begin{aligned}
\psi''_1 &= \psi_1 \cdot \psi'_1
= (18+8i) \cdot (6+2i) = (0+15i) \pmod{23}, \\
\psi''_2 &= \psi_2 \cdot \psi'_2
= (21+11i) \cdot (21+16i) = (12+15i) \pmod{23}.
\end{aligned}$}
\]
The combined ciphertext for the product is then expressed as $(\psi''_1, \psi''_2) = (0+15i, 12+15i)$.

Decryption of this combined ciphertext $(\psi''_1, \psi''_2) = (0+15i, 12+15i)$ using the private key $a=7$ yields $\mu'' = (7+22i) \pmod{23}$. This precisely matches the direct multiplication of the original plaintexts:
\[
\begin{split}
\mu_1 \cdot \mu_2 &= (5+4i) \cdot (3+2i) \pmod{23} \\
 &= \left((15-8) + (10+12)i\right) \pmod{23} \\
 &= (7+22i) \pmod{23}.
\end{split}
\]
This confirms the correct operation and homomorphic property of ElGamal encryption over Gaussian integers.

\subsection{Component-wise Paillier encryption of Gaussian integers}\label{Xsec17-4.2}

While the previously discussed ElGamal scheme was extended to operate natively over Gaussian integers, the additive homomorphic property of Paillier encryption allows for a straightforward application to complex numbers (specifically, Gaussian integers) without requiring a specialized ``complex version'' of the cryptosystem. This is achieved by encrypting the real and imaginary parts of a Gaussian integer independently.

Let a Gaussian integer plaintext message be represented as $\mu = m_R + i m_I$, where $m_R$ and $m_I$ are integers. To encrypt $\mu$ using the standard Paillier scheme with public key $\mathsf{K_{pub}}=(n, g)$, the process is as follows:
\begin{itemize}
 \item Each component is encrypted separately:
 \begin{align*}
 c_R &= \enc_{\mathsf{K_{pub}}}(m_R, r_R) = g^{m_R} \cdot r_R^n \pmod{n^2}, \\
 c_I &= \enc_{\mathsf{K_{pub}}}(m_I, r_I) = g^{m_I} \cdot r_I^n \pmod{n^2},
 \end{align*}
 where $r_R$ and $r_I$ are randomly chosen integers for each encryption, satisfying $0 < r_R, r_I < n$ and $\mathrm{gcd}(r_R, n) = \mathrm{gcd}(r_I, n) = 1$.
 \item The ciphertext for the Gaussian integer $\mu$ is then represented as a pair of Paillier ciphertexts: $(c_R, c_I)$.
\end{itemize}

Decryption of a Gaussian integer ciphertext $(c_R, c_I)$ is performed component-wise using the Paillier private key $\mathsf{K_{priv}}=(L_P, M_P)$: $m_R = \dec_{\mathsf{K_{priv}}}(c_R)$ and $m_I = \dec_{\mathsf{K_{priv}}}(c_I)$.
The decrypted Gaussian integer is then reconstructed as $\mu = m_R + i m_I$.

The primary advantage of this component-wise approach is that it directly inherits the additive homomorphic property of the standard Paillier scheme. If we have two encrypted Gaussian integer messages, $\enc(\mu_1) = (c_{1R}, c_{1I})$ for $\mu_1 = m_{1R} + i m_{1I}$ and $\enc(\mu_2) = (c_{2R}, c_{2I})$ for $\mu_2 = m_{2R} + i m_{2I}$, their sum can be computed in the encrypted domain as follows:
\begin{itemize}
 \item For the real part: $c_{\mathrm{sum},R} = c_{1R} \cdot c_{2R} \pmod{n^2}$.
 \item For the imaginary part: $c_{\mathrm{sum},I} = c_{1I} \cdot c_{2I} \pmod{n^2}$.
\end{itemize}
When the resulting ciphertext $(c_{\mathrm{sum},R}, c_{\mathrm{sum},I})$ is decrypted, the real and imaginary components sum independently:
\begin{align*}
 \dec_{\mathsf{K_{priv}}}(c_{\mathrm{sum},R}) &= (m_{1R} + m_{2R}) \pmod  n, \\
 \dec_{\mathsf{K_{priv}}}(c_{\mathrm{sum},I}) &= (m_{1I} + m_{2I}) \pmod n.
\end{align*}
Thus, the decryption yields the sum of the original Gaussian integers: $\mu_1 + \mu_2 = (m_{1R} + m_{2R}) + i(m_{1I} + m_{2I}) \pmod n$.

This component-wise additive homomorphism is useful for applications such as privacy-preserving aggregation of complex-valued data, where sums of watermarked sensor readings can be computed without revealing individual values. The security of this approach remains equivalent to that of the underlying regular Paillier encryption scheme.

\subsection{Cryptographic strength and parameter selection guidelines}\label{Xsec18-4.3}\label{sec:security_parameter_selection}

To formally establish the security foundations of the proposed complex-domain frameworks, it is essential to define their cryptographic strength and operational lifetimes. This analysis evaluates the underlying group hardness against classical cryptanalysis and provides key-length parameterization trajectories based on global standardization guidelines.

\subsubsection{Theoretical security bounds and group hardness}\label{Xsec19-4.3.1}
Under conventional classical cryptanalysis, mapping the encryption primitives to the Gaussian integer ring $\mathbb{Z}[i]$ preserves the foundational hardness of the Discrete Logarithm Problem (DLP) and the Integer Factorization Problem (IFP). As already noted in \linkref[Section]{\ref{sec:GaussianElGamal}}, the multiplicative cyclic group order $\mathcal{N} = p^2 - 1 = (p-1)(p+1)$ naturally inherits robust resistance to Pohlig-Hellman reduction attacks from the large prime factors of the underlying scalar prime $p$. Because the security of the Gaussian ElGamal variant fundamentally relies on the DLP in $\mathbb{Z}[i]{}_p^*$, classical asymptotic complexity is governed by the General Number Field Sieve (GNFS), matching the resilience of regular integer counterparts. Similarly, for the component-wise Paillier variant, security directly reduces to the composite residuosity problem of the standard integer-domain Paillier scheme over the composite modulus $n = pq$, guaranteeing direct parity against classical attack vectors.

\subsubsection{NIST recommendations and modulus scaling timeline}\label{Xsec20-4.3.2}
The operational security strength of both primitives scales with the bit-length $\eta$ of the public modulus ($p$ or $n$). Following the National Institute of Standards and Technology (NIST) SP 800-57 \cite{barker202recommendation} guidelines, the parameter selection bounds dictate the acceptable protection lifetimes of the cryptosystem:
\begin{itemize}
 \item \textbf{Near-term baseline ($\eta$ = 2048 bits):} A public modulus of 2048 bits provides approximately 112 bits of symmetric security strength. According to NIST SP 800-57 Part 1, this configuration remains approved for applying new cryptographic protection through December 31, 2030. This makes it an ideal engineering baseline for current resource-constrained architectures where minimizing processing overhead and representation volume is critical.
 \item \textbf{Long-term migration ($\eta$ = 3072 bits):} To sustain changing corporate and industrial compliance requirements beyond the 2030 horizon, infrastructures mandate a migration to a minimum key size of 3072 bits. This scales the symmetric security threshold to 128 bits, satisfying the mandatory NIST requirement for secure data encryption effective January 1, 2031, and offering robust protection against advancing classical cryptanalysis.
 \item \textbf{Extended protection threshold ($\eta$ = 4096 bits):} For critical utility or military infrastructures requiring defense bounds extending well past the immediate 2030 NIST transition horizons, scaling the parameter length to 4096 bits shifts the classical security ceiling to approximately 140+ bits of symmetric strength.
\end{itemize}

Whether deploying the native Gaussian ElGamal framework (which yields a pair of complex ciphertexts modulo $p$) or the component-wise Paillier architecture (which encrypts the real and imaginary channels independently modulo $n^2$), the total ciphertext payload scales uniformly as a function of $4\eta$ bits. While selecting $\eta = 2048$ bits serves as our default baseline evaluation parameter throughout the remainder of this work to protect resource-constrained endpoints, the algebraic design remains fully agnostic to the underlying bit-length. The protocol steps adapt dynamically as configurations scale up to 3072-bit or 4096-bit boundaries (or as far as required).

\subsubsection{Quantum vulnerability and pre-quantum lifecycle}\label{Xsec21-4.3.3}
Conversely, under a quantum computing paradigm, the unit groups of these Gaussian quotient rings remain finite abelian structures. As a consequence, they are entirely vulnerable to the polynomial-time evaluations of Shor's algorithm \cite{shor1999polynomial}. The proposed protocols do not claim post-quantum resilience, and their intended protection lifetime is strictly bound to pre-quantum operational environments. Protection against quantum cryptanalysis requires transitioning the core algebraic layers to alternative homomorphic frameworks, such as lattice-based Fully Homomorphic Encryption (FHE) or Ring Learning with Errors (RLWE) primitives, which remains an open path for future research.

\section{Protocols for joint watermarking and encryption on the complex domain}\label{Xsec22-5}
\label{sec:protocols}

This section presents the Hiding in the Imaginary Domain with Data Encryption (\hidden) framework, a novel approach designed to simultaneously embed information (watermarks) into data and encrypt it. Within this general framework, we detail two distinct protocols tailored for different application scenarios and taking advantage of different homomorphic properties.

The first protocol, termed \textbf{\hidden-EG}, uses Gaussian ElGamal encryption to provide joint watermarking and encryption for individual data points. The second one, denoted as the \textbf{\hidden-AggP} protocol, uses Paillier encryption's additive homomorphism to enable joint watermarking, encryption, and privacy-preserving aggregation of data from multiple sources. Each protocol is presented with its initialization and data exchange phases, and also its complexity.

\subsection{System model and security requirements}\label{Xsec23-5.1}
\label{sec:model}

Establishing a rigorous security foundation for dispersed data in distributed environments requires a clear definition of the participating entities and the adversarial assumptions. The \hidden framework is specifically designed to address the vulnerabilities inherent in resource-constrained networks, where data must be transmitted across potentially insecure channels while maintaining perfect reversibility. In this section, we formalize the system model as a precursor to the detailed analysis of specific threat vectors and protocol-specific defenses.

\subsubsection{System entities}\label{Xsec24-5.1.1}
The \hidden framework involves the following primary entities:
\begin{itemize}
 \item \textbf{Sensors} {($\mathrm{S}_j$):} Distributed, resource-constrained devices that generate original scalar data $d_{j,k}$. They perform the reversible embedding of the watermark $w_k$ and the initial encryption steps.
 \item \textbf{Partial Aggregators} {($\mathrm{S}_{\ell}$):} In the \hidden-AggP (Paillier-based) version, certain sensors or intermediate nodes act as aggregators. They use the homomorphic properties of the cipher to sum ciphertexts without accessing the underlying plaintexts.
 \item \textbf{Data Collector (DC):} The central, trusted entity that initializes system parameters, distributes challenge factors $\lambda_k$, and performs the final decryption to recover both the fine-grained (or aggregated) data and the verified watermark.
\end{itemize}

\subsubsection{Security and privacy requirements}\label{Xsec25-5.1.2}
The following properties define the core security goals of the framework.

\begin{defn}(Confidentiality)
The sensor data $d_{j,k}$ and challenge factors $\lambda_k$ must remain protected from unauthorized parties. This is guaranteed by the Indistinguishability under Chosen-Plaintext Attack (IND-CPA) security of the underlying ElGamal and Paillier cryptosystems, as formally proven in \linkref[Section]{\ref{sec:security}}.
\label{Xenun37-6}
\end{defn}

\begin{defn}(Integrity and Accuracy)
The system must ensure that data has not been altered during transmission or aggregation. Accuracy is strictly guaranteed by the perfect reversibility of the complex-domain embedding, while integrity is verified through the successful extraction of the pre-agreed watermark $w_k$.
\label{Xenun38-7}
\end{defn}

\begin{defn}(Data Provenance)
 The framework must provide unforgeable proof of the origin of the data. Since the watermark $w_k$ is secretly agreed upon between the DC and legitimate sensors $\mathrm{S}_j$, its correct extraction serves as cryptographic proof that the data originated from an authorized source.
\label{Xenun39-8}
\end{defn}

\begin{defn}(Privacy)
Privacy is defined as the inability of any entity (including intermediate aggregators) to learn individual sensor readings. Additive homomorphic encryption ensures this property is maintained throughout the network hierarchy.
\label{Xenun40-9}
\end{defn}

\subsubsection{Attack model}\label{Xsec26-5.1.3}\label{sec:attack}
To formalize the security analysis of the framework, we adopt the standard Dolev-Yao threat model \citep{dolev2003security} for the communication channels and classify adversarial capabilities as follows:
\begin{enumerate}
\item \textbf{Entity Trust Assumptions:} The sensors $\mathrm{S}_j$ and Partial Aggregators $\mathrm{S}_{\ell}$ are assumed to be \textit{honest-but-curious}; they execute the protocol correctly but may attempt to glean information about other sensors' plaintexts. The Data Collector (DC) is assumed to be fully \textit{trusted}.
 \item \textbf{Adversary Capabilities:} We consider an adversary ($\mathcal{A}$) with full control over the communication network. $\mathcal{A}$ is classified into three capability tiers:
 \begin{itemize}
 \item \textit{Passive:} $\mathcal{A}$ eavesdrops on the channels to compromise confidentiality (mitigated by the IND-CPA properties of the ciphers).
 \item \textit{Active:} $\mathcal{A}$ can intercept, drop, delay, or arbitrarily modify ciphertexts in transit (e.g., Man-in-the-Middle).
 \item \textit{Adaptive:} Specifically targeting the aggregation scenario, an adaptive $\mathcal{A}$ exploits the inherent malleability of partially homomorphic encryption. They can perform targeted algebraic operations on intercepted ciphertexts without knowing the underlying plaintexts or keys.
 \end{itemize}
 \item \textbf{Threat Vectors:} Within this formalized model, the system is designed to defend against:
 \begin{itemize}
 \item \textbf{Eavesdropping:} Passive interception of ciphertexts to compromise confidentiality.
 \item \textbf{Man-in-the-Middle (MitM):} Active interception and modification of messages between entities.
 \item \textbf{Masquerading:} Attempting to impersonate a legitimate sensor to inject unauthorized data.
 \item \textbf{Replay Attacks:} Re-transmitting valid historical messages to skew current data rounds.
 \item \textbf{False Data Injection (FDI) and Adaptive Attacks:} Injecting malicious readings into the individual or aggregated stream. This includes defending against \textit{Adaptive False Data Injection (Homomorphic Injection)}: an advanced attack where a compromised aggregator or external adversary~$\mathcal{A}$ exploits the cipher's homomorphic properties to inject mathematically tailored false values. As formally analyzed in \linkref[Section]{\ref{sec:structural}}, the complex-domain watermark acts as the primary detection mechanism against this specific vulnerability.
 \end{itemize}
\end{enumerate}

\begin{figure*}[ht]
\centerline{\includegraphics[width=\textwidth]{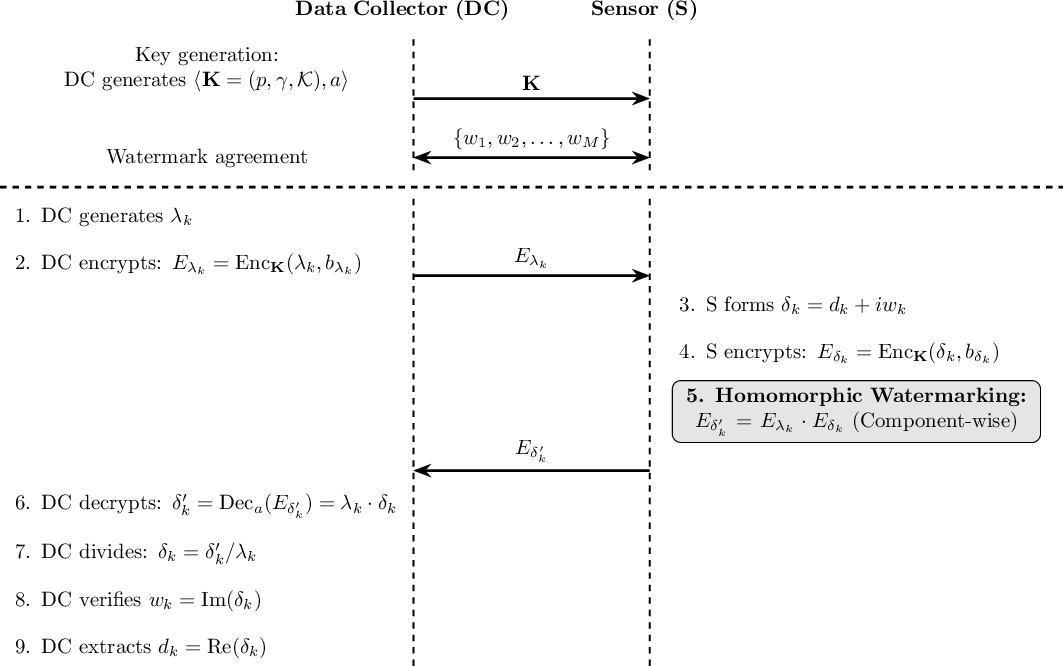}}
\caption{Graphical representation of the \hidden-EG protocol, including the initialization (above the horizontal dashed line) and the data exchange (below the horizontal dashed line) phases.}
\label{fig:elgamal}
\end{figure*}

\subsection{\hidden-EG: joint watermarking and encryption for individual data}\label{Xsec27-5.2}
\label{sec:elgamal_protocol_description}

This section describes a protocol for secure and reversible information hiding between a sensor (S) and the Data Collector (DC). It applies the proposed complex domain framework and Gaussian ElGamal encryption to ensure sensor data ($d_k$) is transmitted confidentially while simultaneously embedding a time-dependent watermark ($w_k$) for verification and integrity.

\begin{remk}\label{rem:ElGamal}
Although the data $d_{j,k}$ and the watermark $w_k$ were assumed to be simply integers (\linkref[Definitions]{\ref{def:data}} and \ref{def:wm}), the constraint of working modulo $p$ for a Gaussian prime satisfying $p\equiv 3 \pmod{4}$ in the \hidden-EG protocol imposes further constraints on the ranges of the data, the watermark, and the challenge factor. If both positive and negative values are allowed, the Gaussian integers $\pmod{p}$ can represent any value in:
\[\resizebox{\columnwidth}{!}{$
{\mathbb{Z}[i]}^*_p=\left\{ x + iy \mid x, y \in \left\{-\frac{p-1}{2}, -\frac{p-1}{2}+1, \dots,\frac{p-1}{2}\right\} \text{ and } (x,y) \neq (0,0) \right\}$.}\]
Hence, the following condition is required to avoid wrap-around:
$\max(|\Re(\lambda_k\delta_k)|, |\Im(\lambda_k\delta_k)|) < p/2.$
\remarkend\label{Xenun23-17}
\end{remk}

The protocol proceeds in two distinct phases: Initialization and watermarking and data exchange. The initialization phase establishes the necessary cryptographic keys and shared secrets between the DC and S:

\textbf{1) Public key distribution: }DC generates a Gaussian ElGamal key pair. It computes and publicly distributes its public key $\mathbf{K}=(p, \gamma, \mathcal{K})$.,

\textbf{2) Watermark agreement:} DC and S agree on a set of shared watermarking keys or a mechanism to generate sequential, time-dependent watermarks: $w_1, w_2, \ldots, w_M$ corresponding to times $t_1, t_2, \ldots, t_M$. Each $w_j$ is an integer represented with $B$ bits (e.g., taking values in the range $0, \ldots, 2^B-1$ for unsigned watermarks, or $-2^{B-1},-2^{B-1}+1,\dots,2^{B-1}-1$ for signed ones).

Then, for the watermarking and data exchange phase, at each designated time $t_k$, sensor data is generated, watermarked, encrypted, and transmitted for verification.

\textbf{1) DC's challenge factor: }DC chooses a Gaussian integer factor $\lambda_k$ with $\Re(\lambda_k)\cdot\Im(\lambda_k)\neq 0$ (from ${\mathbb{Z}[i]}^*_p$). DC then encrypts $\lambda_k$ using its own public key $\mathbf{K}$, yielding the ciphertext $(\psi_{k,1}, \psi_{k,2}) = \enc_{\mathbf K}(\lambda_k, b_{\lambda_k})$, where $b_{\lambda_k}$ is a randomly chosen ephemeral key. DC sends this ciphertext to S.

\textbf{2) Sensor's data and watermark embedding:} S generates its data reading $d_k$ (an integer, typically representing a real-world measurement). S then assigns the pre-agreed watermark $w_k$ to the imaginary part of the data, forming a complex-valued plaintext $\delta_k = d_k + i w_k$. Next, S encrypts this combined plaintext $\delta_k$ using DC's public key $\mathbf{K}$, resulting in the ciphertext $(\psi'_{k,1}, \psi'_{k,2}) = \enc_{\mathbf K}(\delta_k, b_{\delta_k})$, where $b_{\delta_k}$ is another randomly chosen ephemeral key. Making use of the homomorphic property of Gaussian ElGamal, S component-wise multiplies the received challenge ciphertext with its own generated ciphertext:
 \begin{align*}
 \psi''_{k,1} &= \psi_{k,1} \cdot \psi'_{k,1} \pmod{p}, \\
 \psi''_{k,2} &= \psi_{k,2} \cdot \psi'_{k,2} \pmod{p}.
 \end{align*}
 This step simultaneously performs both embedding and encryption, with the additional benefit of homomorphic multiplication. S then sends the resulting combined ciphertext $(\psi''_{k,1}, \psi''_{k,2})$ to the DC.

\textbf{3) DC's decryption and verification:} Upon receiving the combined ciphertext $(\psi''_{k,1}, \psi''_{k,2})$, the DC decrypts it using its private key $a$. Due to the homomorphic property, the decrypted value $\mu_k = \dec_a(\psi''_{k,1}, \psi''_{k,2})$ will correspond to the product of the two original plaintexts: $\mu_k = \lambda_k \cdot \delta_k \pmod{p}$. Given the careful selection of $p$ to be sufficiently large, this modular result $\mu_k$ is guaranteed to be equivalent to the exact, non-modular product $\lambda_k \cdot \delta_k$ as Gaussian integers. To isolate the watermarked data, DC then computes $\delta_k = \mu_k / \lambda_k$ using standard complex number division. Finally, DC extracts the original sensor data $d_k$ from the real part of $\delta_k$ and the embedded watermark $w_k$ from its imaginary part. The data $d_k$ is accepted only if the extracted watermark $w_k$ correctly matches the expected time-dependent watermark; otherwise, the data is discarded, indicating potential tampering or an unauthorized source.

A graphical representation of the protocol is shown in \linkref[Fig.]{\ref{fig:elgamal}}, and a simulation example (with small numbers) is presented in \linkref[Appendix]{\ref{app:sim_eg}}.

\subsubsection{Complexity analysis}\label{Xsec28-5.2.1}

Regarding computational complexity, the number of complex modular exponentiations and integer modular exponentiations, required at each party during the data exchange phase, is analyzed below.

For a Gaussian integer $\chi = x+iy \in {\mathbb{Z}[i]}^*_p$, its modular inverse $\chi^{-1} \pmod{p}$ is typically computed as $\overline{\chi} \cdot (\mathrm{N}(\chi)){}^{-1} \pmod{p}$. This involves obtaining the modular inverse of the integer norm $\mathrm{N}(\chi) = x^2+y^2$ modulo $p$ (an integer modular exponentiation), followed by a complex multiplication (Fermat's Little Theorem):
\[
(x^2+y^2){}^{-1} \pmod p = (x^2+y^2){}^{p-2} \pmod p.
\]

\textbf{At the Sensor:}
 In step 2 of the data exchange phase, S computes $\psi'_{k,1} = \gamma^{b_{\delta_k}} \pmod{p}$ and $\mathcal{K}^{b_{\delta_k}} \pmod{p}$ (for $\psi'_{k,2}$). Both of these are complex modular exponentiations. Therefore, the Sensor performs \textbf{2 complex modular exponentiations} per data exchange.

\textbf{At the Data Collector:}
 In step 1, DC computes $\gamma^{b_{\lambda_k}} \pmod{p}$ and $\mathcal{K}^{b_{\lambda_k}} \pmod{p}$ for the challenge factor's encryption. This amounts to \textbf{2 complex modular exponentiations}.
 In step 3, DC performs the decryption and subsequent isolation of $\delta_k$:
 \begin{itemize}
 \item Decryption: This involves computing $(\psi''_{k,1}){}^a \pmod{p}$, which is \textbf{1 complex modular exponentiation}.
 \item Computation of $\tau_k^{-1}$: To compute the inverse of the shared secret $\tau_k= (\psi''_{k,1}){}^a$, it is calculated as $\overline{\tau_k} \cdot (\mathrm{N}(\tau_k)){}^{-1} \pmod{p}$. This involves \textbf{1 integer modular exponentiation} (for $(\mathrm{N}(\tau_k)){}^{-1} \pmod{p}$).
 \item Computation of $\mu_k$: This is $\psi''_{k,2} \cdot \tau_k^{-1} \pmod{p}$, involving complex multiplication.
 \item Computation of $\delta_k$: As $\mu_k$ is now the exact product $\lambda_k \cdot \delta_k$, DC computes $\delta_k = \mu_k / \lambda_k$ using standard complex number division.
 \end{itemize}
 In summary, per data exchange, the Data Collector performs a total of \textbf{3 complex modular exponentiations} (2 for encrypting $\lambda_k$ and 1 for computing $(\psi''_{k,1}){}^a$) and \textbf{1 integer modular exponentiation} (for the inverse of $\mathrm{N}(\tau_k)$).

\begin{remk}\label{rem:cost_factor}
Regarding the relative computational cost, a complex modular exponentiation is approximately four times as expensive as an integer modular exponentiation of a similar exponent length. This is primarily because each complex modular multiplication, which forms the building block of exponentiation, involves four integer modular multiplications: \[(a+ib)(c+id) \pmod p=((ac -bd)+ i (ad+bc)) \pmod p.\]

It is worth noting that the number of multiplications can be theoretically reduced from four to three using \textit{Gauss's complex multiplication trick}. By defining the intermediate terms $k_1 = c(a + b)$, $k_2 = a(d - c)$, and $k_3 = b(c + d)$, the product can be computed as $\left((k_1 - k_3) + i(k_1 + k_2) \right)\pmod p$.

However, the empirical gain of this optimization is strictly dependent on the execution environment and the underlying programming language. In low-level environments (e.g., C++ or Assembly) where multiplication cycles significantly outweigh addition cycles, this trick can reduce the overhead ratio toward $3\times$. Conversely, in high-level interpreted environments like Python, the overhead of managing additional object allocations for the five required additions and subtractions may offset the savings of one multiplication. Empirical tests conducted in Python on server-grade hardware (AMD EPYC 7B12) with 2048-bit keys confirmed that the standard four-multiplication approach ($\approx 0.18$s) remained more efficient than the Gauss-optimized version ($\approx 0.22$s). Consequently, we retain the $4\times$ factor as a robust and conservative baseline for our performance analysis. Although these baseline timings reflect a server-grade environment to isolate the algebraic overhead, empirical execution data simulated on a constrained IoT edge microcontroller architecture is detailed in \linkref[Section]{\ref{sec:iot_edge_latency}}. \remarkend
\end{remk}

The computational cost for each entity during a single data exchange round is analyzed in terms of equivalent integer modular exponentiations (as defined in \linkref[Remark]{\ref{rem:cost_factor}}). The Sensor performs approximately \textbf{8 equivalent integer modular exponentiations} for encrypting its watermarked data. In contrast, the Data Collector incurs a higher computational burden, totaling approximately \textbf{13 equivalent integer modular exponentiations} per round (and sensor). The majority of this computational load at the DC is primarily due to the operations involved in encrypting the challenge factor ($\lambda_k$) sent to the Sensor, decrypting the combined ciphertext received from the Sensor, and then isolating the original watermarked data. This distribution of workload, where the Sensor's load is comparatively lighter, aligns well with typical sensor network architectures where sensors are often resource-constrained devices, while the data collector usually possesses greater computational power.

Consequently, the transition to the complex domain introduces a predictable cost. As detailed in \linkref[Remark]{\ref{rem:cost_factor}}, complex modular exponentiation is approximately four times as expensive as a standard integer modular exponentiation. Therefore, the sensor's workload effectively quadruples relative to an equivalent regular integer-valued ElGamal baseline. However, this overhead is the strict mathematical requisite to achieve intrinsic data-watermark entanglement. To maintain viability for resource-constrained IoT devices, the \hidden-EG protocol strategically distributes this burden. The sensor's role is functionally streamlined to focus exclusively on the initial algebraic embedding and encryption, while the more demanding tasks of complex division for watermark extraction and logical verification are centralized at the Data Collector. An empirical evaluation of the viability of the encryption on constrained hardware is provided in \linkref[Section]{\ref{sec:iot_edge_latency}}.

Regarding communication overhead, the use of complex-valued data in the \hidden-EG protocol involves the transmission of two components (real and imaginary) per sample. In the encrypted domain, a standard ElGamal ciphertext natively consists of a pair of integers. By extending this to the Gaussian integers, the resulting ciphertext consists of a pair of complex values, effectively doubling the payload to four integer components modulo $p$. This results in a two-fold increase in the total number of bits transmitted compared to a single encrypted regular integer-valued data item. While the data expansion inherent to homomorphic encryption is already significant ---mapping small plaintext integers to large modular values--- the transition to the complex domain adds a further factor of two to the bandwidth requirements. This increased bit-load is the fundamental trade-off for the unique advantages of the \hidden framework, specifically the intrinsic data-watermark entanglement and perfect host data reversibility that are not achievable with traditional regular integer-domain encryption alone.

\subsection{\hidden-AggP: joint watermarking, encryption, and aggregation}\label{Xsec29-5.3}
\label{sec:paillier_protocol_description}

This section describes a protocol enabling privacy-preserving data aggregation from multiple sensors ($\mathrm{S}_j$, $j=1,\dots,N$) by a Data Collector (DC), incorporating reversible information hiding and applying Paillier encryption for additive homomorphism. The protocol ensures that aggregated sensor data is obtained securely, while a time-dependent watermark allows for verification and integrity checking.

The initialization phase establishes the cryptographic keys and shared secrets required for secure communication and data aggregation:

\textbf{1) Paillier public key distribution:} DC generates a Paillier key pair. It computes and publicly distributes its public key $\mathsf{K_{pub}}=(n, g)$, where $n=pq$ for large primes $p,q$, and $g$ is a generator as defined in \linkref[Section]{\ref{sec:paillier}}. DC keeps its corresponding private key $\mathsf{K_{priv}}=(L_P, M_P)$ secret.

\textbf{2) Symmetric key exchange (for DC to Sensor):} Each sensor $\mathrm{S}_j$ ($j=1, \ldots, N$) generates a unique, random symmetric key $K_{\mathrm{sym},j}$ (e.g., an AES key). $\mathrm{S}_j$ then encrypts $K_{\mathrm{sym},j}$ using DC's public Paillier key $\mathsf{K_{pub}}$, resulting in ciphertext $c_j = \enc_{\mathsf{K_{pub}}}(K_{\mathrm{sym},j}, r_j)$, where $r_j$ is a random integer. $\mathrm{S}_j$ transmits $c_j$ to DC. Upon reception, DC decrypts $c_j$ using its private key $\mathsf{K_{priv}}$ to obtain $K_{\mathrm{sym},j}$. This process establishes a unique symmetric key $K_{\mathrm{sym},j}$ shared securely between DC and each individual sensor $\mathrm{S}_j$. These symmetric keys will subsequently be used for secure communication from DC to $\mathrm{S}_j$.

\begin{remk} To prevent ciphertext substitution attacks during this phase, the transmission of the encrypted symmetric key must be protected by an authenticated identity bound. We assume that sensors undergo a secure bootstrapping or enrollment phase prior to deployment, using a factory-provisioned Master Key ($\mathsf{MK}$) to append an $\mathsf{HMAC}$ or $\mathsf{AES\text{-}CMAC}$ tag to the initial key-exchange packet. This enforces data origin authentication, ensuring that the ciphertext originates from a verified sensor and cannot be substituted by an adversary.
\label{Xenun25-19} \remarkend
\end{remk}

\textbf{3) Watermark agreement:} DC and all sensors $\mathrm{S}_j$ collectively agree on a set of shared, sequential, time-dependent watermarks: $w_1, w_2, \ldots, w_M$ corresponding to times $t_1, t_2, \ldots, t_M$. Each $w_k$ (for the $k$th round) is an integer represented with $B$ bits, which may be signed or unsigned as detailed in the \hidden-EG case.

At each designated time $t_k$, sensor data is generated, watermarked, encrypted, and aggregated for privacy-preserving transmission and verification:

\textbf{1) DC's challenge factor distribution:} DC chooses a Gaussian integer factor $\lambda_k$ with $\Re(\lambda_k)\cdot\Im(\lambda_k)\neq 0$. The real and imaginary parts of $\lambda_k$ must be chosen such that the resulting components of $\delta'_{j,k}$ can be represented within Paillier's plaintext modulus $n$. For each sensor $\mathrm{S}_j$ ($j=1, \ldots, N$), DC encrypts $\lambda_k$ using the pre-established unique symmetric key $K_{\mathrm{sym},j}$ shared with that specific sensor. Let this symmetric encryption be $\enc_{K_{\mathrm{sym},j}}(\lambda_k)$ (implying $\lambda_k$ is suitably serialized for symmetric encryption). DC then sends $\enc_{K_{\mathrm{sym},j}}(\lambda_k)$ to each $\mathrm{S}_j$. Each sensor $\mathrm{S}_j$ decrypts the received message using its $K_{\mathrm{sym},j}$ to obtain the plaintext $\lambda_k$.

\textbf{2) Sensor data embedding and initial Paillier encryption:} Each sensor $\mathrm{S}_j$ generates its data reading $d_{j,k}$ (an integer). $\mathrm{S}_j$ then combines the pre-agreed watermark $w_k$ with its data, forming a complex number $\delta_{j,k} = d_{j,k} + i w_k$. Using the received $\lambda_k$ (once decrypted), $\mathrm{S}_j$ then computes the scaled and watermarked complex number $\delta'_{j,k} = \lambda_k \delta_{j,k}$. This $\delta'_{j,k}$ is the value that will be aggregated. $\mathrm{S}_j$ encrypts $\delta'_{j,k}$ component-wise using DC's public Paillier key $\mathsf{K_{pub}}$. Let $\delta'_{j,k} = m_{j,k,R} + i m_{j,k,I}$ (where $m_{j,k,R}$ and $m_{j,k,I}$ are integers). The ciphertext is $(c_{j,k,R}, c_{j,k,I}) = (\enc_{\mathsf{K_{pub}}}(m_{j,k,R}, r_{j,k,R}), \enc_{\mathsf{K_{pub}}}(m_{j,k,I}, r_{j,k,I}))$, where $r_{j,k,R}$ and $r_{j,k,I}$ are fresh random numbers for each encryption.

\textbf{3) Cascading encrypted summation (tree aggregation):} The individual encrypted watermarked data points are then aggregated in a cascading manner, typically following a binary tree structure involving $\lceil \log_2(N) \rceil$ steps:
 \begin{itemize}
 \item Initial layer: Each $\mathrm{S}_j$ that is a leaf node in the aggregation tree generates its ciphertext $(c_{j,k,R}, c_{j,k,I})$ as described in step 2.
 \item Aggregation steps: For each internal node in the aggregation tree, a designated sensor (or an intermediate aggregator) receives two ciphertexts, say $(C_{1R}, C_{1I})=({c_{j_1,k,R},c_{j_1,k,I})}$ and $(C_{2R}, C_{2I})=({c_{j_2,k,R},c_{j_2,k,I})}$, from its children (or from other sensors). Taking advantage of the additive homomorphic property of Paillier encryption, this aggregator computes the sum of the encrypted real parts ($C_{\mathrm{sum},R} = C_{1R} \cdot C_{2R} \pmod{n^2}$) and the sum of the encrypted imaginary parts ($C_{\mathrm{sum},I} = C_{1I} \cdot C_{2I} \pmod{n^2}$). The resulting ciphertext $(C_{\mathrm{sum},R}, C_{\mathrm{sum},I})$ is then passed up the tree. This process continues until a single final encrypted sum is obtained.
 \item Handling $N \neq 2^X$: If the number of sensors $N$ is not a power of 2, some ``blank'' or identity ciphertexts (e.g., encryption of $0+i0$) may be included in the aggregation process to complete the tree structure, or the tree may be handled as an unbalanced structure.
 \end{itemize}
 The final encrypted sum, $\sigma'_{\mathrm{sum},k} = (C_{\mathrm{final},R}, C_{\mathrm{final},I})$, is obtained by the root sensor in the aggregation tree (e.g., $\mathrm{S}_1$ if sensors are ordered by aggregation path) and is then sent to DC.

\textbf{4) DC's decryption and verification: }Upon receiving the final aggregated ciphertext $\sigma'_{\mathrm{sum},k} = (C_{\mathrm{final},R}, C_{\mathrm{final},I})$, DC decrypts both components using its private Paillier key $\mathsf{K_{priv}}$. This yields the real and imaginary parts of the aggregated watermarked data:
 \begin{align*}
 \sigma'_{k,R} &= \dec_{\mathsf{K_{priv}}}(C_{\mathrm{final},R}) = \sum_{j=1}^N m_{j,k,R} \pmod n, \\
 \sigma'_{k,I} &= \dec_{\mathsf{K_{priv}}}(C_{\mathrm{final},I}) = \sum_{j=1}^N m_{j,k,I} \pmod n.
 \end{align*}
 The decrypted value is the aggregated watermarked data $\sigma'_k = \sigma'_{k,R} + i \sigma'_{k,I}$. As derived in \linkref[Section]{\ref{sec:aggregation}}, $\sigma'_k = \lambda_k(S_k+iNw_k)$, where $S_k = \sum_{j=1}^N d_{j,k}$. DC then isolates the true aggregated data $S_k$ and the watermark $w_k$ by computing $(S_k+iNw_k) = \sigma'_k / \lambda_k$ using standard complex number division. Finally, DC extracts $S_k$ from the real part and obtains $w_k$ by dividing the imaginary part by $N$. If the extracted watermark $w_k$ correctly matches the expected time-dependent watermark, the aggregated data $S_k$ is accepted; otherwise, it is discarded, indicating potential tampering or an unauthorized source.

\begin{remk}\label{rem:Paillier}
The component-wise application of Paillier encryption requires the real and imaginary parts of the aggregate $\mu = \lambda_k(S_k + i N w_k)$ to be handled modulo $n$. In this scenario, the additive homomorphic property allows for the accumulation of $N$ sensor readings. To ensure that the aggregate result does not exceed the capacity of the modulus $n$ and avoids modular wrap-around, the parameters must satisfy:
\[
\max(|\Re(\mu)|, |\Im(\mu)|) < n/2.
\]
This condition ensures that both components of the aggregate Gaussian integer can be correctly mapped back from the centered representative system of $\mathbb{Z}_n$ to their exact integer values in $\mathbb{Z}$, which is a prerequisite for the successful application of the perfect reversibility proven in \linkref[Theorem]{\ref{the:rev}}.
\remarkend \end{remk}

\begin{figure*}[ht]
\centerline{\includegraphics[width=\textwidth]{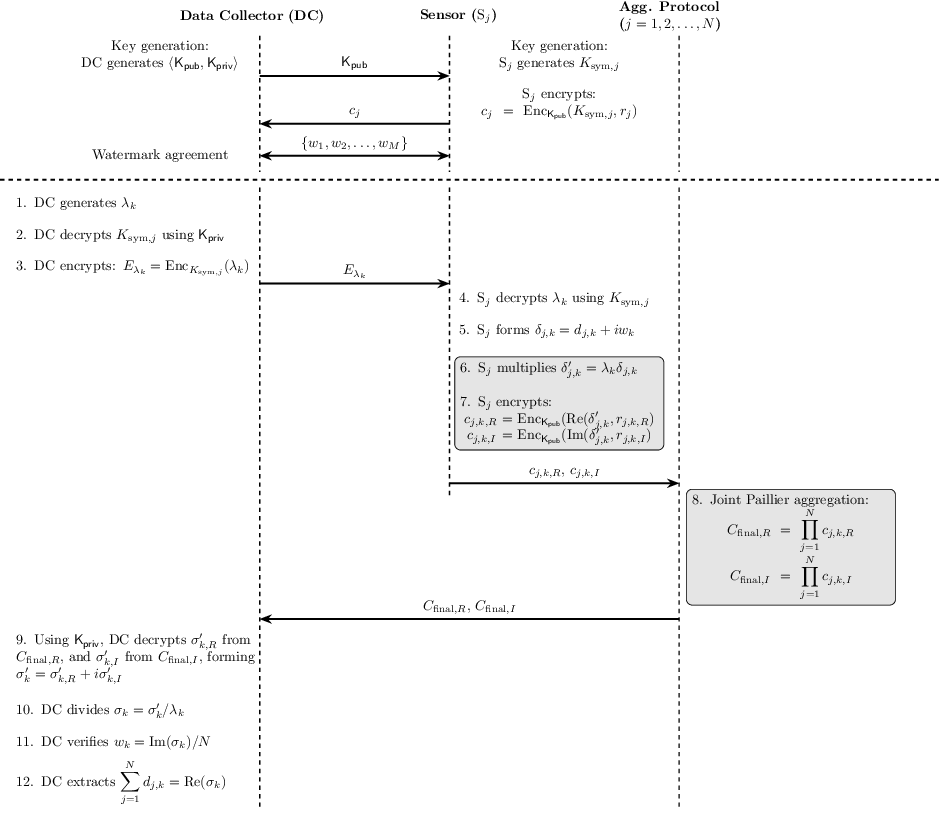}}
\caption{Graphical representation of the \hidden-AggP protocol, including the initialization (above the horizontal dashed line) and the data exchange (below the horizontal dashed line) phases.}
\label{fig:paillier}
\end{figure*}

\begin{figure}[ht]
\centerline{\includegraphics[width=.95\linewidth]{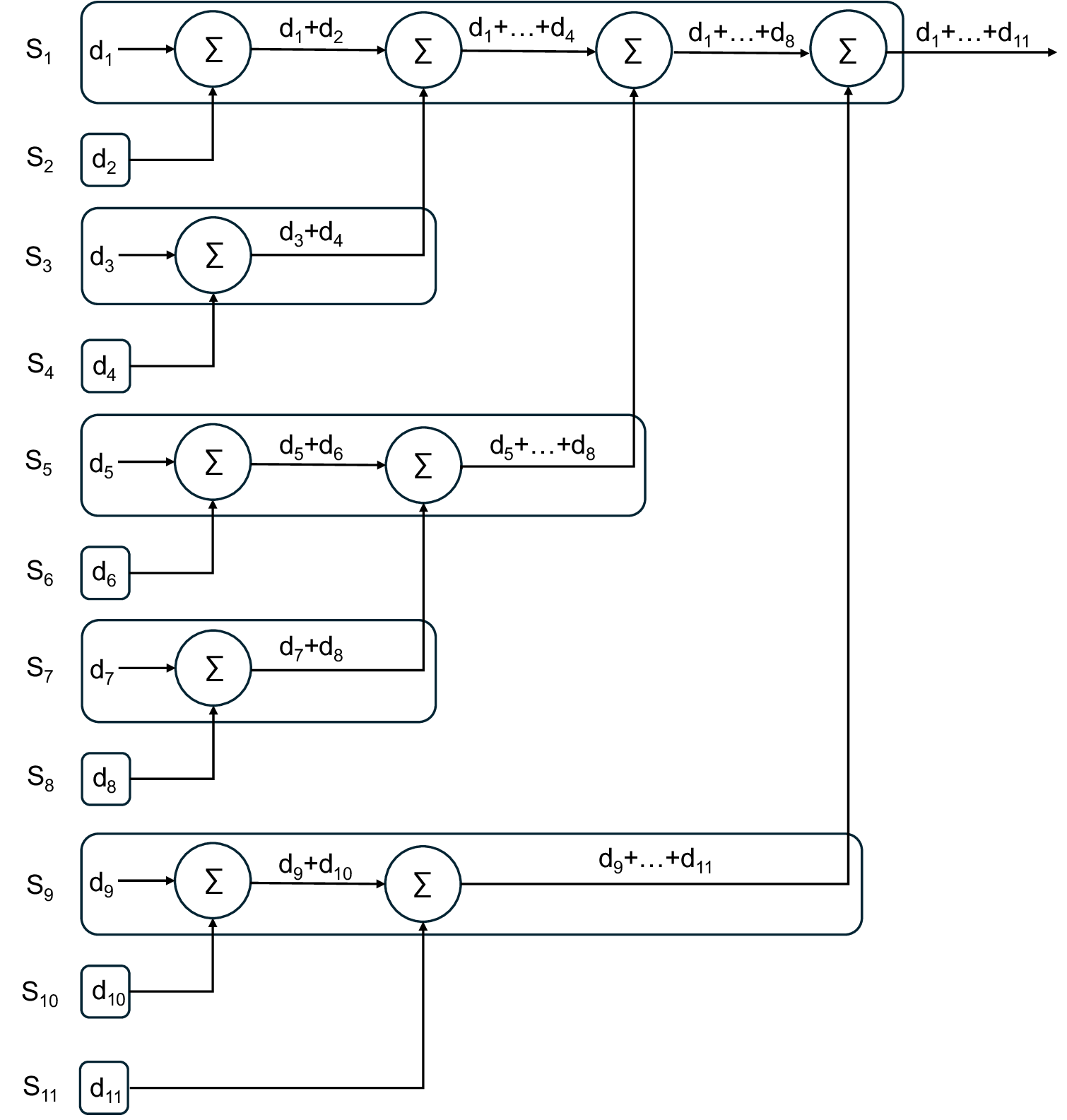}}
\caption{Cascading summation protocol: example for 11 sensors.}
\label{fig:tree_agg}
\end{figure}

A graphical representation of the \hidden-AggP protocol is shown in \linkref[Fig.]{\ref{fig:paillier}}, whereas \linkref[Fig.]{\ref{fig:tree_agg}} illustrates how the cascading summation would work for 11 sensors. In this figure, the $\mathrm{S}_1$ sensor would obtain the final aggregate result to send it to the DC. It must be taken into account that the ``$\Sigma$'' symbols in the figure represent multiplication of ciphertexts for the Paillier cryptosystem. A simulation of the protocol (with small numbers) is provided in \linkref[Appendix]{\ref{app:sim_aggp}}.

\begin{remk}The cascading aggregation protocol shown in \linkref[Fig.]{\ref{fig:tree_agg}} offers the advantage of performing the necessary summation internally within the sensor network without requiring a dedicated external entity. While this approach distributes the computational load, the topology implies a sequential delay of $\lceil\log_2(N)\rceil$ communication steps as partial aggregates propagate. Alternatively, if a participating sensor (e.g., $\mathrm{S}_1$) is sufficiently equipped with storage and computational power, it can act as a hub, receiving all ciphertexts directly from other nodes. While this reduces the transmission latency to a single step, the node must still perform $N-1$ modular multiplications (modulo $n^2$) sequentially to produce the final result. In fact, whether using a tree or a star arrangement, the algebraic work remains consistent. Performing aggregation within the sensor network itself rather than employing an external Data Aggregator (DA) minimizes the attack surface and circumvents the complex security considerations of third-party collusion (e.g., potential collusions between the DA and the DC).
\remarkend
\label{Xenun27-21}
\end{remk}

As discussed in \linkref[Remark]{\ref{rem:provenance}}, this aggregation approach shifts the verification focus from individual-level provenance to group-level integrity.

\begin{remk}\label{rem:provenance}
It is critical to distinguish between the individual provenance provided by \hidden-EG and the aggregate integrity provided by \hidden-AggP. In the \hidden-AggP protocol, the shared secret structure ensures that the final aggregate result is formed by nodes authorized to participate in the network (group-level integrity). However, this does not intrinsically provide non-repudiation for individual sensor inputs; a compromised node could potentially contribute malicious, correctly-watermarked data to the aggregate. This is a deliberate design trade-off prioritizing homomorphic computational efficiency over individual-level data forensic tracking. In scenarios where individual-level non-repudiation is mandatory, \hidden-AggP can be extended with a lightweight, per-node message authentication code (MAC) layer, at the expense of additional transmission overhead. \remarkend
\end{remk}

\subsubsection{Complexity analysis}\label{Xsec30-5.3.1}
\label{sec:agg_complexity}

This section analyzes the computational and communication complexity of the proposed data aggregation protocol. We focus on the operations within a single data exchange round ($t_k$), as initialization costs are typically one-time or amortized over many rounds.

For computational complexity, the dominant costs arise from Paillier encryption and decryption operations, which primarily involve modular exponentiations modulo $n^2$. Other operations, such as symmetric encryption/decryption, complex modular multiplications (including those from homomorphic ciphertext additions), and standard complex divisions, are considered computationally negligible in this high-level\break analysis.

\textbf{At the Sensor} {($\mathrm{S}_j$):}
 Each sensor $\mathrm{S}_j$ performs one initial encryption of its watermarked data $\delta'_{j,k}$. Since this involves encrypting both the real and imaginary components separately using Paillier encryption, each sensor performs \textbf{2 Paillier encryptions} per round. Each Paillier encryption typically involves two modular exponentiations (one for $g^m$ and one for $r^n$), totaling \textbf{4 modular exponentiations modulo $n^2$} per sensor for initial encryption.

 If a sensor also acts as an intermediate aggregator in the tree, it performs additional Paillier ciphertext multiplications (homomorphic additions). While these are critical for the protocol's functionality, their computational cost is significantly lower than modular exponentiations and is thus neglected in this dominant cost analysis.

\textbf{At the Data Collector:}
 DC's primary computational burden occurs in the final decryption and extraction step. DC receives the final aggregated ciphertext, $\sigma'_{\mathrm{sum},k}$, which consists of two Paillier ciphertexts $(C_{\mathrm{final},R}, C_{\mathrm{final},I})$.
 DC performs \textbf{2 Paillier decryptions} (one for the real part and one for the imaginary part). Each Paillier decryption primarily involves one modular exponentiation modulo $n^2$. Therefore, the DC performs approximately \textbf{2 modular exponentiations modulo $n^2$} per round.

\textbf{Network-Wide Aggregation (Aggregator Nodes):}
 The aggregation tree involves $N-1$ aggregation steps in total (for $N$ sensors in a binary tree). Each aggregation step involves two modular multiplications modulo $n^2$ (one for the real component sum and one for the imaginary component sum). These operations, while numerous, are computationally less intensive than modular exponentiations and are therefore not included in this dominant cost count.

The computational overhead for each party during a single data exchange round is analyzed based on the number of dominant modular exponentiations. The Sensor ($\mathrm{S}_j$) performs \textbf{4 integer modular exponentiations modulo $n^2$} for encrypting its watermarked data (two for the real part and two for the imaginary part). In contrast, the Data Collector (DC) incurs a comparatively lighter computational burden, totaling approximately \textbf{2 integer modular exponentiations modulo $n^2$} per round for decrypting the aggregated ciphertext. The network-wide aggregation process, while involving numerous modular multiplications, is not included in this dominant cost count as these operations are less intensive than exponentiations. This distribution of workload, where the Sensor bears a higher computational burden in terms of dominant Paillier exponentiations, is an important consideration for resource-constrained sensor deployments, implying the need for adequate computational capabilities at the sensor level.

Regarding the computational overhead, the transition to the complex domain involves a measurable increase in cryptographic\break operations compared to a hypothetical regular integer-valued baseline. In the \hidden-AggP protocol, the component-wise application of Paillier encryption requires two independent encryptions per data sample, resulting in a 100\% increase in computational cost compared to a single regular integer-valued encryption. It is important to note that in this protocol, the sensors bear the primary computational burden, as Paillier encryption requires four modular exponentiations per sample (two per component) compared to the two decryptions performed by the Data Collector on the final aggregate. Although the total number of modular operations effectively doubles relative to its single-component regular integer-domain equivalent, this overhead is the requisite investment to achieve intrinsic data-watermark entanglement and perfect reversibility. As with the individual encryption protocol, the framework remains viable for IoT ecosystems as long as the encryption task can be executed by the hardware. This is discussed empirically in \linkref[Section]{\ref{sec:iot_edge_latency}}.

On the other hand, the communication complexity is measured by the number of messages exchanged in a single data exchange round ($t_k$):

\begin{itemize}
 \item DC to Sensors ($\lambda_k$ distribution): The DC sends the symmetrically encrypted challenge factor $\enc_{K_{\mathrm{sym},j}}(\lambda_k)$ to each of the $N$ sensors. This accounts for $N$ messages from DC to sensors.
 \item Sensor-to-Sensor and Sensor-to-DC (aggregation): The cascading summation process in a binary tree structure involves messages being passed up the tree. For instance, if $N$ is a power of 2, the first level of aggregation generates $N/2$ messages, the second level $N/4$, and so on, until the last level generates 1 message. This sums up to $N/2 + N/4 + \ldots + 1 = N-1$ messages for the intermediate aggregation steps. A final message is then sent from the root aggregator sensor ($\mathrm{S}_N$) to the DC. Thus, the total number of messages for the aggregation phase (from sensors to DC) is $(N-1) + 1 = N$ messages, regardless of whether $N$ is a power of 2.
\end{itemize}

Combining the DC-to-sensor and the aggregation messages, the overall number of messages exchanged in the network per round is $N + N = 2N$ messages. This indicates an efficient communication pattern, as the total message count scales linearly with the number of sensors.

Regarding communication overhead in the aggregation scenario, the \hidden-AggP protocol requires the transmission of two components per sample. Because Paillier encryption is applied component-wise, each watermarked sample is represented as a pair of independent ciphertexts. Consequently, this results in a two-fold increase in the total number of bits transmitted compared to a single encrypted regular integer-valued data item. While the data expansion inherent to homomorphic encryption is already substantial ---mapping small plaintext integers to large modular values modulo $n^2$--- the transition to the 2D complex domain doubles these bandwidth requirements. As in the individual encryption case, this increased bit-load is the fundamental trade-off necessary to achieve intrinsic data-watermark entanglement and perfect host data reversibility within the encrypted aggregate.

\section{Formal security and privacy analysis}\label{Xsec31-6}\label{sec:security}

In this section, we present a formal theorem-proof evaluation of the proposed protocols. We demonstrate the resilience of the \hidden framework against the attack model defined in \linkref[Section]{\ref{sec:attack}} through rigorous cryptographic proofs.

\subsection{Confidentiality and privacy proofs}\label{Xsec32-6.1}

It is important to clarify that the vulnerability to Gaussian integer factorization or statistical analysis of the complex plane is only a relevant threat vector if the data is transmitted without a dedicated encryption layer. By adopting a layered approach, the security of the data is decoupled: the cryptographic layer (whether homomorphic or symmetric) provides resistance against unauthorized access (confidentiality), while the RDH scheme provides resistance against tampering and unauthorized aggregation (integrity).

\begin{thm}(Resistance to Eavesdropping)\label{the:eaves}
The \hidden framework ensures the confidentiality of $d_k$ (or $d_{j,k}$), $w_k$, and $\lambda_k$ against a passive eavesdropper~$\mathcal{A}$ under the assumption that the Gaussian ElGamal and Paillier cryptosystems are IND-CPA (Indistinguishability under Chosen Plaintext Attack) secure.
\label{Xenun1-1}
\end{thm}

\begin{proof}
Suppose there exists an adversary $\mathcal{A}$ that can recover any part of $d_k$ (or $d_{j,k}$), $w_k$, or $\lambda_k$ from the transmitted ciphertexts with non-negligible advantage. We can construct a simulator $\mathcal{B}$ that uses $\mathcal{A}$ to break the semantic security (IND-CPA) of the underlying cryptosystems.

In \hidden-EG, the final combined ciphertext is $\psi'' = (\gamma^b, \lambda_k \cdot \delta_k \cdot \mathcal{K}^b) \pmod p$. The ephemeral key $b$ results from the combination of individual keys $b = b_{\lambda_k} + b_{\delta_k}$. Since the multiplicative group $\mathbb{Z}[i]{}_p^*$ is cyclic with order $\mathcal{N} = p^2 - 1$, we have $\gamma^{\mathcal{N}} \equiv 1 \pmod p$. Consequently, the combined exponent $b$ is an element of the additive group of exponents $\mathbb{Z}_{\mathcal{N}}$. Each individual key is chosen independently and uniformly from $\{1, \dots, \mathcal{N}-1\}$, ensuring that the combined key $b$ is uniformly distributed in $\mathbb{Z}_{\mathcal{N}}$. If $\mathcal{A}$ can distinguish between the encryptions of two different plaintexts $(\delta, \hat\delta)$, then $\mathcal{B}$ can solve the Decision Diffie-Hellman (DDH) problem in $\mathbb{Z}[i]{}_p^*$. Given that the DLP and DDH are computationally infeasible in $\mathbb{Z}[i]{}_p^*$ for sufficiently large $p$, no such $\mathcal{A}$ can exist.

In \hidden-AggP, the individual watermarked data points $\delta_{j,k}'$ are protected by the IND-CPA security of the Paillier cryptosystem. The challenge factor $\lambda_k$ is transmitted to each sensor $\mathrm{S}_j$ using symmetric encryption with a unique key $K_{\mathrm{sym},i}$, whose secrecy is established during initialization via Paillier encryption. If $\mathcal{A}$ could recover $\lambda_k$ or any component of the sensor data, the simulator $\mathcal{B}$ would be able to break the underlying semantic security of either the Paillier scheme or the symmetric cipher. Given the assumed security of these primitives, no such adversary $\mathcal{A}$ can exist, and confidentiality is maintained.
\end{proof}

\begin{thm}(Aggregation privacy)\label{the:agg}
The \hidden-AggP protocol preserves the privacy of individual sensor readings $d_{j,k}$ against honest-but-curious partial aggregators $\mathrm{S}_{\ell}$.
\label{Xenun2-2}
\end{thm}

\begin{proof}
 In \hidden-AggP, a partial aggregator $\mathrm{S}_{\ell}$ receives pairs of ciphertexts $(c_{j,k,R}, c_{j,k,I})$, where $c_{j,k,R} = \enc_{\mathsf{K_{pub}}}(\Re(\delta'_{j,k}),r_{j,k,R})$ and $c_{j,k,I} = \enc_{\mathsf{K_{pub}}}(\Im(\delta'_{j,k}),r_{j,k,I})$. The aggregator computes the product component-wise: $C_{\mathrm{sum},R} = \prod_j c_{j,k,R} \pmod{n^2}$ and $C_{\mathrm{sum},I} = \prod_j c_{j,k,I} \pmod{n^2}$. Due to the IND-CPA security of the Paillier cryptosystem, the aggregator cannot distinguish between the encryption of a valid component (real or imaginary) and an encryption of a random value in $\mathbb{Z}_n$. Since $\mathrm{S}_{\ell}$ does not possess the private key $\mathsf{K_{priv}}$, the view of $\mathrm{S}_{\ell}$ consists solely of pairs of ciphertexts which reveal zero information about the underlying component values. Consequently, the aggregator is unable to learn any information about the Gaussian integer $\delta'_{j,k}=\lambda_k (d_{j,k} + i w_k)$, ensuring that individual readings $d_{j,k}$ remain private during the aggregation process.
\end{proof}

\subsection{Integrity, provenance, and active attack proofs}\label{Xsec33-6.2}

\begin{thm}(Resistance to replay attacks)\label{the:replay}
An active adversary $\mathcal{A}$ cannot successfully replay a ciphertext (or a component thereof) captured at time $t_{k'}$ to be accepted at time $t_k$ (where $k \neq k'$) with non-negligible probability.
\label{Xenun3-3}
\end{thm}

\begin{proof}
At each time sample $t_k$, the Data Collector expects a specific Gaussian integer result $\delta'_k$ (or $\sigma'_k$) that, upon extraction with the {current} secret challenge factor $\lambda_k$, yields the expected time-dependent watermark $w_k$ (or $N w_k$ in the aggregated case). We distinguish two cases:
 \begin{enumerate}
 \item Individual case (\hidden-EG): If $\mathcal{A}$ replays a ciphertext from $t_{k'}$ at time $t_k$, the DC recovers $\delta'_{k'} = \lambda_{k'} \delta_{k'}$. The extraction yields:
 \[\tilde{\delta}_k = \frac{\delta'_{k'}}{\lambda_k} = \frac{\lambda_{k'}}{\lambda_k} (d_{k'} + i w_{k'}).\]
 Since $\lambda_k$ and $\lambda_{k'}$ are secret and refreshed every round, the resulting ratio $\lambda_{k'}/\lambda_k$ is unknown to $\mathcal{A}$, making $\Im(\tilde{\delta}_k)$ effectively random. The condition $\Im(\tilde{\delta}_k) = w_k$ is satisfied only with probability $2^{-B}$.
 \item Aggregated case (\hidden-AggP): If~$\mathcal{A}$ captures a sensor's component ciphertexts $(c_{j,k,R}, c_{j,k,I})$ at $t_{k'}$ and replays them into the aggregate for $t_k$, the decrypted aggregate sum is:
 \[\tilde{\sigma}_k = \left(\sum_{l \neq j} \lambda_k \delta_{l,k}\right) + \lambda_{k'} \delta_{j,k'}.\]
 The DC extracts the aggregate watermark:
 \[\Im(\tilde{\sigma}_k / \lambda_k) = (N-1)w_k + \Im\left(\frac{\lambda_{k'}}{\lambda_k} \delta_{j,k'}\right).\]
 For the aggregate to be accepted, the condition $\Im(\tilde{\sigma_k}/\lambda_k) = N w_k$ must hold, implying that
 \[\Im\left(\frac{\lambda_{k'}}{\lambda_k} (d_{j,k'} + i w_{k'})\right) = w_k.\]

 The probability of an accidental collision where the resulting imaginary part matches the current expected watermark $w_k$ is exactly $2^{-B}$.
 \end{enumerate}
 In both cases, the attacker must satisfy a condition involving three independent, secret, and time-varying parameters: the old watermark $w_{k'}$, the old challenge $\lambda_{k'}$, and the new challenge $\lambda_k$. Thus, detection occurs with probability $1 - 2^{-B}$ in either scenario.
\end{proof}

\begin{thm}(Resistance to man-in-the-middle (MitM) attacks)\label{the:mitm}
Any unauthorized modification of a legitimate ciphertext by $\mathcal{A}$ will be detected by the DC with overwhelming probability, preserving data integrity.
\label{Xenun4-4}
\end{thm}

\begin{proof}
In the \hidden framework, the DC chooses a modulus $p$ (for \hidden-EG) or $n$ (for \hidden-AggP) sufficiently large to ensure exact recovery of the Gaussian integer value $\delta'_k = \lambda_k \delta_k$ or the aggregate $\sigma'_k = \lambda_k \sigma_k$, respectively, upon decryption of a valid ciphertext. Consider a MitM adversary $\mathcal{A}$ who intercepts a ciphertext of $\delta'_k$ (for \hidden-EG) or $\delta'_{j,k}$ (for \hidden-AggP) and transmits a modified version, resulting in an induced error $\epsilon = e_R + i e_I$ in the plaintext domain after decryption.

 \begin{enumerate}
 \item Individual case (\hidden-EG): The DC extracts:
 \[\tilde{\delta}_k = \frac{\delta'_k + \epsilon}{\lambda_k} = \delta_k + \frac{\epsilon}{\lambda_k}.\]
 Expanding using $\lambda_k^{-1} = (a_k - i b_k)/(a_k^2+b_k^2)$:
 \[
 \tilde{\delta}_k = (d_k + i w_k) + \frac{(e_R a_k + e_I b_k) + i(e_I a_k - e_R b_k)}{a_k^2 + b_k^2}.
 \]
 The extracted watermark is (imaginary part):
 \[
 \tilde{w}_k = w_k + \frac{e_I a_k - e_R b_k}{a_k^2 + b_k^2}.
 \]

 \item Aggregated case (\hidden-AggP): The DC extracts the aggregate
 \[\tilde{\sigma}_k = \frac{\sigma'_k + \epsilon}{\lambda_k} = \sigma_k + \frac{\epsilon}{\lambda_k}.\]
 Since $\sigma_k = S_k + i N w_k$ (\linkref[Section]{\ref{sec:aggregation}}):
 \[
 \tilde{\sigma}_k = (S_k + i N w_k) + \frac{(e_R a_k + e_I b_k) + i(e_I a_k - e_R b_k)}{a_k^2 + b_k^2}.
 \]
 The extracted watermark is (imaginary part divided by $N$):
 \[ \tilde w_k=w_k + \frac{e_I a_k - e_R b_k}{N(a_k^2 + b_k^2)}.\]
 \end{enumerate}

 In both cases, for the modification to go undetected, the induced imaginary shift must be zero, implying $e_I a_k - e_R b_k = 0$. Since $\mathcal{A}$ blindly manipulates ciphertexts without knowledge of $(a_k, b_k)$, the attacker cannot choose a modification that purposely aligns $\epsilon$ with the secret ``slope'' of $\lambda_k$. For a large secret space, the probability of an accidental collision is negligible, ensuring protection against MitM modifications.
\end{proof}

\begin{thm}(Resistance to Masquerading and false data injection (FDI))\label{the:fdi}
The \hidden framework ensures resistance to masquerading and FDI attacks; a malicious entity $\mathcal{A}$ cannot successfully inject fraudulent readings $d_{\mathrm{fake}}$ that the DC accepts as originating from a legitimate sensor $\mathrm{S}_j$.
\label{Xenun5-5}
\end{thm}

\begin{proof}
 An FDI attack differs from MitM modification in that $\mathcal{A}$ attempts to inject an entirely synthetic reading $d_{\mathrm{fake}}$ without access to a valid message's ciphertext. To succeed, $\mathcal{A}$ must generate a ciphertext that decrypts to a specific target value $\delta'_{\mathrm{target}} = \lambda_k(d_{\mathrm{fake}} + iw_k)$.

 As established in the Attack Model, $\mathcal{A}$ does not possess the secret challenge factor $\lambda_k$ nor the current watermark $w_k$. In the \hidden-EG protocol, $\lambda_k$ is transmitted to the sensor $\mathrm{S}_j$ encrypted using the DC's public key; its confidentiality is thus guaranteed by the semantic security of the Gaussian ElGamal scheme. Similarly, in the \hidden-AggP protocol, $\lambda_k$ is protected by a unique symmetric key $K_{\mathrm{sym},j}$ shared between the DC and $\mathrm{S}_j$.

 Consequently, $\mathcal{A}$ is unable to learn $\lambda_k$ by eavesdropping in either protocol. Even if $\mathcal{A}$ intercepts the encrypted challenge and applies it homomorphically in \hidden-EG, they must still blindly guess the expected watermark $w_k$ for the current time frame. With a watermark space of cardinality $2^B$, the probability of $\mathcal{A}$ successfully crafting an injected value that survives the DC's division and watermark extraction test is strictly bounded by:
 \[
 P(\mathrm{success}) \leq \frac{1}{2^B}.
 \]

 Notably, in the specific case of the \hidden-AggP protocol, the attacker's capabilities are further restricted. Because Paillier encryption is only additively homomorphic, an attacker cannot homomorphically multiply two intercepted ciphertexts. To homomorphically scale a forged ciphertext, $\mathcal{A}$ would require the plaintext scalar $\lambda_k$, which is securely transmitted via the symmetric cipher using $K_{\mathrm{sym},j}$. Without access to this plaintext challenge, the homomorphic injection vector is neutralized. Consequently, to successfully forge a reading in the \hidden-AggP scheme, the attacker must blindly guess both the specific multiplicative factor $\lambda_k$ and the expected watermark $w_k$, recovering the stricter joint probability bound of:
 \[
 P(\mathrm{success}_{\mathrm{AggP}}) \approx \frac{1}{|S_\lambda|} \cdot \frac{1}{2^B}.
 \]

 For standard cryptographic parameters, this probability is negligible. Thus, the successful extraction of $w_k$ serves as a proof of origin and authenticity, fulfilling the requirement for data provenance.
\end{proof}

The resistance of \hidden-AggP against an adaptive version of FDI that can take advantage of the vulnerabilities of additive homomorphic encryption in regular integer scalar protocols is discussed in \linkref[Section]{\ref{sec:structural}}.

\begin{thm}(Perfect reversibility and accuracy)\label{the:rev}
The \hidden framework provides zero-distortion reconstruction of the host data if the range bounds $\max(|\Re(\delta'_k)|, |\Im(\delta'_k)|) < p/2$ (for \hidden-EG) and $\max(|\Re(\sigma'_k)|, |\Im(\sigma'_k)|) < n/2$ (for \hidden-AggP) are satisfied.
\label{Xenun6-6}
\end{thm}

\begin{proof}
 The correctness of the reconstruction relies on the modular decryption recovering the exact Gaussian integer result of the watermarking process:
 \begin{enumerate}
 \item Individual case (\hidden-EG): Decryption modulo $p$ in a centered representative system yields the exact value $\delta'_k = \lambda_k(d_k + i w_k)$ if its real and imaginary parts satisfy $|\Re(\delta'_k)| < p/2$ and $|\Im(\delta'_k)| < p/2$.
 \item Aggregated case (\hidden-AggP): Component-wise decryption modulo $n$ recovers the aggregate components exactly if $|\Re(\sigma'_k)| < n/2$ and $|\Im(\sigma'_k)| < n/2$.
 \end{enumerate}
 In both scenarios, once the exact Gaussian integer ($\delta'_k$ or $\sigma'_k$) is recovered, standard complex division by the current secret challenge factor $\lambda_k$ in the field $\mathbb{C}$ yields the exact watermarked data $\delta_k = d_k + i w_k$ or the aggregate $\sigma_k = S_k + i N w_k$. The host data ($d_k$ or $S_k$) is then perfectly reconstructed by taking the real part of the result, ensuring zero distortion.
\end{proof}

In short, the formal evaluation conducted through \linkref[Theorems]{\ref{the:eaves}}--\ref{the:rev} provides a comprehensive verification of the system requirements and attack model established in \linkref[Section]{\ref{Xsec23-5.1}}. The robustness against eavesdropping (\linkref[Theorem]{\ref{the:eaves}}) and the guarantee of aggregation privacy (\linkref[Theorem]{\ref{the:agg}}) satisfy the requirements of confidentiality and privacy. The probabilistic detection of replay (\linkref[Theorem]{\ref{the:replay}}) and the algebraic detection of MitM modifications (\linkref[Theorem]{\ref{the:mitm}}) fulfill the integrity requirement. Furthermore, the analysis of masquerading and FDI resistance (\linkref[Theorem]{\ref{the:fdi}}) validates the data provenance and authenticity properties by binding successful watermark extraction to the shared secrets. Finally, the accuracy proof (\linkref[Theorem]{\ref{the:rev}}) confirms that the protocols achieve perfect reversibility under defined conditions. Consequently, the \hidden framework is formally proven to be secure against both passive and active adversarial capabilities $\mathcal{A}$ within the distributed environment.

\section{Performance and comparative analysis}\label{Xsec34-7}\label{sec:performance}
In this section, we provide a rigorous comparative evaluation of the \hidden framework against the most relevant state-of-the-art protocols combining reversible data hiding (RDH) and homomorphic encryption. To the best of our knowledge, the literature addressing the joint application of RDH and partially homomorphic encryption specifically for individual, regular integer scalar-valued data in dispersed IoT environments is highly constrained. Currently, the P3HF high-frequency scheme \citep{kabir2024privacy} ---which embeds 1 bit per two data samples--- and the ReWaP protocol \citep{kabir2025rewap} represent the most direct and advanced solutions in this niche. Therefore, we select them as our primary baselines, as they define the current standard against which our complex-domain approach must be measured.

It is important to note that our comparative analysis strictly focuses on frameworks applying Homomorphic Encryption (HE). While symmetric encryption-based RDH schemes exist, they inherently lack the algebraic properties required to perform computations ---such as privacy-preserving data aggregation--- directly on the ciphertext. In symmetric architectures, intermediate nodes must decrypt the data to aggregate it, fundamentally breaking the end-to-end confidentiality model required by our target scenarios. Consequently, symmetric schemes operate under entirely different functional capabilities and trust models, rendering direct quantitative comparisons with HE-based aggregation frameworks mathematically ill-matched. As noted in \linkref[Remark]{\ref{rem:factor}}, while the proposed RDH scheme technique can be coupled with lightweight symmetric ciphers for extreme resource constraints, our system-level evaluation focuses on the advanced HE baselines that match our framework's homomorphic capabilities.

To establish a standardized comparative framework and address the requirements for a practical performance assessment, and provide a concrete proof-of-concept,\footnote{The Python implementation of the \hidden-EG protocol and its Gaussian integer arithmetic is publicly available at: \url{https://daniellerch.me/stegolab-en/\#watermarking}.} we assume a uniform implementation environment consistent with modern cryptographic and IoT security requirements:
\begin{itemize}
 \item Data representation: All protocols are evaluated using 32-bit long integers for both host data and watermarks, representing a common standard for high-precision sensor readings.
 \item Security parameters: We adopt 2048-bit security parameters (for the Paillier modulus $n$ or ElGamal prime $p$) to align with NIST standards for near-term and long-term cryptographic security.
 \item Protocol normalization: In the non-aggregated analysis (\linkref[Section]{\ref{sec;NonAggregated}}), the scalar baselines are adapted to the ElGamal cryptosystem. This normalization is essential to demonstrate that the observed security and capacity differences stem from the structural algebraic decoupling of the complex domain rather than the specific properties of the underlying cipher.
\end{itemize}

\begin{remk}It is worth noting that if the host data were assumed to natively span the entire available modular range (e.g., from $-(p-1)/2$ to $(p-1)/2$ for the ElGamal plaintext space, or $-(n-1)/2$ to $(n-1)/2$ for Paillier), the relative data expansion introduced by the encryption and embedding process would be mathematically negligible. However, using such a baseline for evaluation would be fundamentally unfair and misrepresentative of practical IoT deployments. In real-world sensing applications, data measurements occupy a very small numerical range (e.g., 16-bit or 32-bit integers). The massive bounds of $p$ and $n$ (typically 2048 bits) are dictated strictly by the security parameters required to resist modern cryptanalysis, rather than by the intrinsic magnitude of the physical data. Therefore, our benchmark intentionally evaluates the system under the realistic assumption of small plaintexts, explicitly acknowledging the resulting ciphertext growth as a necessary cryptographic requirement upon which the structural bit-doubling of the complex domain is applied. \remarkend
\label{Xenun29-23}
\end{remk}

While specific hardware-dependent benchmarks are transient, the fundamental computational overhead of the evaluated schemes is governed by the number of dominant modular exponentiations. In our comparative analysis, we use equivalent integer modular exponentiations as the primary metric for computational cost. This provides a hardware-agnostic performance guarantee, acknowledging that complex modular arithmetic in the \hidden-EG framework is approximately four times more intensive than regular integer-based equivalents, while the dual application of standard ciphers in \hidden-AggP (e.g., Paillier) incurs exactly twice the cost of a regular integer-based counterpart. The following analysis demonstrates how the \hidden framework leverages this controlled overhead to deliver superior capacity and several orders of magnitude higher resilience against False Data Injection (FDI) attacks.

Before detailing the specific performance metrics for the individual and aggregated scenarios, it is essential to contextualize the architectural entities involved. Differences in network topology heavily influence computational distribution, message complexity, and the overall attack surface across the compared protocols.

The native P3HF protocol proposes a four-tier architecture: Sensors (Smart Meters), an Encryption Server (ES) located at the edge, a Data Aggregator (DA), and a Control Center (CC). By offloading the homomorphic encryption to the ES, P3HF minimizes the computational burden on the sensors. However, this introduces a severe communication bottleneck: to ensure that a sensor can cryptographically verify its data was not maliciously altered or dropped by the edge server, the message flow must logically proceed from the Sensor to the ES, back to the Sensor, and finally to the DA. For $N$ sensors, this yields a message complexity of $3N$. In contrast, the ReWaP protocol uses a three-tier architecture (Sensors, DA, CC), where encryption is executed directly by the sensors, eliminating the ES but retaining the intermediate aggregator. Beyond the communication overhead, introducing intermediate entities like an ES or DA inherently expands the system's attack surface and introduces complex new trust assumptions.

To ensure a rigorous, fair, and meaningful comparative baseline, our evaluation mathematically normalizes all protocols to a strict two-tier architecture (Sensors and a single Data Collector/CC). By stripping away the ES and DA, we force all schemes to execute their core operations natively on the sensors. Consequently, in the sequel, the computational and communication overheads evaluated for both P3HF and ReWaP assume they operate strictly within this normalized two-entity configuration. We emphasize that this methodological decision is not designed to favor the \hidden framework. If evaluated under a three- or four-tier architecture, the \hidden protocols would equally benefit from offloading computations to an ES or offloading homomorphic additions to a DA, yielding analogous comparative insights. Instead, by establishing this uniform two-tier baseline, we eliminate architectural variables and isolate the true fundamental computation and communication overheads of each scheme. This guarantees a strict fair comparison of the core cryptographic mechanics and security properties, evaluating the frameworks under their most demanding physical configurations while minimizing the theoretical attack surface.

\begin{table*}[ht]
\centering
\caption{Comparative performance analysis ($\abs{p} \sim 2048$ bits). Baselines are normalized adaptations of regular scalar methods to the ElGamal system.}
\label{tab:comp_eg}
\begin{tabular}{|l|cc|c|}
\hline
\multirow{2}{*}{\textbf{Metric}} & \multicolumn{2}{c|}{\textbf{Normalized baselines$^\dagger$}} & \textbf{Proposed} \\ \cline{2-3}
 & \textbf{P3HF-EG} & \textbf{ReWaP-EG} & \textbf{\hidden-EG} \\ \hline
Watermark capacity ($B$) & 1 bit (per sample pair) & 16 bits & \textbf{32 bits} \\ \hline
FDI success probability ($2^{-B}$) & $0.5$ (per sample pair) & $1.53 \cdot 10^{-5}$ & $\mathbf{2.33 \cdot 10^{-10}}$ \\ \hline
Entanglement mechanism & Positional (LSB) & Positional (LSB) & \textbf{Algebraic diffusion} \\ \hline
Ciphertext size (total) & 4096 bits & 4096 bits & \textbf{8192 bits} \\ \hline
IEEE 802.11 total size & 592 bytes & 592 bytes & \textbf{1104 bytes} \\ \hline
6LoWPAN frame count & 6 frames & 6 frames & \textbf{11 frames} \\ \hline
6LoWPAN total size & 698 bytes & 698 bytes & \textbf{1360 bytes} \\ \hline
Overhead vs. host data (32-bit)$^\star$ & 128$\times$ & 128$\times$ & \textbf{256$\times$} \\ \hline
Computational cost (Sensor) & 2 Modular exp. & 2 Modular exp. & \textbf{8 Equiv. modular exp.} \\ \hline
Computational cost (DC) & 4 Modular exp. & 4 Modular exp. & \textbf{13 Equiv. modular exp.} \\ \hline
\end{tabular}
\begin{flushleft}
\footnotesize{
$^\dagger$ Hypothetical adaptations for comparative parity; they are not native ElGamal implementations. \\
$^\star$ Evaluated as the ratio of ciphertext bits to the 32-bit original host data measurement, providing a normalized metric independent of the watermark size.}
\end{flushleft}
\end{table*}

\subsection{Non-aggregated protocols}\label{Xsec35-7.1}
\label{sec;NonAggregated}

To provide a consistent baseline, we evaluate the \hidden-EG protocol against the regular integer-valued scalar RDH methods of \cite{kabir2024privacy} and \cite{kabir2025rewap}, assuming their adaptation to the ElGamal cryptosystem over a 2048-bit prime $p$ to meet modern security requirements. This allows for a direct evaluation of the algebraic mapping efficiency independent of the encryption layer. For this reason, in this section, we use the acronyms ``P3HF-EG'' and ``ReWaP-EG'' to denote the necessary adaptation to ElGamal cryptography.

\subsubsection{Capacity and FDI resilience}\label{Xsec36-7.1.1}
In regular integer-valued scalar methods like P3HF-EG and ReWaP-EG, the data and watermark must share a single 32-bit space, using either difference expansion (P3HF) or data expansion (ReWaP-EG): $d' = qd + w$. In ReWaP-EG, allocating 16 bits for the host data $d$ restricts the watermark $w$ to a maximum of 16 bits. In contrast, \hidden-EG uses the imaginary domain to provide a dedicated 32-bit channel for the watermark, independent of the host data's range.

As shown in \linkref[Table]{\ref{tab:comp_eg}}, this decoupling allows \hidden-EG to provide a 32-bit watermark, significantly enhancing security. The probability of an attacker successfully masquerading or injecting a fraudulent reading is reduced from $2^{-16} \approx 1.53 \cdot 10^{-5}$ in ReWaP-EG to the worst-case success bound of $2^{-32} \approx 2.33 \cdot 10^{-10}$ in \hidden-EG, as formalized in \linkref[Theorem]{\ref{the:fdi}}.

Furthermore, \hidden-EG benefits from \textit{intrinsic algebraic entanglement}; the secret challenge factor $\lambda_k$ ensures that both data and watermark are diffused across both components of the Gaussian integer. Even under a threat model where an active attacker fully observes the communication channels and attempts a homomorphic multiplication attack using the intercepted encrypted $\lambda_k$, this complex-domain entanglement prevents the trivial positional separation (e.g., bit-shifting) possible in regular integer methods, strictly enforcing the worst-case $2^{-32}$ security bound against targeted data forgery.

\subsubsection{Expansion, computation, and communication overhead}\label{Xsec37-7.1.2}\label{sec:expansionEG}

By normalizing the encryption primitives to a 2048-bit ElGamal modulus, we evaluate the data expansion directly relative to the original 32-bit host measurement payload to avoid misleading denominator scaling. For individual non-aggregated monitoring, the standard scalar ElGamal baseline produces a 4096-bit ciphertext pair (128$\times$ overhead), while \hidden-EG produces an 8192-bit ciphertext over the Gaussian integer group (256$\times$ overhead). Concurrently, as detailed in \linkref[Table]{\ref{tab:comp_eg}}, this structural transformation increases the cryptographic workload: sensor-side execution scales from 2 baseline modular exponentiations to 8 equivalent modular exponentiations, while data center decryption and validation processing scale from 4 to 13 equivalent modular exponentiations.

To evaluate downstream network impacts under standard IoT infrastructure constraints, we contrast two transmission scenarios: high-bandwidth IEEE 802.11 and low-power 6LoWPAN over IEEE 802.15.4. Under IEEE 802.11, both the 512-byte scalar ElGamal baseline ciphertext and the 1024-byte (1~KB) \hidden-EG payload fit seamlessly within the standard MTU and are transmitted as single, atomic network packets. Accounting for a standard 24-byte MAC header, an 8-byte LLC/SNAP header, an uncompressed 40-byte IPv6 header, and an 8-byte UDP header yields a fixed 80-byte header overhead per packet. Consequently, the baseline requires the transmission of 592 total bytes over the air, whereas \hidden-EG requires 1104 total bytes, representing a clear 86.5\% increase in total over-the-air byte transmission for the unfragmented Wi-Fi scenario.

Conversely, under 6LoWPAN, the payloads must be fragmented to navigate the rigid 127-byte physical layer frame limit. To maximize transmission efficiency and prevent redundant cryptographic overhead, link-layer encryption is disabled (0-byte AES-CCM overhead) since payload-level data confidentiality is already guaranteed natively by our protocol. Deducting a standard 25-byte MAC header leaves exactly 102 bytes of available capsule space per frame.

The fragmentation sublayer constructs the transmission stream using highly specific byte-boundary constraints to allow stateful packet reassembly at the Edge Server. The initial frame requires a 4-byte 6LoWPAN fragmentation header and 7 bytes for the compressed IPv6/UDP network headers, leaving 91 bytes available. However, 6LoWPAN dictates a strictly mandatory 8-byte data alignment boundary (an integer multiple of octets) for its internal Datagram Offset tracking. Consequently, the combined network headers and data payload block must round down to the nearest multiple of 8, capping the total network datagram chunk at 96 bytes ($96 / 8 = 12$ exact blocks); this leaves exactly 89 bytes for the actual ciphertext data ($96 - 7 = 89$). Subsequent intermediate frames use a 5-byte fragmentation header containing the offset pointer, leaving 97 bytes of physical room. Enforcing the 8-byte boundary alignment limits these intermediate payloads to exactly 96 bytes of pure ciphertext per frame. Finally, the terminal residue frame carries the remaining trailing bytes left over from the mathematical division and is exempt from the 8-byte alignment rule since no subsequent data offset needs to be computed.

Applying this precise construction matrix across both configurations results in the deterministic frame distributions detailed in \linkref[Table]{\ref{tab:fragmentation_details_eg}}. For the standard ElGamal 512-byte scalar baseline, the ciphertext is distributed across exactly 6 physical frames, transmitting 698 total bytes over the air. In comparison, the 1024-byte payload of \hidden-EG requires exactly 11 physical frames, resulting in 1360 total bytes sent over the air. Consequently, \hidden-EG incurs an 83.3\% increase in packet overhead (11 frames vs. 6 frames) and a 94.8\% increase in raw byte overhead (1360 bytes vs. 698 bytes). It is critical to recognize that this increase represents a serialization latency and radio on-time trade-off for resource-constrained nodes.

In summary, a generalized communication overhead increase of 83\% to 95\% across typical network scenarios, alongside the expanded processing load compared with regular scalar baselines, represents the clear operational premium required by the complex-domain paradigm. This expanded footprint is the necessary price to pay for the multi-dimensional protection offered by \hidden-EG. By operating over the Gaussian integer group, the protocol successfully embeds robust watermark tracking directly within the encrypted payload without altering the core cryptographic layer. Ultimately, while this trade-off results in higher serialization latency, increased computational complexity, and more radio on-time for the sensor nodes, it transitions the network from basic data confidentiality to a comprehensive, tamper-resilient security solution fully compatible with modern IoT architectures.

\begin{table*}[ht]
\centering
\caption{Comprehensive 6LoWPAN frame construction and header (Hdr) overhead breakdown for ElGamal-based schemes}
\label{tab:fragmentation_details_eg}
\resizebox{\textwidth}{!}{\begin{tabular}{lcccccc}
\hline
\textbf{Frame configuration} & \textbf{Count} & \textbf{MAC Hdr} & \textbf{6LoWPAN Hdr} & \textbf{Network Hdr} & \textbf{Data payload} & \textbf{Physical size} \\ \hline
\multicolumn{7}{l}{\textit{\textbf{Proposed protocol} (\hidden-EG) — 1024 bytes total ciphertext}} \\
Initial frame  & 1 & 25 B & 4 B & 7 B & 89 B & 125 B \\
Intermediate frames & 9 & 25 B & 5 B & 0 B & 96 B (each) & 126 B (each) \\
Final residue  & 1 & 25 B & 5 B & 0 B & 71 B & 101 B \\
\textbf{Total over-the-air} & \textbf{11} & \textbf{275 B} & \textbf{54 B} & \textbf{7 B} & \textbf{1024 B} & \textbf{1360 B} \\ \hline
\multicolumn{7}{l}{\textit{\textbf{Regular scalar baseline} (P3HF-EG/ReWaP-EG) — 512 bytes total ciphertext}} \\
Initial frame  & 1 & 25 B & 4 B & 7 B & 89 B & 125 B \\
Intermediate frames  & 4 & 25 B & 5 B & 0 B & 96 B (each) & 126 B (each) \\
Final residue  & 1 & 25 B & 5 B & 0 B & 39 B & 69 B \\
\textbf{Total over-the-air} & \textbf{6} & \textbf{150 B} & \textbf{29 B} & \textbf{7 B} & \textbf{512 B} & \textbf{698 B} \\ \hline
\end{tabular}}
\end{table*}

\begin{remk}\label{rem:migration_impact}
Following the compliance roadmap established in \linkref[Section]{\ref{sec:security_parameter_selection}}, migrating the public modulus to $\eta = 3072$ bits to meet the January 1, 2031 NIST deadline directly scales the transmission footprint of the protocol. For \hidden-EG, the absolute ciphertext size doubles the scalar baseline to expand from 1024 bytes ($8192$ bits) to exactly 1536 bytes ($12,288$ bits). Under high-bandwidth IEEE 802.11 infrastructures, this 1.5~KB payload fits within the native 2304-byte Wi-Fi maximum frame body capacity, executing as a single, unfragmented atomic packet. Conversely, under low-power IEEE 802.15.4 networks, a 1536-byte payload combined with the UDP and network layers exceeds the 1280-byte IPv6 MTU ceiling enforced by the 6LoWPAN adaptation interface. Because UDP lacks native transport-layer segmentation mechanisms, the network layer must perform IPv6 fragmentation, splitting the datagram across two distinct IP fragments (restricting the ciphertext payloads to 1224 bytes and 312 bytes, respectively, to accommodate the 8-byte IPv6 Fragment Header). The 6LoWPAN sublayer then processes each IP fragment independently, enforcing the mandatory 8-byte data alignment rules to fracture the transmission across a combined total of exactly 17 over-the-air link-layer frames (comprising 13 frames for the primary segment and 4 frames for the residual segment). This 54.5\% surge in fragmentation volume (scaling from 11 frames to 17 frames) underscores the significant serialization and radio on-time trade-offs required to achieve post-2030 classical cryptographic compliance.
\remarkend \end{remk}

\subsubsection{Discussion of non-aggregated protocols results}\label{Xsec38-7.1.3}
The performance metrics in \linkref[Table]{\ref{tab:comp_eg}} illustrate that the \hidden-EG protocol offers a superior security-to-overhead ratio compared to regular integer-valued scalar approaches. In traditional regular integer-valued methods like ReWaP-EG, the host data and watermark are constrained to share a single 32-bit register. To preserve the host data's magnitude, the watermark is typically relegated to the lower 16 bits, which limits the security space and increases vulnerability to masquerading or replay attacks.

The \hidden-EG protocol, by contrast, takes advantage of the complex domain to provide two independent 32-bit channels. This allows for a full 32-bit watermark without any interference with the host data's numerical range. This decoupling results in a security resilience of $2^{-32} \approx 2.33 \cdot 10^{-10}$, providing five orders of magnitude more protection than the 16-bit regular integer scalar alternative ($2^{-16} \approx 1.53 \cdot 10^{-5}$) at the price of an increased relative expansion ratio from $128\times$ to $256\times$.

\begin{table*}[ht]
\centering
\caption{Comparative performance for aggregated protocols ($N_{\max}=100$, $\abs{n}\sim2048$ bits)}
\label{tab:comp_agg}
\resizebox{\textwidth}{!}{\begin{tabular}{|l|c|c|c|}
\hline
\textbf{Metric} & \textbf{P3HF \citep{kabir2024privacy}} & \textbf{ReWaP \citep{kabir2025rewap}} & {\boldmath \textbf{\hidden-AggP}} \\ \hline
Watermark capacity ($B$) & 1 bit (per sample pair) & 9 bits & \textbf{25 bits} \\ \hline
Entanglement mechanism & Positional (LSB) & Positional (LSB) & \textbf{Algebraic diffusion} \\ \hline
FDI success probability & $0.5$ (per sample pair) & $2^{-9} \approx 1.95 \cdot 10^{-3}$ & $\mathbf{2^{-25}\cdot2^{-48} \approx 1.06 \cdot 10^{-22}}$ \\ \hline
Ciphertext size (total) & 4096 bits & 4096 bits & \textbf{8192 bits} \\ \hline
IEEE 802.11 total size & 592 bytes & 592 bytes & \textbf{1104 bytes} \\ \hline
6LoWPAN frame count & 6 frames & 6 frames & \textbf{11 frames} \\ \hline
6LoWPAN total size & 698 bytes & 698 bytes & \textbf{1360 bytes} \\ \hline
Overhead vs. host data (32-bit)$^\star$ & 128$\times$ & 128$\times$ & \textbf{256$\times$} \\ \hline
Computational cost (Sensors)\textsuperscript{\textbf{\dag}} & 2 Modular exp. & 2 Modular exp. & \textbf{4 Modular exp.} \\ \hline
Computational cost (DC)\textsuperscript{\textbf{\dag}} & 1 Modular exp. & 1 Modular exp. & \textbf{2 Modular exp.} \\ \hline
Homomorphic FDI injection resistance & \checkmark & $\times$ & \ding{52}\\ \hline
\end{tabular}}
\begin{flushleft}
\footnotesize{$^\star$ Evaluated as the ratio of ciphertext bits to the 32-bit original host data measurement, providing a normalized metric independent of the watermark size.} \\
\footnotesize{\textsuperscript{\dag} Total dominant modular exponentiations modulo $n^2$ per round.}
\end{flushleft}
\end{table*}

Beyond capacity, \hidden-EG provides \textit{intrinsic algebraic entanglement}. Unlike scalar methods where the watermark is often positionally separated (e.g., as LSBs), the use of the complex challenge factor $\lambda_k$ ensures that the host data and watermark are mathematically diffused across both components of the Gaussian integer. This algebraic bond prevents an adversary from easily isolating or stripping the watermark from the carrier. Finally, it is important to note that, at our core baseline parameter setting ($\eta = 2048$ bits), the 8192-bit (1\thinspace KB) ciphertext footprint remains fully compatible with the Maximum Transmission Unit (MTU) of standard IoT communication protocols, such as IEEE 802.11 or 6LoWPAN. This confirms the framework's practical viability for individual sensor transmissions, as the baseline payload fits within a single data packet.

However, as shown in \linkref[Table]{\ref{tab:comp_eg}}, these advantages come at a documented computational and transmission cost. Specifically, the sensor-side CPU demand is approximately $4\times$ higher than scalar methods, and the Data Collector (DC) experiences a $3.25\times$ increase in computational overhead. Furthermore, the absolute size of the transmitted data is\break doubled (8192 bits vs. 4096 bits). We argue that in the intended context of dispersed, low-frequency data readings, these additional costs are well-compensated by the leap in cryptographic-strength integrity. While a regular scalar approach is more lightweight, it cannot provide the same level of FDI resilience or the structural protection against watermark isolation that \hidden-EG offers. Furthermore, as detailed in \linkref[Remark]{\ref{rem:migration_impact}}, this overhead structure carries over predictably into the mandatory post-2030 key migration timeline ($\eta = 3072$ bits). Although the expanded 1536-byte payload triggers a two-tier encapsulation penalty that forces exactly 17 over-the-air fragmented frames within low-power 6LoWPAN link layers, it remains seamlessly absorbed as a single unfragmented datagram by high-bandwidth IEEE 802.11 networks. This structural adaptability confirms that the framework can be flexibly scaled to meet shifting regulatory protection timelines without modifying the core complex-domain architecture.

\subsection{Aggregated protocols}\label{Xsec39-7.2}
In this section, we evaluate the \hidden-AggP protocol against the ReWaP protocol \citep{kabir2025rewap}, a state-of-the-art scalar RDH scheme for privacy-preserving aggregation. For this analysis, we assume a typical deployment with a maximum of $N_{\max}=100$ devices and a standard 2048-bit Paillier modulus $n$. In \hidden-AggP, this setting allows 32 bits for the representation of data ($d_{j.k}$) and 32 bits for $Nw_k$. Since $N$ can need up to 7 bits ($N\leq100)$, this leaves 25 bits for the watermark.

To ensure highly efficient hardware-level execution both at the sensor node and during the extraction phase at the Data Center (DC), \hidden-AggP is designed to keep all plaintext operations within native 64-bit signed integer boundaries (for the multiplication of $\lambda_k$ and $\delta_{j,k}$).
This requires that the homomorphically aggregated sum of $N_{\max}$ complex components does not overflow 63 bits of magnitude. We analyze the worst-case absolute expansion of the complex multiplication: $\lambda_k \delta_{j,k} = (a_k d_{j,k} - b_k w_k) + i(a_k w_k + b_k d_{j,k})$. Because the inputs can be negative, the terms may add constructively. Consequently, both the real and imaginary components share the same maximum absolute bound: $M_{\max} \approx 2^{B_\lambda} (2^{32} + 2^{25})$, where $B_\lambda$ is the bit-depth of the challenge components $a_k$ and $b_k$.

Because the 32-bit host data $d_{j,k}$ dominates the 25-bit watermark $w_k$, the cross-multiplication term evaluates to approximately $32.01$ bits of magnitude. Aggregating $100$ devices scales this sum by a factor of $\log_2(100) \approx 6.64$ bits. Thus, the maximum allowable bit-depth for the challenge components is strictly derived as:
\[
 B_\lambda \leq 63 - 32.01 - 6.64 = 24.35 \text{ bits}.
\]

By safely allocating 24 bits to $a_k$ and 24 bits to $b_k$, the expanded complex components compute natively (in 64-bit integers) prior to Paillier encryption. In this way, when the DC decrypts the 2048-bit aggregated ciphertext, the recovered plaintexts can be cast directly to native 64-bit integers, entirely bypassing the computational overhead of software-based big-integer division during data and watermark recovery.

\subsubsection{Capacity and FDI resilience}\label{Xsec40-7.2.1}
In regular integer-valued scalar RDH methods like ReWaP, the host data $d$ and the aggregated watermark $Nw$ must share a single bit-space. In a typical 32-bit integer implementation, allocating 16 bits for the host data and accounting for aggregation growth ($\lceil \log_2(N_{\max} )\rceil = 7$ bits) leaves a watermark capacity of approximately $B = 9$ bits. This results in a False Data Injection (FDI) success probability of $2^{-9} \approx 1.95 \cdot 10^{-3}$.

In contrast, \hidden-AggP uses the imaginary domain as a dedicated 32-bit channel for the watermark. As discussed above, after accounting for the 7-bit expansion from aggregating $N_{\max}=100$ readings, the watermark retains 25 bits of capacity. Furthermore, as established in the bit-allocation architecture, the native 64-bit boundary constraint limits the complex challenge components $a_k$ and $b_k$ to 24 bits each. Because \hidden-AggP uses a symmetric cipher to protect the complex challenge $\lambda_k$, homomorphic multiplication attacks (see \linkref[Section]{\ref{sec:structural}}) are neutralized. Consequently, an adversary attempting an FDI attack must simultaneously guess the 25-bit watermark and the 48-bit complex key space ($|S_\lambda| = 2^{48}$). As formalized in \linkref[Theorem]{\ref{the:fdi}}, this strictly suppresses the FDI success probability to an exceptional bound of $2^{-25} \cdot 2^{-48} = 2^{-73} \approx 1.06 \cdot 10^{-22}$. This represents a massive, multi-order-of-magnitude improvement in cryptographic integrity over regular scalar baselines.

\subsubsection{Expansion, computation, and communication overhead}\label{Xsec41-7.2.2}\label{sec:expansionAggP}

The transition to a 2D algebraic structure inherently introduces a well-defined trade-off in both data expansion and computational complexity. As detailed in \linkref[Table]{\ref{tab:comp_agg}}, \hidden-AggP doubles the raw ciphertext size to 8192 bits and doubles the number of dominant modular exponentiations modulo $n^2$ compared to the scalar Paillier baselines, requiring 4 operations at the sensor nodes and 2 at the data center.

Due to the exact mathematical equivalence in ciphertext volume (512 bytes for the scalar baselines and 1024 bytes for the proposed \hidden-AggP protocol), the downstream communication footprint maps identically to the ElGamal network evaluation detailed in\break \linkref[Section]{\ref{sec:expansionEG}}. Under IEEE 802.11, \hidden-AggP incurs an 86.5\% over-the-air byte increase (1104 bytes vs. 592 bytes), while under the 6LoWPAN fragmentation matrix, it replicates the identical 83.3\% packet increase (11 frames vs. 6 frames) and 94.8\% raw byte overhead penalty (1360 bytes vs. 698 bytes). Ultimately, this generalized 83\% to 95\% network premium, alongside the increased computational load, represents the necessary operational price to pay for the vastly expanded watermark capacity, inseparable algebraic diffusion, and robust homomorphic FDI injection resistance provided by the complex-domain paradigm.

Notably, because the absolute ciphertext payload of \hidden-AggP scales as a function of $4\eta$ bits ---matching the transmission volume of \hidden-EG exactly--- the post-2030 parameter migration consequences and the 17-frame network-layer IPv6 fragmentation penalties detailed in \linkref[Remark]{\ref{rem:migration_impact}} apply analogously to this aggregated architecture.

\subsubsection{Discussion of aggregated protocols results}\label{Xsec42-7.2.3}
As shown in \linkref[Table]{\ref{tab:comp_agg}}, the advantages of the complex-domain approach become even more pronounced in the aggregated scenario. In regular integer-valued scalar methods such as ReWaP, the watermark capacity is not only constrained by the initial 32-bit register limit but is further degraded by the bit-growth required to accommodate the summation of $N$ watermarks. For a typical 100-device deployment, this competition for bit-space restricts the effective watermark to approximately 9 bits.

The \hidden-AggP protocol mitigates this limitation by using the imaginary domain as a dedicated channel. While it doubles relative expansion ratio compared to the regular scalar baselines ($256\times$ vs. $128\times$), it provides nearly triple the watermark capacity (25 bits) after accounting for aggregation growth. Combined with the necessity to guess the 48-bit complex challenge space, this results in a success probability for False Data Injection (FDI) attacks of $2^{-73} \approx 1.06 \cdot 10^{-22}$, which is 19 orders of magnitude lower than the $1.95 \cdot 10^{-3}$ offered by regular integer scalar alternatives.

These results demonstrate that \hidden-AggP offers significantly higher security efficiency for the same relative data expansion. However, we acknowledge that this enhanced resilience carries a specific resource cost: the sensor-side computational demand is doubled (4\thinspace vs. 2 modular exponentiations), and the total transmitted ciphertext size is twice that of scalar methods (8192 bits vs. 4096 bits). As established in \linkref[Section]{\ref{sec:expansionEG}}, we argue that these costs are justifiable in the context of critical IoT infrastructures where the baseline 1~KB ciphertext remains fully compatible with the MTU of protocols like 6LoWPAN or IEEE 802.11, and where the ``price'' of a nineteen-order-of-magnitude increase in integrity protection is a worthwhile investment for dispersed, low-frequency data readings. In addition, because the absolute ciphertext payload of \hidden-AggP scales as a function of $4\eta$ bits ---matching the transmission volume of \hidden-EG exactly--- the post-2030 parameter migration consequences and the 17-frame network-layer IPv6 fragmentation penalties detailed in \linkref[Remark]{\ref{rem:migration_impact}} apply analogously to this aggregated architecture.

A critical advantage of the \hidden-AggP protocol is its \textbf{superior scalability in multi-sensor environments,} particularly when compared to state-of-the-art baselines such as ReWaP. In our framework, the maximum number of aggregated devices, $N$, is not a hard cryptographic constraint but rather a highly flexible parameter determined by the bit allocation for the aggregated watermark ($N \cdot w$). As established in our evaluation, a standard 32-bit integer representation provides a vast operational space. For instance, accommodating a massive deployment of $N = 4096$ ($2^{12}$) devices still preserves 20 bits for the individual watermark $w$. Because the capacity in the \hidden framework is algebraically decoupled from the host data range, the \hidden-AggP protocol can support an aggregation network up to $2^{16}$ times larger than that of ReWaP, which is strictly bottlenecked by the limited capacity of its spatial embedding domain. This flexibility makes the proposed framework exceptionally well-suited for massive, large-scale IoT deployments.

The scalability of P3HF is, in principle, better than that of ReWaP, since it allows for an unlimited number of devices as far as $N$ is an odd number. If $N$ were not odd, this could be circumvented by adding a ``dummy'' sensor injecting ``0'' readings with the appropriate watermark bit. However, as shown in \linkref[Table]{\ref{tab:comp_agg}}, the FDI detection probability of P3HF is very low (just 0.5), which cannot compete with that of ReWaP or \hidden-AggP, making it unsuitable for low frequency scenarios.

\subsection{Structural security and malleability resilience}\label{Xsec43-7.3}\label{sec:structural}
A fundamental limitation of scalar RDH-HE schemes is their vulnerability to targeted homomorphic manipulation. In a standard integer mapping where data $d$ and watermark $w$ are partitioned (e.g., $m = d \cdot 2^{16} + w$), like in ReWaP \citep{kabir2025rewap}, the spatial separation is preserved through the homomorphic process, leading to distinct vulnerabilities.

A similar attack would not be possible in P3HF \cite{kabir2024privacy}, as this scheme includes an additional cryptographic layer (symmetric encryption through XOR in all communications). However, it still suffers from a low probability of FDI detection.

\subsubsection{Malleability in scalar additive schemes}\label{Xsec44-7.3.1}
The most severe vulnerability exists in aggregated protocols based on the Paillier cryptosystem, such as ReWaP. Since the aggregator performs a summation, the scalar ``slots'' are preserved. An active Man-in-the-Middle (MitM) attacker can intercept \textbf{any} individual sensor ciphertext or partial aggregated result $C$ and compute:

\[
C' = C \cdot \enc_{\mathsf{K_{pub}}}\left(d_{\mathrm{fake}} \cdot 2^{16}\right) \pmod{n^2},
\]
Upon decryption, the result is $D(C') = (d + d_{\mathrm{fake}}) \cdot 2^{16} + w$, \textbf{preserving the correct watermark value.} This attack is effective at any stage:
\begin{itemize}
 \item At the sensor level: A MitM can forge a specific reading before it reaches the aggregator.
 \item At the aggregator level: A MitM can inject a global bias into the sum of hundreds of sensors.
\end{itemize}
In all cases, the watermark sum $Nw$ remains perfectly intact. The scalar architecture effectively creates a ``blind spot'' in the integrity check, where the host data can be manipulated independently of the security tag.

\subsubsection{Fragility in scalar multiplicative schemes}\label{Xsec45-7.3.2}
In non-aggregated scenarios using standard multiplicative ElGamal, a direct ``injection'' is not possible because the multiplicative property scales the entire plaintext by some factor. While this prevents the ``stealthy'' injection of a specific bias without altering the watermark, it demonstrates the \textit{fragility} of the mapping. The integrity of the data is purely dependent on the noise-tolerance of the watermark channel. If the implementation uses \textit{Exponential ElGamal} to support additive homomorphic operations, it immediately inherits the same injection vulnerability as the Paillier-based scalar schemes.

\subsubsection{The \hidden defense: algebraic diffusion}\label{Xsec46-7.3.3}
In contrast, the proposed RDH scheme replaces positional separation with \textit{algebraic entanglement} in the complex domain. By using a secret complex challenge $\lambda_k$, the data $d$ and watermark $w$ are mathematically fused into a Gaussian integer. Because $\lambda_k$ is secret and involves both rotation and scaling, an adversary cannot distinguish a ``data channel'' from a ``watermark channel'' in the ciphertext.

Any homomorphic operation (whether additive or multiplicative) results in \textit{cross-domain contamination}. During decryption, the complex-domain processing causes any injected noise to be mathematically diffused across both the recovered data and the recovered watermark. This ``all-or-nothing'' structural bond ensures that any unauthorized modification to the carrier inevitably corrupts the watermark, providing a non-malleable integrity guarantee that the regular integer-based scalar architectures cannot provide.

\subsubsection{Security analysis under insider compromise}\label{Xsec47-7.3.4}
To formally bound the non-malleability claims of the \hidden-AggP protocol, we examine the cryptographic resilience of the aggregated stream against a compromised network participant under three distinct adversary knowledge states:

\begin{enumerate}
 \item \textbf{Case 1: Full secret exposure ($\mathcal{A}$ knows both $\lambda_k$ and $w_k$):} If a malicious insider successfully extracts both its node-specific challenge $\lambda_k$ and its localized watermark seed $w_k$, it possesses the complete cryptographic state required to forge the readings and bypass watermark detection. The adversary can perfectly construct a fake payload that maintains local geometric alignment. We explicitly define this as the theoretical upper boundary of our scheme; preventing data forgery by an adversary who holds the full valid secret key space is outside the functional scope of reversible information hiding and requires orthogonal endpoint security mechanisms (e.g., physical tamper-resistance or hardware security modules).

 \item \textbf{Case 2: Naive material injection ($\mathcal{A}$ knows $\lambda_k$ but not $w_k$):} If the adversary compromises the challenge factor $\lambda_k = a_k + ib_k$ but lacks knowledge of the underlying localized watermark value $w_k$, any attempt to inject a false payload with a guessed or randomized watermark component ($\Delta_d + i \cdot \Delta_{\text{rand}}$) will disrupt the global aggregation space. Because the attacker must blindly guess the active watermark configuration to avoid corrupting the macro-invariant, the probability of executing this injection seamlessly is strictly bounded by the net watermark capacity (e.g., $B = 25$ bits after accounting for aggregation growth), yielding a success probability of only $2^{-25} \approx 2.98 \cdot 10^{-8}$. With an overwhelming probability of $1 - 2^{-25}$, the localized alteration propagates chaotically into the aggregate components, causing the extracted watermark at the DC to fail validation, spoiling the entire batch and triggering an intrusion response.

 \item \textbf{Case 3: Strategic watermark annihilation ($\mathcal{A}$ knows $\lambda_k$ but not $w_k$):} To cleverly circumvent their ignorance of $w_k$, a sophisticated adversary may attempt to inject a pure real-valued data bias ($\Delta_d + i \cdot 0$) using their known challenge $\lambda_k$. Since the complex embedding tightly entangles the components, multiplying the challenge by a zero-valued imaginary offset completely annihilates node $u$'s (the compromised node) own valid watermark contribution $w_k$ from the homomorphic sum during aggregation. Upon decryption, the DC recovers a global aggregate watermark of exactly:
 \[
 \sum_{j\neq u} w_k = (N - 1)w_k.
 \]
 Consequently, this strategic attack creates a localized aggregated watermark deficit of exactly $-1$ unit. In standard secure IoT infrastructures where the DC actively monitors session states and maintains an exact count of reporting cryptographic devices ($N$), this structural mismatch guarantees immediate detection, isolating the forgery attempt. Consequently, the success of this strategy is strictly conditional on the adversary's ability to simultaneously spoof the network layer and deceive the DC regarding the exact count of active session participants ($N$).

 \item \textbf{Case 4: Hybrid eavesdropping-injection attack ($\mathcal{A}$ knows $\lambda_k$ but not $w_k$) via MitM:} A distinct, highly sophisticated scenario occurs if the adversary expands their operational posture beyond the compromised endpoint. Armed with the stolen interval challenge $\lambda_k$, the adversary deploys a Man-in-the-Middle (MitM) architecture on the communication lines of an entirely uncompromised node $v$. By intercepting node $v$'s ciphertext, the attacker homomorphically adds a structured complex bias offset equivalent to $\lambda_k(\Delta + i \cdot 0)$ separately to the encrypted real and imaginary components. Upon global decryption at the DC, this structured addition injects the targeted real-data bias $\Delta$ while adding exactly zero to the imaginary domain. Because node $v$'s authentic watermark contribution $w_k$ remains undisturbed within the underlying ciphertext structure, the global aggregate watermark evaluates perfectly to $N \cdot w_k$, bypassing the aggregate integrity verification.

 We emphasize that while this attack is theoretically viable, it represents a significantly elevated adversarial complexity compared to regular scalar integer-partitioned designs (e.g., ReWaP). In regular scalar architectures, a network MitM requires \textit{zero} secret parameters or cryptographic components to inject a stealthy bias. In contrast, the algebraic entanglement of \hidden forces an external network adversary to first successfully execute an endpoint node breach to extract $\lambda_k$ before any targeted homomorphic injection can succeed. This effectively shifts the system's security boundary from a simple network-layer vulnerability to a complex, multi-stage Advanced Persistent Threat (APT) scenario.
\end{enumerate}

\subsection{IoT edge feasibility and latency tolerance}\label{Xsec48-7.4}
\label{sec:iot_edge_latency}

To explicitly bridge theoretical computational costs with practical IoT deployment realities, the framework's baseline execution was evaluated using a standard ESP32 edge microcontroller architecture (Xtensa dual-core 32-bit LX6 microprocessor at $240$ MHz with $520$ KB SRAM). To achieve an accurate architectural representation, the compiled Arduino (C++) binary was executed within an instruction-level hardware emulator. More precisely, this emulation evaluated the latency of the encryption of the \hidden-EG protocol (Gaussian integer ElGamal) using robust $2048$-bit cryptographic keys. As established, this complex-domain computation requires approximately $4\times$ the effort of a regular integer ElGamal encryption. Despite the massive scale of the underlying $2048$-bit integer arithmetic, the encryption yielded a conservative emulated execution time of $6.14$ s.

To estimate the performance of the \hidden-AggP variant, we can relate it to the same baseline. A standard 2048-bit Paillier encryption operates over a larger $n^2$ modulus space (4096 bits), making it roughly $3\times$ as expensive as a regular 2048-bit ElGamal encryption. Because \hidden-AggP requires two independent Paillier encryptions (one for the real part and one for the imaginary part), its total cost equates to approximately $6\times$ a regular ElGamal encryption. Comparing these ratios ($6\times$ vs.\ $4\times$), the \hidden-AggP protocol incurs roughly a $50\%$ computational overhead relative to \hidden-EG, scaling the estimated edge encryption latency to approximately $9.2$ s.

While physical bare-metal hardware execution inherently bypasses emulator overhead and would further reduce these latencies, even these conservative bounds ($6$ to $9$ s) definitively demonstrate the framework's real-world viability. In dispersed IoT networks and Advanced Metering Infrastructure (AMI), smart sensors typically transmit readings at low frequencies (e.g., intervals ranging from $15$ to $60$ min). Consequently, a maximum computational latency of $9$ s is entirely masked by the extensive idle time between transmissions, ensuring the complex-domain embedding remains highly practical without disrupting network operations.

\subsection{Summary of comparative evaluation}\label{Xsec49-7.5}
The comparative analysis presented in this section highlights a fundamental trade-off between computational efficiency and structural security. \linkref[Tables]{\ref{tab:comp_eg}} and {\ref{tab:comp_agg}} confirm that the \hidden framework requires a higher resource investment than recent scalar-based alternatives: specifically, a $4\times$ (for the ElGamal version) or $2\times$ increase in sensor-side modular exponentiations and a doubling of the transmitted ciphertext size (for \hidden-AggP) at the baseline parameter setting ($\eta = 2048$ bits).

However, this overhead is directly translated into two critical advantages that regular integer-valued scalar architectures cannot provide:
\begin{enumerate}
 \item Channel independence and capacity: By using the complex domain, \hidden provides an increase in security resilience against False Data Injection (FDI) by a factor between $10^5$ (for the \hidden-EG worst-case bound) and $10^{19}$ (for the \hidden-AggP protocol). While regular integer methods are forced to squeeze watermarks into limited bit-slots alongside host data, the complex mapping allows for high-precision integrity tags that do not degrade the host signal's range or accuracy.
 \item Non-malleable integrity: As demonstrated in \linkref[Section]{\ref{sec:structural}}, aggregated regular integer-valued scalar schemes suffer from a structural ``blind spot'' due to positional separation. The \hidden framework's use of a secret complex challenge $\lambda_k$ ensures \textit{algebraic entanglement}, making it impossible for a Man-in-the-Middle to homomorphically manipulate host data without corrupting the watermark.
\end{enumerate}

In the context of modern IoT deployments, such as smart grids or environmental monitoring, data is typically collected at low frequencies. In these scenarios, the baseline 1~KB ciphertext footprint is easily accommodated by standard IEEE 802.11 or 6LoWPAN MTUs, but requiring physical-level fragmentation in the latter. Furthermore, as parameter lengths scale up to meet post-2030 regulatory guidelines ($\eta = 3072$ bits), the resulting 1536-byte payload remains fully viable; it is seamlessly handled as an unfragmented datagram by high-bandwidth Wi-Fi links or split across predictable, IPv6 network-layer and 6LoWPAN link-layer fragmentation boundaries over low-power radios. The second-scale computational cost in really constrained hardware (like ESP32 microcontrollers) is negligible compared to the long-term protection of data integrity. We conclude that for low-frequency measurements in critical infrastructures, where stealthy data forgery is a primary threat, the \hidden framework provides a superior and more practical security-to-overhead ratio than the existing RDH-HE scalar protocols with regular integer values.

\section{Conclusion}\label{Xsec50-8}
\label{sec:conclusion}

This paper introduces Hiding in the Imaginary Domain with Data Encryption (\hidden), a novel framework for RDH that operates uniquely within the complex domain. By intrinsically mixing data and watermark through complex number arithmetic, the \hidden approach offers perfect reversibility and allows for a highly scalable watermark capacity that is algebraically decoupled from the host data's range, directly addressing key limitations of existing RDH techniques for individual and dispersed data. Two distinct protocols embodying this framework are proposed: the \hidden-EG protocol, providing joint RDH and encryption for individual data points by using the extension of ElGamal encryption to Gaussian integers, and the \hidden-AggP protocol, which enables privacy-preserving data aggregation by taking advantage of a novel component-wise application of Paillier encryption to complex-valued data.

The proposed \hidden framework and its instantiations offer novel solutions for ensuring data integrity and authenticity in scenarios involving sensitive, distributed, and potentially resource-constrained environments like Wireless Sensor Networks and IoT ecosystems. By integrating advanced cryptographic primitives with innovative complex-domain embedding, the proposed methods provide strong security guarantees beyond simple data hiding, including confidentiality and privacy-preserving aggregation. This work demonstrates a significant step towards building more trustworthy data pipelines from diverse sources, where the provenance and integrity of information are strong requirements.

Rigorous comparative evaluation against recent state-of-the-art baselines establishes the practical and theoretical advantages of this approach. By achieving true algebraic entanglement, the framework neutralizes structural malleability vulnerabilities ---such as homomorphic injection attacks--- inherent to scalar positional embedding. This structural security is bolstered by a significant expansion of the watermark channel, which decouples the integrity tag from the host data's range. Consequently, the framework achieves an exponential improvement ($10^5$ to $10^{19}$ under reasonable settings) in False Data Injection (FDI) resilience, as the security bound scales with the combined bit-depth of the decoupled watermark and the secret complex challenge factor. Furthermore, empirical performance analysis, based on both hardware-agnostic cryptographic metrics and direct IoT edge hardware emulation (ESP32), confirms that the computational overhead of the complex domain is well within the operational limits of standard sensor duty cycles, ensuring real-world deployment feasibility.

Moreover, the architecture proves resilient to long-term regulatory parameter scaling. Transitioning to the post-2030 NIST-recommended $\eta = 3072$ bits compliance boundary preserves the core complex-domain algebraic designs while demonstrating high structural adaptability across diverse network infrastructures, seamlessly traversing high-bandwidth IEEE 802.11 links as atomic datagrams or resolving into predictable, fragmented structures over low-power 6LoWPAN layers.

As directions for future work, one key area involves exploring the inclusion of other homomorphic encryption schemes, such as BGN or FHE schemes, to operate efficiently within the complex domain or over Gaussian integers. Such extensions could unlock new homomorphic capabilities (e.g., multiple multiplications on ciphertexts) or improved efficiency for complex-valued computations. In addition, this could provide post-quantum security, an aspect that is not covered by ElGamal or Paillier.

Another future research direction is the potential application of the \hidden approach to other domains, such as media information or non-dispersed time series data. By adapting integer-based media formats to the complex domain for embedding, this method could offer a novel approach to traditional image, audio, or video watermarking challenges, taking advantage of the inherent entanglement properties and high capacity of the \hidden scheme.

\subsection*{Acknowledgments}
The author acknowledges the funding obtained by the ``SECURING'' (PID2021-125962OB-C31) and ``SAFE'' (PID2024-156914OB-C4) grants funded by the Ministry of Science and Innovation through the State Research Agency (AEI) and the European Regional Development Fund (ERDF), as well as the ARTEMISA International Chair of Cybersecurity (C057/23) and the DANGER Strategic Project of Cybersecurity (C062/23), both funded by the Spanish National Institute of Cybersecurity through the European Union -- NextGenerationEU and the Recovery, Transformation and Resilience Plan.




\appendix

\section{Modular arithmetic over Gaussian integers}
\label{app:modular}

\def\thetable{A.\arabic{table}}
\setcounter{table}{0}
\def\thefigure{A.\arabic{figure}}
\setcounter{figure}{0}

It is worth pointing out that Gaussian prime numbers have some differences compared to the regular integer case. The obvious one is that Gaussian primes can have both real and imaginary parts. The second main difference is that not all integer primes are also Gaussian primes. For example, prime factorizations exists for both 2 and 5 in $\mathbb Z[i]$: $2=(1+i)(1-i)$ and $5=(2+i)(2-i)$. The integer primes $p$ that are also Gaussian primes satisfy the following condition $p\equiv 3 \pmod{4}$ as proven, for example, in \citep{conrad2016gaussian}.

\begin{figure}[ht]
\centerline{\includegraphics[width=.75\columnwidth]{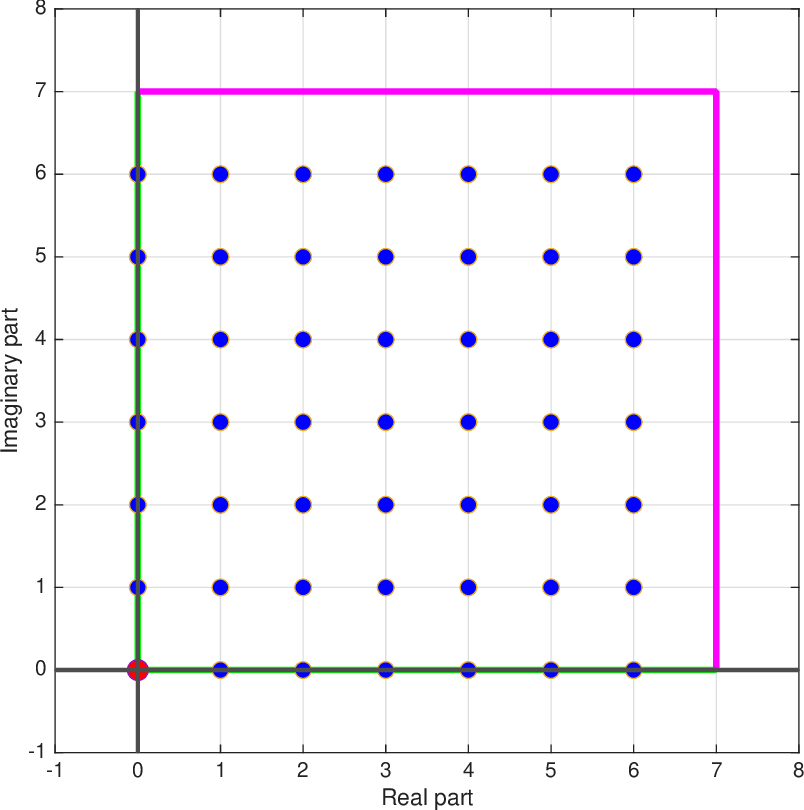}}
\caption{Representatives of Gaussian integers $\pmod{7}$, or $\mathbb Z[i]{}_7$.}
\label{fig:mod7}
\end{figure}

Regarding modular arithmetic, we need to extend the definition of integer division to the complex domain. For general Gaussian integers $
\alpha,\beta\in \mathbb Z[i]$, the Gaussian integer division is defined as $\alpha=\kappa \beta+\rho$, for some residual $\rho$ with $\mathrm{N}(\rho)<\mathrm{N}(\beta)$, where $\mathrm{N}(\cdot)$ stands for the norm of a complex number, which is defined as follows:
\[
\mathrm{N}(a+ib)=\abs{a+ib}^2=a^2+b^2.
\]
This definition preserves the intuition of integer division for Gaussian integers since, for each number $\alpha$ and each divider $\beta$ in Gaussian integers, a quotient $\kappa$ and a remainder $\rho$ can be found such that $\alpha=\kappa \beta+\rho$, and $\rho$ is, in some sense, smaller than the divider $\beta$. Here, the norm of a complex number is the criterion for determining whether a number is smaller than another.

With this definition of the integer division, we can apply the same rules of modular arithmetic as for standard integers, i.e., for $\alpha,\beta,\kappa,\rho\in \mathbb Z[i]$, if $\alpha=\kappa\beta+\rho$, we have $\alpha \equiv \rho \pmod{\beta}$, since $\beta | (\alpha-\rho)$.

It is well known that, in regular integers, for a prime number $p\in\mathbb Z$, the elements of the cyclic group $\mathbb Z_p$ (or $\mathbb Z/p\mathbb Z$) are $0,1,\dots,p-1$, while the multiplicative cyclic group $\mathbb Z^*_p$ has the same elements, excluding 0. The order of the corresponding cyclic groups $\mathbb{Z}[i]{}_{\pi}$ and ${\mathbb{Z}[i]}^*_{\pi}$ is not so straightforward when $\pi$ is a Gaussian prime. For Gaussian integers, given a prime number $\pi\in\mathbb Z[i]$, the norm $\mathrm N(\pi)$ determines the order of the additive cyclic group $\mathbb{Z}[i]{}_{\pi}$, whereas $\mathrm N(\pi)-1$ is the order of the multiplicative cyclic group ${\mathbb{Z}[i]}^*_{\pi}$ (because of the exclusion of 0). Proofs for these facts can also be found in \citep{conrad2016gaussian}.

As a practical example, the selection of representatives for the cyclic group ${\mathbb{Z}[i]}^*_{\pi}$ for $\pi=7$, which is both an integer and a Gaussian prime, is discussed below. First, we need to take into account that many possible selections of representatives exist, similar to the regular integer case. For example, 13 and 6 refer to the same equivalence class in both ${\mathbb Z}^{*}_7$ and ${\mathbb Z[i]}^*_7$, since 13 and 6 differ in a multiple of 7 (in fact, exactly 7). The only relevant consideration for the selection is not to include more than one representative for the same class. A method for such a selection is described in \citep{conrad2016gaussian} both when the divider has real and imaginary parts different from zero, and when the imaginary part is zero (the case of the Gaussian prime $\pi=7$).

The procedure is illustrated in \linkref[Fig.]{\ref{fig:mod7}}: We take $7$ (the divider) and $7i$ (by multiplying by $i$) and draw a square which has 7 and $7i$ as the left and lower edges (shown in green color in the picture). The square is completed with the purple edges that connect 7 and $7i$ with the vertex $7+7i$. The representatives of $\mathbb Z[i]{}_7$ are the Gaussian integers that lie within the square and the ones which are on the green edges, but excluding the ones on the purple edges, since they represent the same equivalence class as the elements on the green ones.

\begin{figure}[ht]
\centerline{\includegraphics[width=.75\columnwidth]{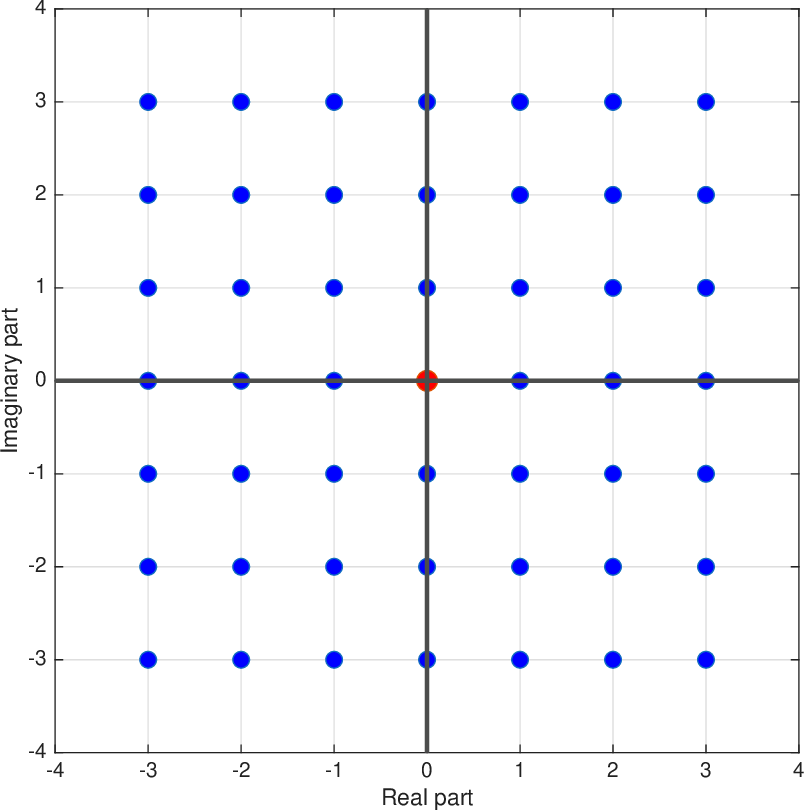}}
\caption{``Canonical'' representatives of Gaussian integers $\pmod{7}$, or $\mathbb Z[i]{}_7$.}
\label{fig:mod7can}
\end{figure}

With this particular selection, the set of values of the cyclic multiplicative group ${\mathbb{Z}[i]}^*_7$ can be easily represented as
\[ {\mathbb{Z}[i]}^*_7=\left\{ x + iy \mid x, y \in \{0, 1, \dots, 6\right\} \text{ and } (x,y) \neq (0,0) \}, \]
with no equivalent elements. This leads to the following observations:
\begin{itemize}
 \item The order of the additive cyclic group ${\mathbb Z[i]}_p$, for $p\in\mathbb Z$ (i.e., $p$ is a regular integer) is $p^2$.
 \item If $p\in \mathbb Z$ is both a regular prime and a Gaussian prime, i.e., when $p\equiv 3\pmod{4}$, then ${\mathbb Z[i]}^*_p$ is a multiplicative cyclic group of order $p^2-1$.
\end{itemize}

This set of representatives is particularly convenient to carry out modular arithmetic operations, since $(a+ib) \pmod{7}$ is simply $a \pmod{7}+ib\pmod{7}$. This means that we can just apply the modular operator to the real and the imaginary parts separately, which is easy to handle with most of the existing programming languages.

This selection of representatives is not ``canonical'', in the sense that it includes some elements whose norms are greater than $\mathrm{N}(7)=7^2=49$. For example $\mathrm{N}(5+5i)=5^2+5^2=50$.
A more ``canonical'' set of representatives for ${\mathbb{Z}[i]}^*_7$ would be the following (\linkref[Fig.]{\ref{fig:mod7can}}):
\[ {\mathbb{Z}[i]}^*_7=\left\{ x + iy \mid x, y \in \{-3, -2, \dots, 3\} \text{ and } (x,y) \neq (0,0) \right\}, \]
which is obtained by shifting the values of \linkref[Fig.]{\ref{fig:mod7}} by $-3$ units in each axis. This new selection of representatives is intuitively closer to that usually made for the cyclic groups defined using standard integers, but the one shown in \linkref[Fig.]{\ref{fig:mod7}} is more convenient for computational operations. In any case, both are completely equivalent. For example, $5+5i \equiv -2-2i \pmod{7}$, where the former is one of the points represented in \linkref[Fig.]{\ref{fig:mod7}} and the latter only appears in \linkref[Fig.]{\ref{fig:mod7can}}. Note that the norm of $-2-2i$ is $2^2+2^2=8$, which, of course, is lower than $N(7)=49$. Hence, although $5+5i$ has a norm greater than $\mathrm{N}(7)$, an equivalent value can easily be found with a norm below that of the modulus. The norms of all values represented in \linkref[Fig.]{\ref{fig:mod7can}} are lower than 49.

\begin{remk}When the data to be represented is signed (including samples with both positive and negative components), the representation of \linkref[Fig.]{\ref{fig:mod7can}} is mandatory. In such a case, a simple shift can map every value of \linkref[Fig.]{\ref{fig:mod7}} to its corresponding value of \linkref[Fig.]{\ref{fig:mod7can}} (or conversely).
\label{Xenun31-25} \linebreak \remarkend
\end{remk}

\section{Numerical protocol simulations}\label{app:numerical_simulations}

\def\thetable{B.\arabic{table}}
\setcounter{table}{0}
\def\thefigure{B.\arabic{figure}}
\setcounter{figure}{0}

To provide concrete clarity regarding the specific operational mechanics of the Gaussian integer arithmetic used within the \hidden framework, this appendix presents step-by-step numerical simulations of the protocol flows depicted in \linkref[Figs.]{\ref{fig:elgamal}} and \ref{fig:paillier}. Small cryptographic parameters are used to allow the reader to easily verify the calculations.

\subsection{\hidden-EG protocol flow}\label{Xsec51-B.1}\label{app:sim_eg}
This simulation traces the specific interactive protocol flow of the \hidden-EG scheme (\linkref[Fig.]{\ref{fig:elgamal}}), seamlessly mapping the foundational numerical example from \linkref[Section]{\ref{sec:numericalEG}} ($p = 23$) onto the protocol's operational steps. All arithmetic operations are performed in the complex ring $\mathbb{Z}[i]{}_{23}$. Consistent with the notation established in \linkref[Section]{\ref{sec:numericalEG}}, all intermediate values are represented strictly within the standard positive residue class $\{0, 1, \dots, 22\}$ rather than the centered representation $\{-11, \dots, 11\}$ (as formally detailed in \linkref[Appendix]{\ref{app:modular}}), recognizing that both mappings are algebraically equivalent.

\paragraph*{Initialization}
The Data Collector (DC) generates the ElGamal keys over $\mathbb{Z}[i]{}^*_{23}$:
\begin{itemize}
 \item Prime modulus: $p = 23$.
 \item Generator: $\gamma = 1 + 2i$.
 \item Private key: $a = 7$.
 \item Public key: $\mathcal{K} \equiv \gamma^a \pmod{23} \equiv (1+2i){}^7 \equiv (6 + 2i) \pmod{23}$.
\end{itemize}
The DC sends $\mathbf{K} = (p, \gamma, \mathcal{K})$ to the Sensor (S), and both parties agree on a target watermark $w_k = 4$ for this transmission.

\paragraph*{Step 1. DC generates $\lambda_k$}
The DC generates a random complex challenge factor to obscure the data during homomorphic transmission. Let $\lambda_k = 3 + 2i$.

\paragraph*{Step 2. DC encrypts $E_{\lambda_k} = \text{Enc}_{\mathbf{K}}(\lambda_k, b_{\lambda_k})$}
The DC encrypts $\lambda_k$ using the public key $\mathcal{K}$ and an ephemeral key $b_{\lambda_k} = 7$ (matching $\mu_2$ from \linkref[Section]{\ref{sec:numericalEG}}):
\begin{itemize}
 \item $E_{\lambda_k, 1} \equiv \gamma^7 \equiv (1+2i){}^7 \equiv (6 + 2i) \pmod{23}$.
 \item $E_{\lambda_k, 2} \equiv \lambda_k \cdot \mathcal{K}^7 \equiv (3+2i) \cdot (6+2i){}^7 \equiv (21 + 16i) \pmod{23}$.
\end{itemize}
The DC transmits $E_{\lambda_k} = (6+2i, 21+16i)$ to the Sensor.

\paragraph*{Step 3. S forms $\delta_k = d_k + iw_k$}
The sensor obtains reading $d_k = 5$ and constructs the complex plaintext (matching $\mu_1$ from \linkref[Section]{\ref{sec:numericalEG}}): $\delta_k = 5 + 4i$.

\paragraph*{Step 4. S encrypts $E_{\delta_k} = \text{Enc}_{\mathbf{K}}(\delta_k, b_{\delta_k})$}
Using ephemeral key $b_{\delta_k} = 5$, the sensor encrypts $\delta_k$:
\begin{itemize}
 \item $E_{\delta_k, 1} \equiv \gamma^5 \equiv (1+2i){}^5 \equiv (18 + 8i) \pmod{23}$.
 \item $E_{\delta_k, 2} \equiv \delta_k \cdot \mathcal{K}^5 \equiv (5+4i) \cdot (6+2i){}^5 \equiv (21 + 11i) \pmod{23}$.
\end{itemize}
Yielding $E_{\delta_k} = (18+8i, 21+11i)$.

\paragraph*{Step 5. Homomorphic Watermarking $E_{\delta'_k} = E_{\lambda_k} \cdot E_{\delta_k}$}
The sensor multiplies the two ciphertexts component-wise to embed the DC's scaling factor homomorphically:
\begin{itemize}
 \item $E_{\delta'_k, 1} \equiv (6+2i) \cdot (18+8i) \equiv (0 + 15i) \pmod{23}$.
 \item $E_{\delta'_k, 2} \equiv (21+16i) \cdot (21+11i) \equiv (12 + 15i) \pmod{23}$.
\end{itemize}
The sensor transmits the mixed ciphertext $E_{\delta'_k} = (0+15i, 12+15i)$ back to the DC.

\paragraph*{Step 6. DC decrypts $\delta'_k = \text{Dec}_a(E_{\delta'_k})$}
The DC uses its private key $a=7$ to decrypt the received ciphertext.
\begin{itemize}
 \item Reconstruct the shared secret $\tau_k \equiv (0+15i){}^7 \pmod{23}$:
 \[\tau_k \equiv 15^7 \cdot i^7 \pmod{23}.\]
 Since $15^7 \equiv 11 \pmod{23}$ and $i^7 = i^4 \cdot i^3 = -i \equiv 22i \pmod{23}$, we get:
 \[\tau_k \equiv 11 \cdot 22i = 242i \equiv 12i \pmod{23}.\]

 \item Calculate $\tau_k ^{-1} \pmod{23}$ by multiplying the numerator and denominator by the positive complex conjugate of $12i$ (which is $-12i \equiv 11i \pmod{23}$):
 \[\tau_k ^{-1} \equiv \frac{11i}{(12i)(11i)} \equiv \frac{11i}{132i^2} \equiv \frac{11i}{-132} \pmod{23}.\]
 Since $-132 = -6 \cdot 23 + 6$, the denominator reduces to $6 \pmod{23}$.
 \[\tau_k ^{-1} \equiv 11i \cdot 6^{-1} \pmod{23}\]
 Applying Fermat's Little Theorem ($x^{-1} \equiv x^{p-2} \pmod{p}$), the inverse of $6 \pmod{23}$ is $6^{21} \equiv 4 \pmod{23}$:
 \[\tau_k ^{-1} \equiv 11i \cdot 4 = 44i \equiv 21i \pmod{23}.\]

 \item Decrypt the scaled plaintext:
 \[\begin{split}
 \delta'_k &\equiv E_{\delta'_k, 2} \cdot \tau_k ^{-1} \equiv (12+15i)(21i) \\ &= 252i + 315i^2 = (-315 + 252i) \pmod{23}.\end{split}\]
 Reducing modulo 23 (where $-315 = -14 \cdot 23 + 7 \equiv 7$ and $252 = 10 \cdot 23 + 22 \equiv 22$), the DC successfully recovers:
 $\delta'_k = 7 + 22i$.
\end{itemize}

\paragraph*{Step 7. DC divides $\delta_k = \delta'_k / \lambda_k$}
The DC removes its complex scaling factor $\lambda_k = 3+2i$ via standard complex division over the Gaussian integers. Because the homomorphic operations preserve the exact algebraic structure without wrap-around, no modular arithmetic is required for this final extraction:
\[
\begin{split}
\delta_k &= \displaystyle\frac{7+22i}{3+2i} = \frac{(7+22i)(3-2i)}{(3+2i)(3-2i)} = \frac{21 - 14i + 66i + 44}{3^2+2^2} \\ &= \frac{65 + 52i}{13}.    
\end{split}
\]
Performing the standard division yields exactly:
$\delta_k = 5 + 4i$.

\paragraph*{Steps 8 \& 9. Verification and Extraction}
The DC separates the real and imaginary components of $\delta_k$:
\begin{itemize}
 \item Verification (Step 8): $w_k = \Im(\delta_k) = 4$ (Matches the expected watermark).
 \item Extraction (Step 9): $d_k = \Re(\delta_k) = 5$ (Recovers the original sensor reading).
\end{itemize}

\subsection{Aggregated scenario: \hidden-AggP protocol flow}\label{Xsec52-B.2}\label{app:sim_aggp}
This simulation traces the specific aggregated protocol flow of the \hidden-AggP scheme (\linkref[Fig.]{\ref{fig:paillier}}). To maintain consistency, this simulation uses the numerical parameters established in the example of \linkref[Section]{\ref{sec:numerical}}, featuring three sensing devices ($N=3$) and a Data Collector (DC).

Because \hidden-AggP separates the real and imaginary components prior to encryption, it relies on standard integer additive homomorphism and does not require Gaussian primes. The DC selects standard primes $p=11$ and $q=13$ ($n=143$) to comfortably accommodate the final aggregated sum without modular wrap-around. The ciphertext space operates modulo $n^2 = 20449$.

\paragraph*{Initialization}
\begin{itemize}
 \item DC generates the paillier keypair $\langle \mathsf{K_{pub}}, \mathsf{K_{priv}} \rangle$:
 \begin{itemize}
 \item $n = 143$, $L_P = \mathrm{lcm}(10, 12) = 60$, generator $g = n+1 = 144$.
 \item To compute $M_P$, the DC first evaluates $L(g^{L_P} \pmod{n^2})$. Using the binomial expansion of $(1+n){}^{L_P}$, this elegantly simplifies to
 \[L(1 + L_P \cdot n) = \frac{(1 + 60 \cdot 143) - 1}{143} = 60.\]
 \item $M_P = 60^{-1} \pmod{143}$. Using the Extended Euclidean Algorithm, $M_P = 31$.
 \item Thus, $\mathsf{K_{pub}} = (143, 144)$ and $\mathsf{K_{priv}} = (60, 31)$.
 \end{itemize}
 \item DC sends $\mathsf{K_{pub}}$ to the sensors $S_j$ ($j \in \{1,2,3\}$).
 \item Each $S_j$ generates a symmetric key $K_{\text{sym}, j}$, encrypts it using $\mathsf{K_{pub}}$, and sends it to the DC.
 \item DC and S agree on a target watermark $w_1 = 4$.
\end{itemize}
\paragraph*{Steps 1--4. Secure transmission of scaling factor}
The DC generates a complex scaling factor $\lambda_1 = 3+2i$ (Step 1). Using $\mathsf{K_{priv}}$, the DC recovers each sensor's symmetric key $K_{\text{sym}, j}$ (Step 2). The DC symmetrically encrypts $E_{\lambda_1} = \text{Enc}_{K_{\text{sym}, j}}(\lambda_1)$ and transmits it (Step 3). Each sensor $S_j$ decrypts the message to securely recover $\lambda_1 = 3+2i$ (Step 4).

\paragraph*{Step 5. S forms $\delta_{j,k} = d_{j,k} + iw_k$}
The three devices obtain their readings ($d_{1,1}=5$, $d_{2,1}=8$, $d_{3,1}=17$) and construct their complex plaintexts:
\begin{itemize}
 \item $S_1: \delta_{1,1} = 5 + 4i$.
 \item $S_2: \delta_{2,1} = 8 + 4i$.
 \item $S_3: \delta_{3,1} = 17 + 4i$.
\end{itemize}

\paragraph*{Step 6. $S_j$ multiplies $\delta'_{j,k} = \lambda_k \delta_{j,k}$}
Each sensor algebraically scales its data using the shared $\lambda_1$:
\begin{itemize}
 \item $S_1: \delta'_{1,1} = (3+2i)(5+4i) = 7 + 22i$.
 \item $S_2: \delta'_{2,1} = (3+2i)(8+4i) = 16 + 28i$.
 \item $S_3: \delta'_{3,1} = (3+2i)(17+4i) = 43 + 46i$.
\end{itemize}

\paragraph*{Step 7. $S_j$ encrypts real and imaginary parts}
Each device encrypts the real ($\Re$) and imaginary ($\Im$) components of its scaled data into two distinct Paillier ciphertexts using $\mathsf{K_{pub}}$ and random ephemeral keys $r_{j,k,R}$ and $r_{j,k,I}$:
\begin{itemize}
 \item $S_1$: $c_{1,1,R} = \enc_{\mathsf{K_{pub}}}(7)$ and $c_{1,1,I} = \enc_{\mathsf{K_{pub}}}(22)$.
 \item $S_2$: $c_{2,1,R} = \enc_{\mathsf{K_{pub}}}(16)$ and $c_{2,1,I} = \enc_{\mathsf{K_{pub}}}(28)$.
 \item $S_3$: $c_{3,1,R} = \enc_{\mathsf{K_{pub}}}(43)$ and $c_{3,1,I} = \enc_{\mathsf{K_{pub}}}(46)$.
\end{itemize}

\paragraph*{Step 8. Joint Paillier aggregation}
An intermediate node (e.g., the sensor $\mathrm{S}_1$) collects the ciphertexts and exploits Paillier's additive homomorphism by multiplying them modulo $n^2$:
\begin{itemize}
 \item $C_{\text{final},R} = \prod_{j=1}^{3} c_{j,1,R} \pmod{20449}$ \quad (Yielding an encryption of $7+16+43=66$).
 \item $C_{\text{final},I} = \prod_{j=1}^{3} c_{j,1,I} \pmod{20449}$ \quad (Yielding an encryption of $22+28+46=96$).
\end{itemize}
The node forwards $(C_{\text{final},R}, C_{\text{final},I})$ to the DC.

\paragraph*{Step 9. DC decrypts using $\mathsf{K_{priv}}$}
The DC receives the aggregated ciphertexts and applies the Paillier decryption algorithm to recover the components:
\begin{itemize}
 \item $\sigma'_{1,R} = \dec_{\mathsf{K_{priv}}}(C_{\text{final},R}) = 66$.
 \item $\sigma'_{1,I} = \dec_{\mathsf{K_{priv}}}(C_{\text{final},I}) = 96$.
\end{itemize}
The DC forms the complex scaled aggregate: $\sigma'_1 = 66 + 96i$.

\paragraph*{Step 10. DC divides $\sigma_k = \sigma'_k / \lambda_k$}
The DC removes the scaling factor via standard complex division:
\[
\begin{split}
\sigma_1 &= \frac{66+96i}{3+2i} = \frac{(66+96i)(3-2i)}{(3+2i)(3-2i)} = \frac{390 + 156i}{13} \\ &= 30 + 12i.    
\end{split}
\]

\paragraph*{Steps 11 \& 12. Verification and Extraction}
\begin{itemize}
 \item Verification (Step 11): The DC verifies the watermark by dividing the imaginary part by $N=3$: $w_1 = \Im(30+12i)/3 = 12/3 = 4$ (Matches the expected watermark).
 \item Extraction (Step 12): The DC extracts the aggregated sensor data from the real part: $\sum d_{j,1} = \Re(30+12i) = 30$ (Correctly matches $5 + 8 + 17$).
\end{itemize}

\bibliographystyle{elsarticle-harv}
\bibliography{ref}

\end{document}